\documentclass{eas}

\usepackage{pdfpages}%[draft]
\usepackage{graphicx}
\usepackage{epstopdf}	
\usepackage[mathcal]{eucal}
\usepackage{amssymb} 
\usepackage{amsmath}
\usepackage{natbib}
\usepackage{hyperref}
\hypersetup{linkcolor=blue,colorlinks=true,citecolor=red}
\usepackage{makecell}
\usepackage{rotating}

\newcommand{\apjl} {ApJ}
\newcommand{\apj} {ApJ}
\newcommand{\apjs} {ApJS}
\newcommand{\mnras} {MNRAS}
\newcommand{\aap} {A\&A}
\newcommand{\pasp} {PASP} 
\newcommand{\pasj} {PASJ}
\newcommand{\solphys}  {Solar Physics}

\newcommand{\araa} {ARA\&A}

\newcommand{\apss} {Ap\&SS}

\newcommand{\aplett} {Astrophys. Lett.}
\newcommand{\nat} {Nature}
\newcommand{\memsai} {MmSAI}

\long\def\jumpover#1{{}}

\newcommand{\derivp} [2] {\frac {\partial #1 } {\partial #2} }
\newcommand{\deriv} [2] {\frac {{\rm d} #1 } {{\rm d} #2} }

\newcommand{\eq}[1] {Eq.~(\ref{#1})}

\newcommand{\eqn} [1] {
\begin{align}#1
\end{align}}
\newcommand{\eqna} [1] {
\begin{eqnarray}#1
\end{eqnarray}}

\def\dLrms{\left(  {{\delta L} / {L }} \right )_{\rm rms}}
\def\teff{T_{\rm eff}}

\def\pmax{ {\cal P}_{\rm max} }
\def\vmax{ {V}_{\rm max} }
\def\dLmax{\left(  {{\delta L} / {L }} \right )_{\rm max}}
\def\numax{ {\nu}_{\rm max} }
\def\pmax{ {\cal P}_{\rm max} }
\def\tauth{{ \tau_{\rm th} }}
\def\tauc{{ \tau_{\rm c} }}
\def\nuc{{ \nu_{\rm c} }}
\def\Eosc{{E_{\rm osc}}}
\def\nablaad{{\nabla_{\rm ad}}}
\def\BV{{Brunt-V\"ais\"al\"a}}
\def\gsim{\hspace{0.3em}\raisebox{0.4ex}{$>$}\hspace{-0.75em}\raisebox{-.7ex}{$\sim$}\hspace{0.3em}}

\begin{document}

%%-----------------------------
%%      the top matter
%%-----------------------------
\title{Stellar oscillations. II The non-adiabatic case}
\author{R. Samadi}\address{ LESIA, Observatoire de Paris, PSL Research University, CNRS, Universit\'e Pierre et Marie Curie,
	Universit\'e Paris Diderot,  92195 Meudon, France}
\author{K. Belkacem$^1$}
\author{T. Sonoi$^1$}

\begin{abstract} 
A leap forward has  been performed due to the space-borne missions, MOST, CoRoT and {\it Kepler}. They provided a wealth of observational data, and more precisely oscillation spectra, which have been (and are still) exploited to infer the internal structure of stars. While an adiabatic approach is often sufficient to get information on the stellar equilibrium structures it is not sufficient to get a full understanding of the physics of the oscillation. Indeed, it does not permit one to answer some fundamental questions about the oscillations, such as:  What are the physical mechanisms responsible for the pulsations inside stars? What determines the amplitudes? To what extent the adiabatic approximation is valid? All these questions can only be addressed by considering the energy exchanges between the oscillations and the surrounding medium. 

This lecture therefore aims at considering the energetical aspects of stellar pulsations with particular emphasis on the driving and damping mechanisms. To this end, the full non-adiabatic equations are introduced and thoroughly discussed. Two types of pulsation are distinguished, namely the self-excited oscillations that result from an instability and the solar-like oscillations that result from a balance between driving and damping by turbulent convection. For each type, the main physical principles are presented and illustrated using recent observations obtained with the ultra-high precision photometry space-borne missions (MOST, CoRoT and {\it Kepler}). 
Finally, we consider in detail the physics of scaling relations, which relates the seismic global indices with the global stellar parameters and gave birth to the development of statistical (or ensemble) asteroseismology. Indeed, several of these relations rely on the same cause: the physics of non-adiabatic oscillations.  
\end{abstract}
\maketitle

\newpage
\tableofcontents
\newpage

\section{Introduction}

\begin{figure}
        \begin{center}
        \resizebox{0.8\hsize}{!}{\includegraphics  {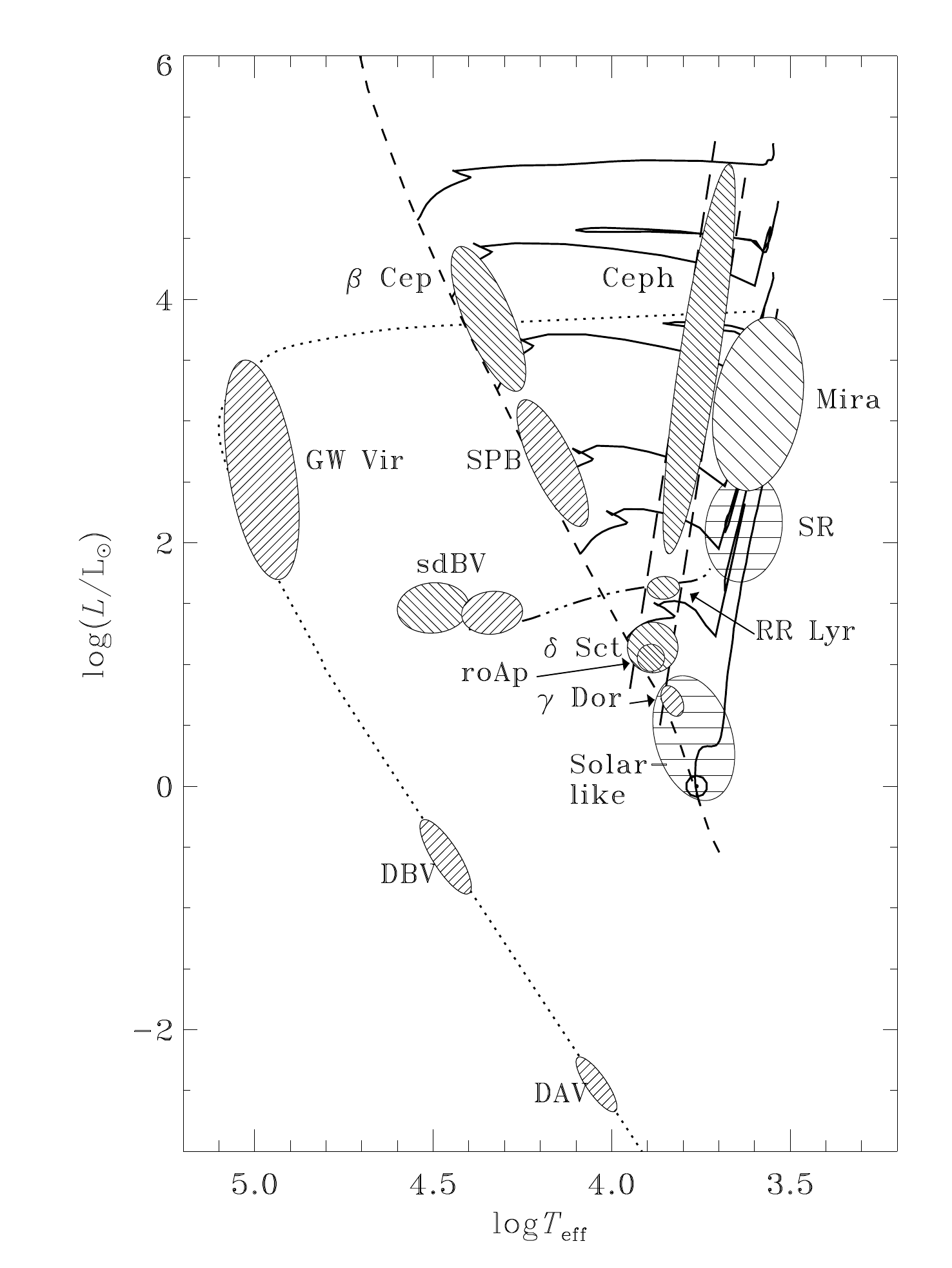}}
        \end{center}    
        \caption{Location of different classes of pulsating stars. Credits: J. Christensen-Dalsgaard.    }
        \label{pulsation-HR}
        \end{figure}

Stellar oscillations are commonly treated in the adiabatic limit, \emph{i.e.} without considering the energy exchanges between the oscillations and the equilibrium medium (for details see Mosser, this volume). This assumption is in general sufficiently accurate to infer the inner structure of stars. Nevertheless, it prevents one from determining if a star pulsates or not and more crucially what are the physical mechanisms at work. A brief look at Fig.~\ref{pulsation-HR} shows that all stars are not pulsating but only stars lying in specific regions of the Hertzsprung-Russel (HR) diagram.  

The first issue is thus to determine what are the mechanisms able to excite modes up to detectable amplitudes. Subsequently, the location of these mechanisms inside the stars must be considered, as well as the way the amplitudes and the lifetimes of the oscillations are setting up. Finally, it is worth determining what can we learn on stellar physics from the non-adiabatic processes. These are --~ among others ~-- the set of fundamental questions that non-adiabatic considerations about stellar pulsations help us to answer.

In this framework, this lecture aims at addressing  energetic aspects of stellar pulsations.  It extends the lecture on adiabatic oscillations (see Mosser, this volume) and assumes that the underlying theoretical backgrounds are mastered.

The lecture is split in four parts: in the first one (Sect.~\ref{Preliminary statements}), we establish under which conditions modes can no longer be treated using the adiabatic approximation and where, inside the star, the departure from adiabaticity becomes important. Finally, we introduce in this section the mode stability criteria, which will enable us to distinguish the two different classes of oscillations. 

Due to their large amplitudes (few milimagnitude up to few magnitude in terms of intensity fluctuations), \emph{unstable} or \emph{self-excited} oscillations were the first to be detected. For instance, Mira, Cepheids, RR~Lyrae and $\delta$~Scuti stars exhibit such a type of pulsations and are often referred to as \emph{classical pulsators}. This type of oscillation is addressed in Sect.~\ref{self-excited} with particular emphasis on the driving mechanisms. The second class of pulsations are the \emph{solar-like} oscillations, which are \emph{stable and stochastically excited} by convection. Historically, they were first  detected in the Sun, but not before the sixties because of their very small amplitudes (few ppm in intensity and few tens of cm/s in velocity). The mechanisms at the origin of their driving and damping, involve complex and subtle coupling between pulsation and turbulent convection. These mechanisms  are addressed in Sect.~\ref{stochastically_excited}.

Since the launch of the space-borne photometry missions  CoRoT and \emph{Kepler}, solar-like oscillations have been observed in a huge number of stars. However, it is not possible to perform a detailed seismic analysis for each star. It motivated the development of \emph{Ensemble Asteroseismology} that consists in extracting in a massive way  seismic indices that characterise at first order their seismic spectra. Among these indices, some of them are related to the mode  amplitude, lifetime and frequency $\numax$ at which the mode height is maximum. They are obviously linked to the energetic aspects of the oscillation and thus to non-adiabatic processes. Observations have permitted to show that these quantities obey characteristic scaling relations that depend on a limited number of global parameters (e.g $\teff$, gravity ...etc). Sect.~\ref{scaling} addresses these non-adiabatic scaling relations so as to explain their origin and to emphasise their potential in the framework of ensemble asteroseismology.

This lecture is largely inspired by the very good books written by \citet{Cox80}, \citet[vol.~2 Chap.~27]{Cox68}, and \cite{Unno89}, where non-adiabatic aspects are addressed in great details and in a very didactic way. We also recommend to read the excellent review by \citet[][]{Gautschy95} \cite[see also][]{Gautschy96}. All these references were, however, written well before the area of the space-borne ultra-high precision photometry missions  MOST, CoRoT and Kepler. Therefore, this lecture will emphasise on  results obtained with these missions concerning non-adiabatic aspects of stellar oscillations.

Finally, some topics related to non-adiabatic aspects such as amplitude limitation and mode selection will not be addressed in this lecture. The reader is referred to the reviews by \citet{Dziembowski93b} and \citet{Smolec14}. For the issue of mode identification we suggest to read M.-A. Dupret's PhD thesis \citep{Dupret02}. 

\section{Preliminary statements}
\label{Preliminary statements}

We first define the set of linearized  equations verified by both adiabatic and non-adiabatic pulsations (Sect.~\ref{non-adiabatic-equations}). 
As we will then see, departure from the adiabatic assumption closely depends on the relative importance of  two relevant time-scales, the modal  period  and the thermal time-scale (see Sect.~\ref{timescales}). These time-scales enable us to identify the regions in the star where departure from adiabaticity are important, hence where mode driving and damping can in principle occur. In those regions, depending on the importance of the driving (w.r.t the damping),  oscillations can  become \emph{unstable}. We will then introduce in Sect.~\ref{Mode stability criteria} a first simple criteria to distinguish unstable modes ({\it i.e.} self-excited modes) from \emph{stable} modes ({e.g.} solar-like modes). Finally,  we perform in Sect.~\ref{The zoo} a very quick overview of the various classes of pulsators. 

\subsection{From adiabatic to non-adiatic oscillations}
\label{non-adiabatic-equations}

As soon as we deal with  oscillations with small amplitudes, we can consider the linearized equations of mass conservation,
\eqn{
{{\delta \rho} \over \rho} = - \vec \nabla . \delta \vec r \, ,
\label{conservation}
}
the momentum equation\footnote{Note that rotation, magnetic field, and molecular viscosity has been neglected.} 
\eqn{
\derivp {^2 \delta \vec r} { t^2} = - \vec  \nabla \psi^\prime - {{\vec \nabla P^\prime} \over {\rho}} + {\rho^\prime \over \rho } \vec \nabla \psi \, ,
\label{momentum}
}
the poisson equation
\eqn{
\nabla^2 \psi^\prime = 4 \pi G \rho^\prime \,,
\label{poisson}
}
where $\rho$ is the density, $P$ is the total pressure, $\delta \vec r$ is the mode displacement, $\psi$ is the gravitational potential, $()^\prime$ refers to Eulerian perturbations, and $\delta$ to Lagrangian ones.
To close the system one has to consider the perturbed equation of state
\eqn{
{ {\delta P} \over P } = \Gamma_1 { {\delta \rho \over \rho }} + {\rho \over P } \left ( \Gamma_3 -1  \right )  T \delta s
\label{thermo}
}
where $s$ is the specific entropy, $\Gamma_1 = \left ( { {d \ln P} \over {d \ln \rho} } \right ) _{\rm ad}$, and $  \Gamma_3 -1 =  \left ( { {d \ln T} \over {d \ln \rho} } \right ) _{\rm ad}$ are the usual adiabatic exponents.  

When there is no energy exchange between the oscillation and the background, the perturbation of the specific entropy vanishes (\emph{i.e.},  $\delta s = 0$), so that \eq{thermo} simplifies to 
\eqn{
{ {\delta P} \over P} = \Gamma_1 { {\delta \rho} \over \rho } \, .
\label{adiabatic}
}
Complemented with boundary conditions,  Eqs.~(\ref{conservation})-(\ref{poisson}) together with \eq{adiabatic} correspond to a 4th-order eigenvalue problem. The solutions are the well-known adiabatic eigenmodes $\vec \xi (r)$ and real eigenfrequencies $\omega$. The corresponding adiabatic mode displacement is expressed as
\eqn{
\delta \vec r = {1 \over 2} \left ( \vec \xi(r) \,   e^{- i \omega t} + c.c. \right ) \, ,
\label{displacement}
}
where $c.c.$ denotes complex conjugate.

Equation.~(\ref{adiabatic}) applies as soon as the modes do not exchange energy with the medium over one pulsation cycle. This is obviously not the case  everywhere inside the star because a mode must be excited and thus energy exchanges with the background must occur.  Nevertheless, we will see in the next section that \eq{adiabatic} holds almost everywhere inside the star and this is a sufficient approximation to derive the mode eigenfrequencies. 

In the general case, however, one must consider the energy equation
\eqn{
T \derivp{ \delta s } {t} = \delta \left ( \epsilon - \deriv{L}{m} \right ) \, ,
\label{energy}
}
where $\epsilon$ is the rate of production of thermonuclear energy, $L$ is the star luminosity at a radius $r$, $m$ is the mass enclosed in a sphere of radius $r$. 
Complemented with boundary conditions,  Eq.~(\ref{conservation})-(\ref{poisson}) together with the energy equation \eq{energy} correspond to an eigenvalue problem whose solutions are the complex eigenmodes of the form
\eqn{
\vec \xi (r,t) = \vec \xi(r) e^{- i \omega t } \, e^{\gamma t} \,,
\label{displacement_damping}
}
where $\omega$ is the oscillation frequency (real) and $\gamma$ is the growth or damping rate.

\subsection{Relevant time-scales}
\label{timescales}

We now introduce two time-scales that will allow us to distinguish between the regions where modes propagate almost adiabatically and the regions where they efficiently exchange energy with the background (and hence can be damped or excited).  

The first relevant time-scale is the modal period $\Pi$. In the asymptotic  regime, for high radial orders, it can be shown that $\Pi$ scales approximately as
\eqn{
\Pi \propto 2 \int_0^R { {{\rm d} r} \over c_s} \, ,
}
where $c_s$ is the sound speed and $R$ the star radius. In that case, $\Pi$  is approximately the time for an acoustic wave for crossing the stellar diameter. Assuming that the stellar stratification can be treated using homology relations \citep[e.g][]{Cox68,Kippenhahn90}, one  shows  that $\Pi$ finally scales as \citep[see, {e.g.},][and references therein]{Kevin12c}
\eqn{
\Pi \propto \left (  {{G M} \over {R^3}}  \right)^{-1/2} \, ,
\label{GMoR3}
}
where $M$ is the total mass of the star, $R$ the total radius, and $G$ the gravitational constant. % , and the term $ {{ M} \over {R^3}} $ is nothing else than the mean stellar density. 

\begin{figure}
        \begin{center}
          \resizebox{0.75\hsize}{!}{\includegraphics  {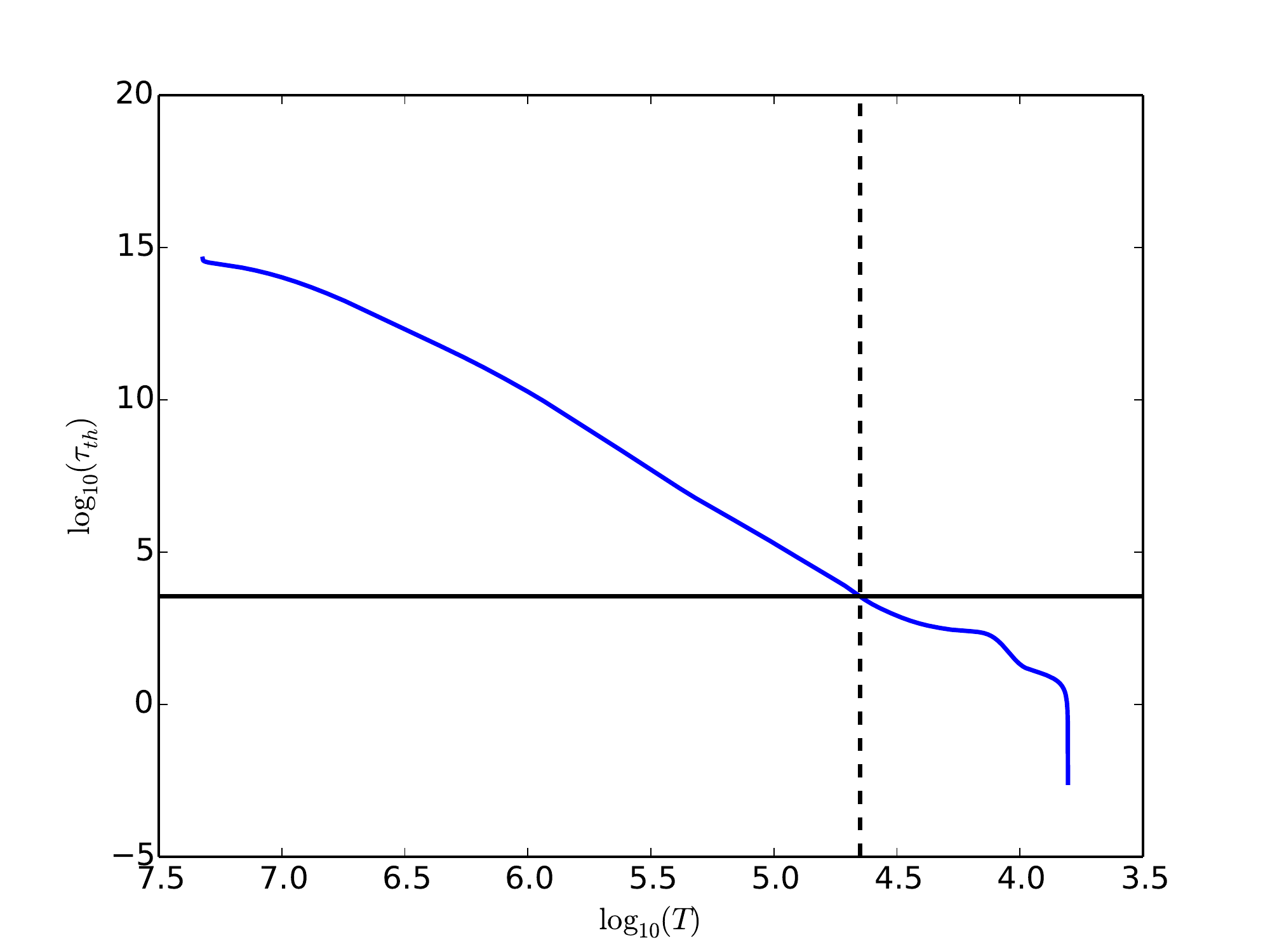}}
          \resizebox{0.75\hsize}{!}{\includegraphics  {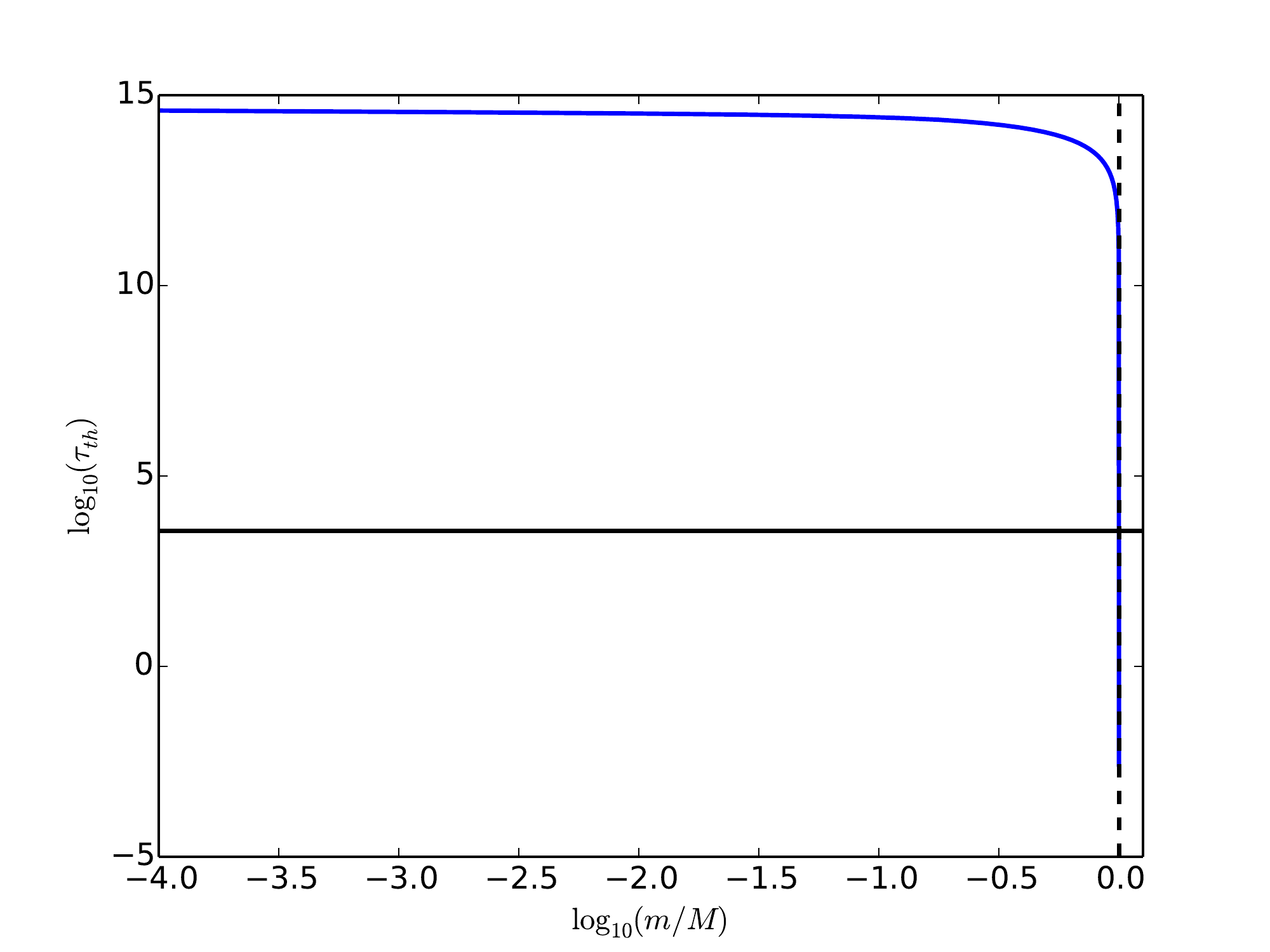}}
        \end{center}    
        \caption{{\bf Top:} $\log_{10} (\tauth)$  (in seconds) as a function of $\log_{10} (T)$ (in K) for an A-type main sequence stellar model. The horizontal black line corresponds to a period $\Pi$ of one hour and the vertical dashed line represents the layer where $\tauth = \Pi$. {\bf Bottom:} $\log_{10} (\tauth)$  as a function of $\log_{10} (m/M)$.}
        \label{fig:tauth_vs_logT}
        \end{figure}

The second time-scale is the so-called thermal-time scale, which represents the characteristic time over which a given shell  loses its energy. For sake of simplicity, we consider a region wihtout nuclear reactions so that the perturbation of the production rate of thermonuclear energy vanishes. Consequently, 
%We thus consider the energy equation neglecting perturbation of the production rate of thermonuclear energy, that is
\eqn{
T \derivp{ \delta s } {t} = - \delta \derivp{L}{m} \, .
\label{energy2}
}
The time derivative in the LHS of \eq{energy2} permits us to identify dimensionally the thermal time-scale $\tauth$ as  
\eqn{
{ { T \Delta s} \over {\tauth} } \approx  { {c_v  \Delta T} \over {\tauth} } \approx { L \over {\Delta m }} \, ,
\label{tau_th}
}
where $c_v$ is the heat capacity at constant volume, $\Delta m$ is the mass of a given shell,   $\Delta s$ the variation of entropy due to the oscillation within that mass shell and  $\Delta T$ the corresponding temperature variation. According to \eq{tau_th}, the thermal  time-scale $\tauth$ can be roughly approximated by the ratio between the heat capacity of the mass shell and the energy  lost per unit time by this shell, that is
\eqn{
 {\tauth} =  { {\rm Heat ~ capacity} \over {\rm Energy~lost~by~unit~time } } \, ,
\label{tau_th2}
}
leading to the following estimate of $  {\tauth}$
\eqn{
 {\tauth} = { { c_v  T \Delta m } \over  {L } }  \, .
\label{tau_th3}
}
The derivation of \eq{tau_th} from \eq{energy2} is not valid when the mass shell $\Delta m$ is large, which is the case when one considers the inner layers of the star. A more general definition is then  to consider the thermal time-scale at a given layer of mass $m$ given by the following integral form \citep[for further details, {e.g.},][]{Dziembowski81,Cox80,Pesnell86}
\eqn{
{\tauth} (m)  = \int_0^m  { {c_v T} \over {L } } \, {\rm d} m  \, . 
\label{tau_th4}
}
Figure~\ref{fig:tauth_vs_logT} illustrates the variation of ${\tauth}  $ in an A-type main-sequence star. 
In the region where $\tauth \gg \Pi$, the mode has no time to loose or gain energy during a pulsation cycle. In those regions, the pulsation is quasi-adiabatic. In the surface layers, we have $\tauth \ll \Pi$, the mode has time to gain or loose energy and the departure from the adiabatic  assumption \eq{adiabatic} becomes important. 
Note that the layer where $\tauth \approx \Pi$ is named the \emph{transition region}. It delimits the inner layers where the pulsations are adiabatic and the upper layers where they are strongly non-adiabatic. 

It is clear from  Fig.~\ref{fig:tauth_vs_logT} that in the major part of the star $\tauth $ is much higher than the typical  periods of acoustic modes. The region where $\tauth \gtrsim \Pi$ represents the overwhelming majority of the star mass. In other words, in the major part of the star, oscillations can be treated in the adiabatic limit.

For adiabatic pulsation, the variationnal principle rigorously holds and one can show that the frequency $\omega_{\rm ad}= 2 \pi / \Pi_{\rm ad}$  verifies the following relation derived from the variational principle \citep[see, {e.g.},][]{Unno89}
\eqn{
 \omega_{\rm ad}^2 = \int_0^M \vec \xi_{\rm ad}^* . \vec {\cal L}_{\rm ad} \left  (\vec \xi_{\rm ad}  \right ) \, {\rm d}m \,,
\label{variational}
} 
where $\omega_{\rm ad}$ is the pulsation frequency, $\vec \xi_{\rm ad}$ the adiabatic eigenmode and $\vec {\cal L}_{\rm ad}$ the adiabatic wave operator.
The integral in the RHS of \eq{variational} can be split into an integral over the adiabatic regions where $\tauth > \Pi_{\rm ad }$ and an integral over the outer layers where $\tauth < \Pi_{\rm ad }$, such that
\eqn{
\omega_{\rm ad}^2 =  \int_0^{m_t} \vec \xi_{\rm ad}^* . \vec {\cal L}_{\rm ad} \left  (\vec \xi_{\rm ad}  \right ) \, dm +  \int_{m_t}^{M} \vec \xi_{\rm ad}^* . \vec {\cal L}_{\rm ad} \left  (\vec \xi_{\rm ad}  \right ) \, dm  \, ,
}
where $m_t$ is the mass below the transition region, {\it i.e.} below the region where $\tauth = \Pi_{\rm ad }$.
Since the non-adiabatic layers represent a very tiny fraction of the stellar mass, we have in good approximation
\eqn{
\omega_{\rm ad}^2  \simeq  \int_0^{m_t} \vec \xi_{\rm ad}^* . \vec {\cal L}_{\rm ad} \left  (\vec \xi_{\rm ad}  \right ) \, dm \; .
}
This tells us that the  frequency $\omega_{\rm ad}$ of an adiabatic pulsation is mostly fixed by the properties of the inner layers where modes behave  adiabatic. The non-adiabatic layers hardly influence the frequency. 
In other words, mode frequencies obtained in the adiabatic approximation reflects the properties of the  overwhelming majority of the star.

\subsection{Mode stability criteria}
\label{Mode stability criteria}

As already stressed, in the regions where $\tauth \lesssim \Pi$ ({\it i.e.} in the non-adiabatic regions), the mode  has time to have --~ in principle  ~-- a net gain or loss of energy during a pulsation cycle. While a loss damps the mode,  a gain obviously drives it.
In some layers inside the star driving dominates while in some others the damping dominates.   However, only the net result matters for the stability of the mode.  To determine if the net result is a gain (or a loss), we define the growth rate as
\eqn{
\gamma = { { \left < \deriv{W} {t} \right>_{\rm cycle}}  \over  { \left <   \Eosc \right>_{\rm cycle} } } \, ,
\label{growth_rate}
}
where ${\rm d}W / {\rm d}t$ is the instantaneous power supplied  to or released from the mode ({\it i.e.} the work received or provided by the mode per unit of time), $\Eosc$ is the total energy of the mode, and  $\left <  ~~ \right>_{\rm cycle}  $ represents an average over a pulsation cycle.
Hence, the mode is dominated by driving for  $\gamma > 0$, or  by damping for $\gamma < 0$. 
In a linear regime, we can show that the mode displacement will grow (or decay) as given by \eq{displacement_damping}. 
If $\gamma > 0$,  $\tau = 1 / \gamma$ is the growth time and if $\gamma < 0$, $\eta = -\gamma$ is named the damping rate with  $\tau = 1 / \eta$  the mode lifetime  (or the e-folding time). 

A mode with $\gamma > 0$ will be \emph{unstable} because its amplitude will grow until the linear approximation is no longer valid. Some complex non-linear mechanisms will limit its amplitude \citep[see, {e.g.},][]{Dziembowski93b,Smolec14}. These modes are considered to be \emph{self-excited}  because as we will see later (see Sect.~\ref{self-excited}) the driving operates as a response to the oscillation itself. They are characteristic of the so-called classical pulsators (Mira, Cepheids, RR~Lyrae and $\delta$~Scuti stars) and similar classes of pulsators identified more recently ({e.g.} $\beta$~Ceph, SPB, $\gamma$~Doradus ...). 
On the contrary, a mode  with $\gamma < 0$ will be \emph{stable}.  Solar-like oscillations are \emph{stable} because their amplitudes are a balance between driving and damping. Their excitation is due to  a stochastic driving mechanism that occurs on a very different time-scale than the damping mechanisms. These \emph{stochastically-excited} pulsations will be extensively treated in  Sect.~\ref{stochastically_excited}.

\subsection{The zoo of pulstating stars}
\label{The zoo}

There are various types of pulsating stars, which differ from each other not only in the nature of their oscillations ({e.g.} $p$ modes versus $g$ modes) and in the range of excited pulsation periods but also in the origin of their driving mechanism.  
Figure~\ref{pulsation-HR} shows the location of various classes of pulsating stars in the HR diagram. Their main characteristics are summarized in Table~\ref{tab-pulsating-stars}. For an extensive review about the different  classes of pulsating stars, please refer to \citet{Gautschy96}. 

\begin{sidewaystable}%[anticlockwise]
\begin{center}
\begin{tabular}{c|cccc}
Class name &  \makecell{Type\\of pulsation} & \makecell{Period or\\frequency range} & \makecell{Excitation~mechanism} & \makecell{First~discovery~and comments}  \\\hline \hline
Mira & $p$ modes & $\gtrsim $ 80~days & $\kappa$-mechanism & by the priest David Fabricius in 1596 \\
Cepheids &  $p$ modes & 2-50 ~ days & $\kappa$-mechanism & by Pigot  in 1784 \\
RR Lyrae &  $p$ modes & 0.2 - 1 ~days & $\kappa$-mechanism & by Fleming in 1899\\
$\delta$ Scuti &  $p$ modes & 30~mn - 1 ~ days &  $\kappa$-mechanism & by Wright in 1900, originally named ``dwarf Cepheids''\\
$\gamma$~Doradus &  $g$ modes & 0.3- 3~days & Convective flux blocking &  \makecell{by \citet{Cousins63},\\ identified as a new class by \citet{Balona94}}\\
Red-giant stars & $p$ modes &  1 - 200$\mu$ Hz &  Stochastically-excited & by \citet{Frandsen02} in $\xi$~Hay\\
Solar-like pulsators &  $p$ modes & 0.5- 5 mHz &  Stochastically-excited & \makecell{detected first in the Sun by  \citet{Leighton62},\\ latter in Procyon by \citet{Martic99}}\\
$\beta$~Cephei & $p$ modes & 2 - 7~hours & $\kappa$-mechanism &  by \citet{Frost1902}\\
\makecell{Slowly Pulsating\\B (SPB)} &  $g$ modes & 1 - 4 days&  $\kappa$-mechanism   & by \citet{Smith77}\\
\makecell{Subdwarf B stars\\ - EC 14026} &  $p$ modes & 40 - 400~s &  $\kappa$-mechanism   & by \citet{Kilkenny97}\\
\makecell{Subdwarf B stars\\ - ``Betsy'' stars} &  $g$ modes & 30' - 150'~s &  $\kappa$-mechanism   & by \citet{Green03}\\
\makecell{White dwarfs\\ - DOV} &  $g$ modes &  400 - 1~000~s &  $\kappa$-mechanism   & by \citet{McGraw79}\\
\makecell{White dwarfs\\ - DBV} &  $g$ modes & 140 - 1~000~s & $\kappa$-mechanism   & by \citet{Winget82}\\
\makecell{White dwarfs\\ - DAV} &  $g$ modes & 100 - 1~000~s & Convective driving  &  \makecell{by \citet{Winget82},\\originally named the ZZ Ceti stars}
\end{tabular}
\end{center}
\caption{Main characteristics of the different classes of pulsating stars. }
\label{tab-pulsating-stars}
\end{sidewaystable}

\section{Self-excited oscillations}
\label{self-excited}
\subsection{Work integral approach}

The growth rate can be determined by solving the non-adiabatic equations for (linear) pulsations, \emph{i.e.} the set of Eq.~(\ref{conservation})-(\ref{thermo}) together with the equation of energy, \eq{energy}.  
However, the stability criteria is more easily determined on the basis of the \emph{work integral} approach, which permits us to easily highlight the driving and damping mechanisms. It must, however, be noted that this approach is only rigorously valid when the departure from adiabatic pulsation is weak, which is the case for most of the pulsators (except the case of strange modes which will be addressed in section~\ref{strange_modes}).

According to \eq{growth_rate},  $\gamma$  is by definition directly proportional to the power supplied to or released from the mode by some external forces  during a cycle. 
The \emph{work integral} approach then consists in calculating the time derivative of the work performed by external forces on the oscillation, \emph{i.e.} 
\eqn{
\deriv{W}{t} = {\rm Forces} \times {\rm Mode~velocity} \, .
}
The only forces considered here are the gravity and the pressure gradient. For the sake of simplicity we restrict ourselves to the radial modes.  Accordingly, we have
\eqn{
\deriv{W}{t} = \int_0^M \left(  - \frac{G m}{r^2}  - 4 \pi r^2 \derivp{P } {m} \right ) \,\dot{r}  \, {\rm d}m  \, , 
\label{work_integral}
}
where the first term in the RHS is the gravity and the second term the pressure gradient, and $ \dot{r} = \derivp{r}{t}$ is the mode velocity. 
The second term in the integral is integrated by parts, it gives
\eqn{
- \int_0^M 4 \pi r^2 \derivp{P } {m}  \dot{r} \, {\rm d}m  = - \left [ 4 \pi r^2 \dot{r}\,   P \right ]_0^M + \int_0^M P \derivp{4 \pi r^2 \dot{r}} {m} \, {\rm d} m \,
\label{work_integral-interm1}
}
The first term vanishes at the center and at the surface provided that $P \rightarrow 0$ at the surface.
From the mass conservation equation, \eq{conservation}, we have the relation
\eqn{
\derivp{}{t} \frac{1}{\rho} = \derivp{}{t} \left [  \derivp{4 \pi r^3/3} {m}\right ] = \derivp{4 \pi r^2 \dot{r}}{m} \, .
\label{work_integral-interm2}
}
Finally, integrating the first integral in \eq{work_integral} and using \eq{work_integral-interm1} and \eq{work_integral-interm2},   one has
\eqn{
\deriv{W}{t} = \deriv{}{t} \left [  - \int_0^M \frac{Gm}{r} dm \right ] + \int_0^M P \derivp{}{t} \left ( \frac{1}{\rho}\right ) \, dm  \, . 
\label{work_integral2}
}
We are interested in  the average of $dW/dt$ over a puslation cycle, 
\eqn{
\left <  \deriv{W}{t} \right > = \frac{1}{\Pi} \oint  \left (  \deriv{W}{t}  \right )  {\rm d} t \, .
}
Integrated over a cycle, the first term in the RHS of \eq{work_integral2} vanishes, \eq{work_integral} thus becomes 
\eqn{
\left <  \deriv{W}{t} \right > = \frac{1}{\Pi} \oint  {\rm d}t \, \int_0^M P \derivp{}{t} \left ( \frac{1}{\rho}\right ) \, {\rm d}m  \, . 
\label{work_integral3}
}
The integrated quantity $ \derivp{}{t} \left ( \frac{1}{\rho}\right )$ represents the ``PdV~work''.  Consequently, 
for a given infinitesimal mass shell ${\rm d}m$, one can distinguish two cases:
\begin{itemize}
\item $\oint P d V >0$  : the mode has a \emph{net gain} of energy during the cycle, the layer is a \emph{driving} layer; 
\item  $\oint P d V <0$ : the mode has a \emph{net loss} of energy during the cycle, the layer is a \emph{damping} layer. 
\end{itemize}

\begin{figure}
        \begin{center}
        \resizebox{10cm}{!}{\includegraphics  {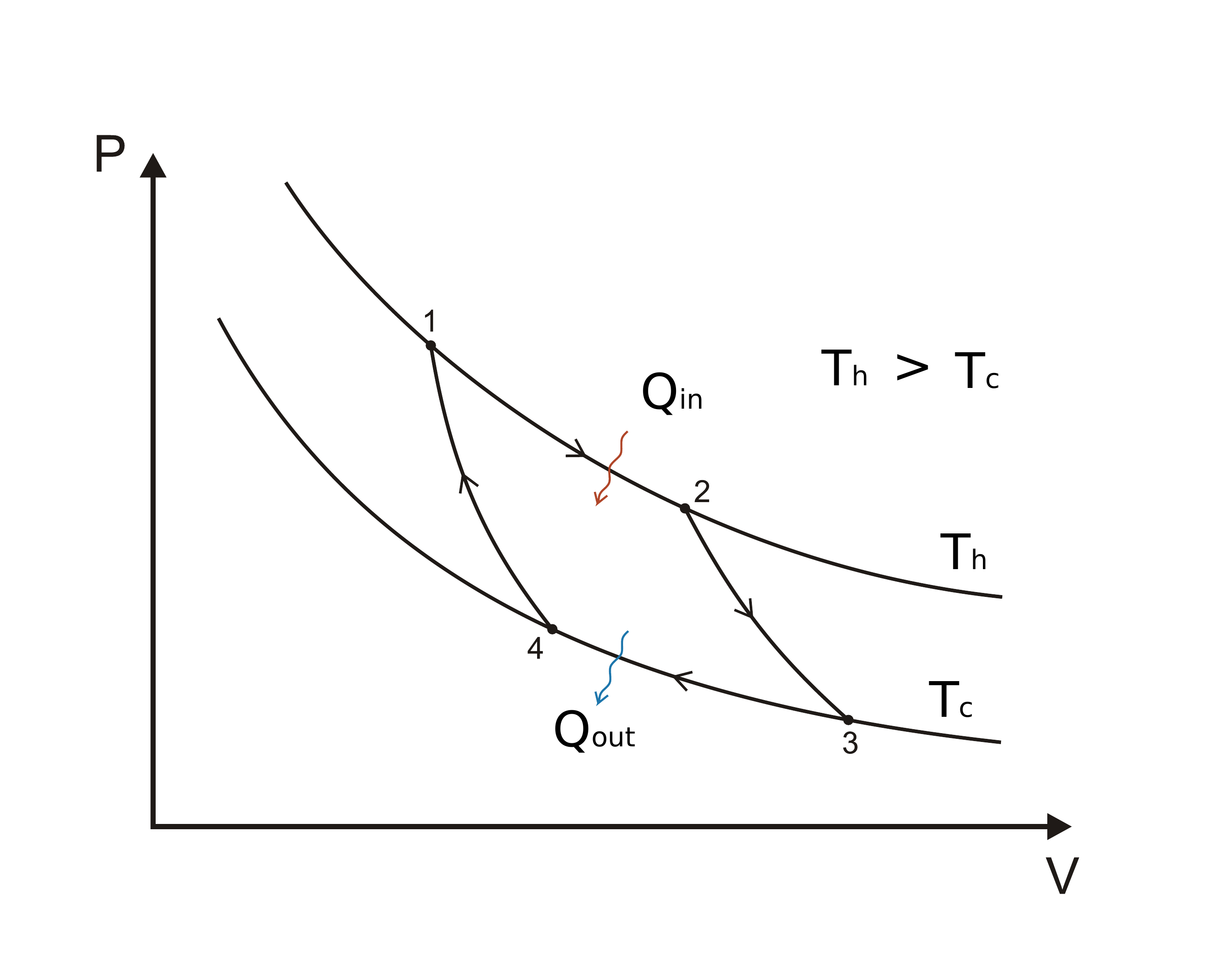}}
        \end{center}    
        \caption{Carnot cycle on PV diagram. Step 1 to 2 corresponds to an isothermal expansion at hot temperature phase, step 2-3 to an adiabatic expansion, step 3-4 to an isothermal contraction at cold temperature, and step 4-1 to an adiabatic contraction. Figure reproduced from Wikipedia \url{http://en.wikipedia.org/wiki/Carnot_cycle}}
        \label{carnot-cycle}
        \end{figure}

Mode driving in stars corresponds  to a Carnot-type heat-engine mechanism, which is illustrated in Fig.~\ref{carnot-cycle}. The pulsation cycle can be decomposed in four steps:
\begin{itemize}
\item the isothermal expansion (step 1 to 2): the  heat $Q_{\rm in}$ is received from the medium at the hot temperature ($T_h$);
\item adiabatic expansion  (step 2 to 3): adiabatic cooling of the gas;
\item the isothermal compression (step 3 to 4): the heat $Q_{\rm out}$  is released to the medium at the cold temperature ($T_c$);
\item adiabatic compression  (step 4 to 1): adiabatic heating of the gas; 
\end{itemize}
The integral $\oint P {\rm d} V$ represents  the area of the P~dV work over the cycle. We have an effective driving ({\it i.e.} $\oint P {\rm d} V>0$) if $Q_{\rm in} > \left | Q_{\rm out} \right |$.

\citet{Eddington1926} was the first to suggest such a type of mechanism, which he named the ``valve mechanism'', as a possible explanation for the mode driving in Cepheids. Since his pioneer work the question was  where in the star this valve mechanism operates. 

\begin{figure}
        \begin{center}
        \resizebox{0.7\hsize}{!}{\includegraphics  {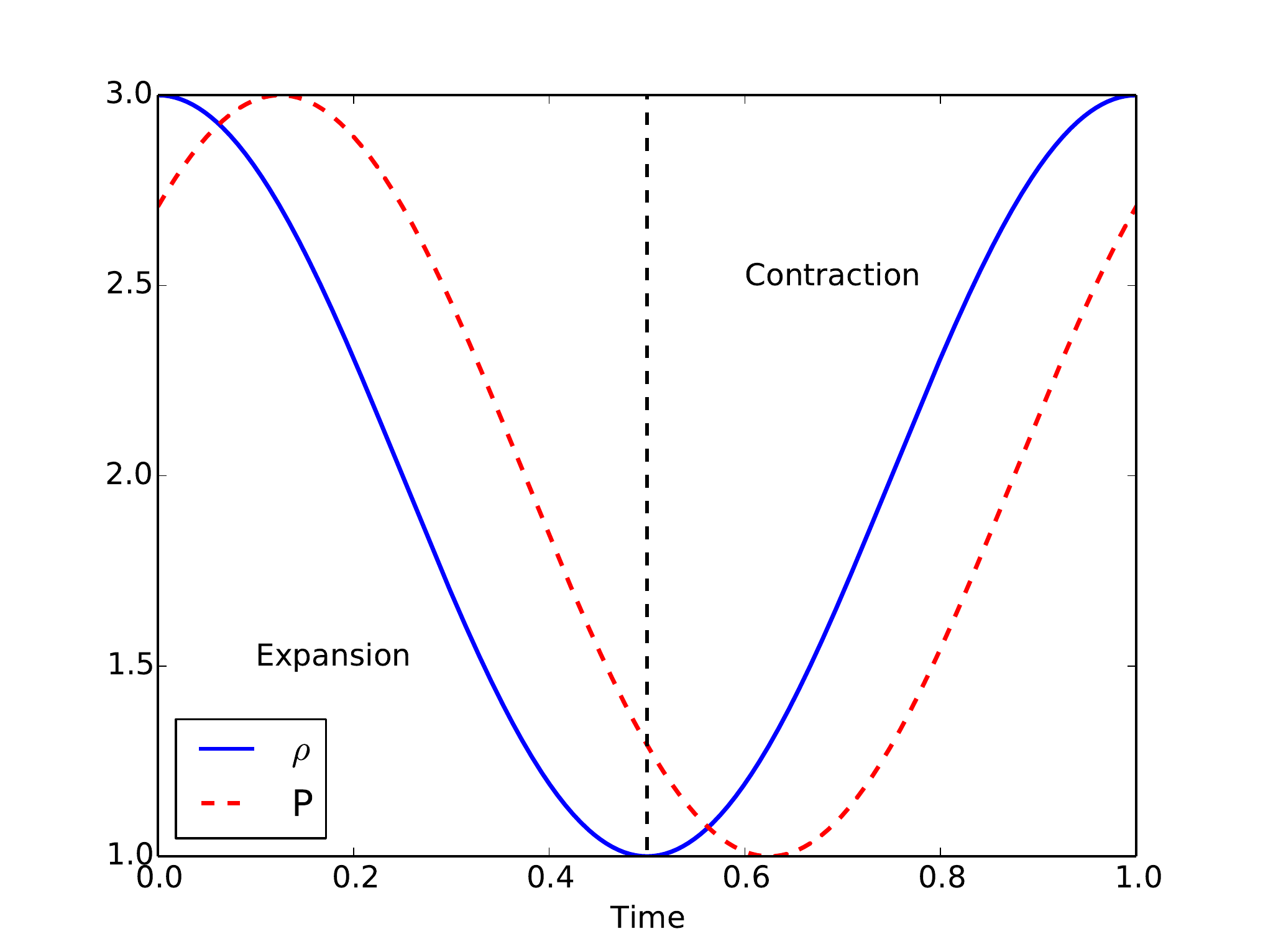}}
        \resizebox{0.7\hsize}{!}{\includegraphics  {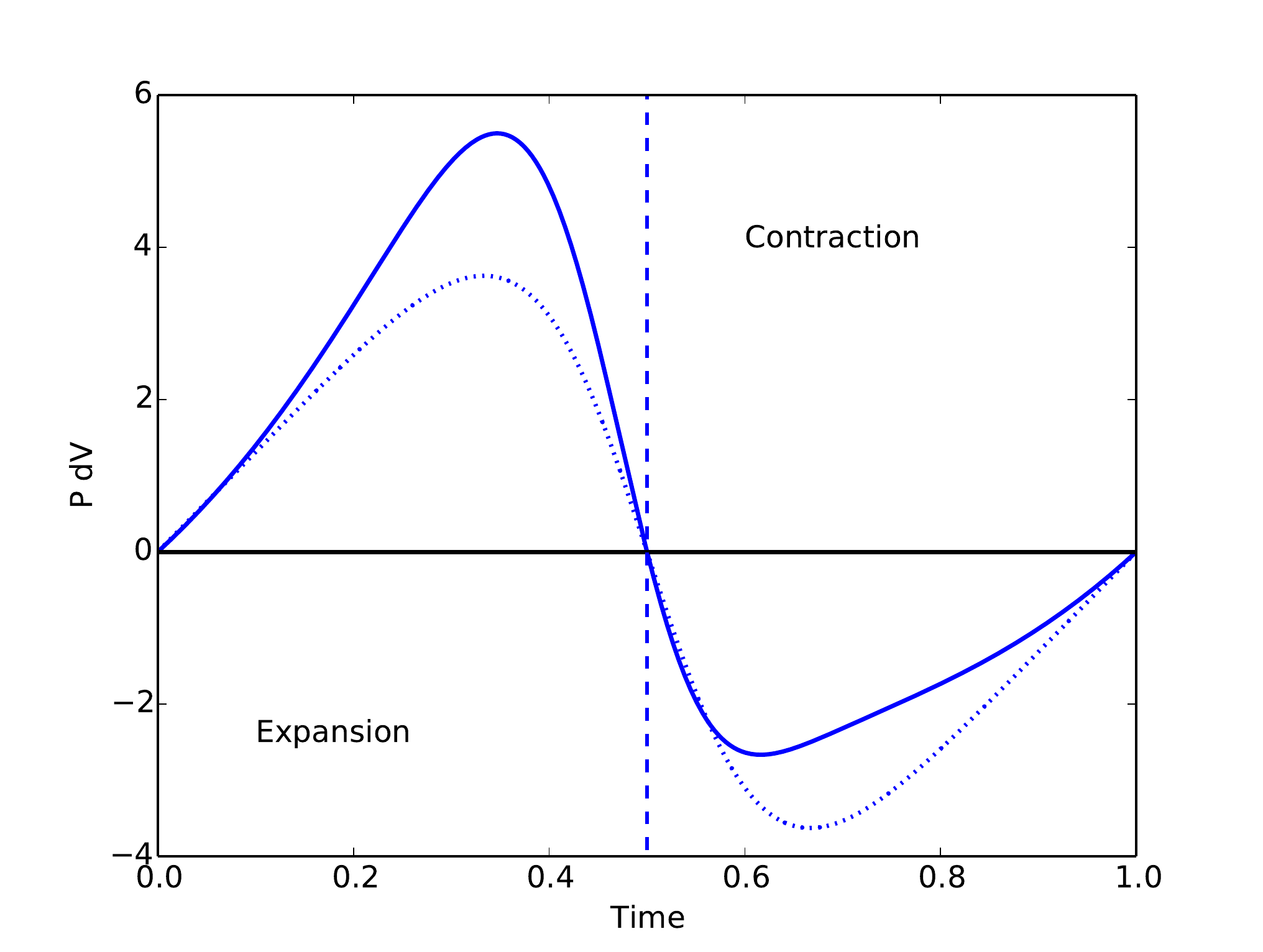}}
        \end{center}    
        \caption{{\bf Top:} Density (blue solid line) and pressure (red dashed line) as  functions of time. {\bf Bottom:} $ P d V$ as a function of time.  The dotted line corresponds to a purely adiabatic transformation.  }
        \label{phase-lag}
        \end{figure}

It is also useful to view the driving and damping as the result of a phase-lag between pressure fluctuations and density fluctuations. 
To do this, we consider the energy equation, \eq{energy}, which we recast as 
\eqn{
\derivp{\ln P}{t} = \Gamma_1 \derivp{\ln \rho} {t} + \frac{\rho}{P} \left ( \Gamma_3 - 1 \right )  T \derivp{s}{t} \, .
}
At maximum compression ({\it i.e.} at the high temperature), we have $\derivp{\ln \rho }{t} = 0$ and accordingly 
\eqn{
\derivp{\ln P}{t} = \frac{\rho}{P} \left ( \Gamma_3 - 1 \right )  T \derivp{s}{t} \, .
}
Now, if $\derivp{s}{t} >0$ at maximum compression,  the pressure maximum occurs \emph{after} the maximum density. There is a \emph{positive phase-lag} between pressure fluctuations and density fluctuations, which results in a net gain of energy (positive $\oint P d V$). This is illustrated in Fig.~\ref{phase-lag}. 
For a purely adiabatic pulsation, $\derivp{s}{t} =0$, and consequently there is no phase-lag between $\rho$ and $P$, hence no net work.

\subsection{Driving criteria}
\label{driving_criteria}

Given the general expression of the work integral, \eq{work_integral3}, we derive an expression for the  growth-rate assuming now linear pulsation (small perturbations) and a weak departure from adiabatic pulsation.
We start from the definition of the growth rate, \eq{growth_rate}.  
The mode energy (kinetic +potential) averaged over a cycle can be written as
\eqn{
\left < E_{\rm osc} (t) \right > _{\rm cycle} = \int_0^M \left <  \left \|  \derivp{} {t} \delta \vec r (t)\right \|^2 \right > _{\rm cycle} \, {\rm d}m \, .
\label{E_osc}
}
With the help of \eq{displacement} and assuming $\gamma \ll \omega$ ({\it i.e.} weak departure from adiabatic pulsation), we can  recast \eq{E_osc}    as 
\eqn{
\left <  E_{\rm osc} (t) \right > _{\rm cycle} = \omega^2 \, \int_0^M   \left <  \left \|   \delta \vec r (t) \right \|^2 \right > _{\rm cycle} \, {\rm d}m = \omega^2 \, I \, ,
\label{E_osc2}
}
where $I = \int_0^M   \left \|    \vec \xi  (r) \right \|^2  \, {\rm d}m $  is the mode inertia.
The variation of the mode energy by definition equals the time derivative of  $W$ and we can show that
\eqn{
 \left <  \deriv{}{t} W \right > _{\rm cycle} = \left <  \deriv{}{t} E_{\rm osc}  \right > _{\rm cycle} = 2 \, \gamma \, \left <  E_{\rm osc}  \right > _{\rm cycle} 
\label{dE_osc_dt}
 }
 Using \eq{dE_osc_dt} and \eq{E_osc2}  we obtain
\eqn{
\gamma = \frac{1}{2} \frac{ \left <  \deriv{}{t} E_{\rm osc}  \right > _{\rm cycle}     } { \left <  E_{\rm osc}  \right > _{\rm cycle} } =  \frac{1}{2} \frac{   \left <  \deriv{}{t} W \right > _{\rm cycle}  } {\omega^2 \, I } \, .
 }
This equation shows that the growth rate is inversely proportional  to the square of mode frequency and the mode inertia, $I$.  

We go back now to our general expression for the work integral, \eq{work_integral3}. Since we consider linear pulsations, we have to perturb the work integral. 
We first consider the perturbed version of the energy equation, \eq{energy}, that is 
\eqn{
\derivp{}{t} \left (   \frac{\delta P} {P}\right ) = \Gamma_1 \derivp{}{t} \left (   \frac{\delta \rho} {\rho}\right ) + \left ( \Gamma_3 - 1 \right) \, \frac{\rho} {P} \left ( \delta \epsilon - \deriv{}{m} \delta L\right ) \, .
\label{energy_linear}
}
 Substituting \eq{energy_linear} into \eq{work_integral3} and keeping terms up to the second order, yields\footnote{First-order terms vanish. For more details, see \citet[][vol.~2]{Cox68}}
 \eqn{
 \gamma = \frac{1}{2 \omega^2 \, I \, \Pi} \int_0^\Pi dt \, \int_0^M \, \left (\Gamma_3-1 \right ) \, \frac{\delta \rho}{\rho} \left ( \delta \epsilon - \deriv{}{m} \delta L \right ) \, dm\  \, . 
 \label{growth_rate2}
 }
Assuming now complex eigenfunctions of the form $\delta X = \delta X_0 \, e^{- i \omega t + \gamma t}$ where $X$ is a given perturbed quantity, permits us to recast \eq{growth_rate2} as
\eqn{
\gamma = \frac{1}{2 \omega^2 \, I } {\cal R}_{\rm e} \left [ \int_0^M \, \left (\Gamma_3-1 \right ) \,  \left ( \frac{\delta \rho}{\rho} \right )^* \, \left ( \delta \epsilon - \deriv{}{m} \delta L\right )     \right ] \, {\rm d}m\, , 
 \label{growth_rate3}
}
where ${\cal R}_{\rm e}$ stands for the real part of a complex quantity and the symbol $*$ refers to the complex conjugate.  

A mode is unstable (driven) when $\gamma >0$. Hence according to \eq{growth_rate3}, at maximum compression ({\it i.e.} when $\delta \rho / \rho$ is maximum and positive) driving occurs in a given shell of mass ${\rm d}m$ if 
\eqn{
 \delta \epsilon - \deriv{}{m} \delta L  > 0 \, .
\label{eq_driving_criteria}
}
The sign of $\gamma$ finally results from a balance between driving and damping regions. 
Hence, to have an effective driving the condition given by \eq{eq_driving_criteria} is not sufficient and one has to consider three additional requirements \citep[see also][]{Pamyatnykh99}:
\begin{itemize}
\item  eigenmode with large amplitude in the driving region ;
\item slowly varying eigenmodes in order to avoid cancellation effects ;  
\item matching of the mode period with the thermal-time scale in the driving region, {\it i.e} $\Pi \approx \tauth$. 
\end{itemize}

The last condition deserves some explanations. It is obvious that driving is inefficient in the adiabatic region, {\it i.e.} in the region where  $\Pi \ll \tauth$ (see Sect.~\ref{timescales}). It turns out to be  also inefficient in the strongly non-adiabatic region, {\it i.e.} where  $\tauth \ll \Pi $. Indeed, neglecting the rate of production of thermonuclear energy, one has  the relation
\eqn{
\deriv{\delta L}{m}  = - T \deriv{\delta s}{t}  \, .
}
When  $\tauth \ll \Pi$, the medium adapts instantaneously to any perturbation such that $\deriv {}{t} \delta s \approx 0$ during a  pulsation cycle. In such condition we have a ``freezing'' of the flux variation, which results in flattening of the luminosity perturbation, {\it i.e.} $ \deriv{}{m} \delta L \approx 0$.
As a consequence, driving (and damping) can only be efficient in the transition region,  {\it i.e.}  where  $\Pi \approx \tauth$. 

{To determine on the basis of \eq{growth_rate3} the stability of the mode, it is in general  qualitatively sufficient to consider the adiabatic eigenfunctions (quasi-adiabatic approach).  Indeed, departure from adiabatic oscillation is weak for most pulsations because $\gamma \ll \omega $, or equivalently the mode e-folding  time, $\gamma^{-1}$,  is much longer than the mode period, $\Pi$.

\subsection{The excitation mechanisms}

Equation (\ref{growth_rate3}) immediately highlights  the possible driving role of $\epsilon$ (which is named  $\epsilon$-mechanism). 
To have more insights into the other excitation mechanisms, we first decompose the luminosity perturbation, $ \delta L$, as   $ \delta L = \delta L_{\rm r} + \delta L_{\rm c}$ where $ L_r$ is the radiation component of the luminosity and $L_c$ is the convective one. 
Now, using the diffusion approximation\footnote{This approximation is  valid in optically thick layers only.}
\eqn{
L_{\rm r} = 4 \pi r^2 F_{\rm r}  = \frac{16 \, \pi \, a \, c \, r^2 \, T^3} {3 \,  \rho \, \kappa}  \nabla T \, ,
\label{diffusion_approx}
}
one can write $ \delta L_r $ as 
\eqn{
\frac{\delta L_r} {L_r} =  \deriv{r} {\ln T} \, \deriv{}{r} \, \left ( \frac{\delta T}{T} \right ) - \frac{\delta \kappa} {\kappa} + 4  \frac{\delta T}{T} + 4 \frac{\delta R}{R}  \, , 
\label{delta_r}
}
where $\kappa$ is the Rosseland mean opacity, $F_{\rm r}$ is the radiative flux, $c$ is the speed of light, and $a$ is the radiation constant.
The first term in the RHS of \eq{delta_r} leads to damping, the second one is responsible to the  $\kappa$-mechanism, the third one to the $\gamma$-mechanism, and finally the last one to the  $r$-radius effect. The latter is linked to an increase of the star surface during expansion \citep{Baker66} and is shown  to be always negligible \citep[see, {e.g.},][and references therein]{Pamyatnykh99}.   
The perturbation of the convective component of the luminosity, $\delta L_c$ results in two characteristic excitation mechanisms: the  convective flux blocking mechanism and the convective driving\footnote{Not to be confused with the stochastic excitation by turbulent convection, which will be addressed in Sect.~\ref{mode_driving}}.
We will overview all these mechanisms except the  $r$-radius effect. 

\subsubsection{The $\epsilon$-mechanism}
\label{epsilon-mechanism}

The rate of production of thermonuclear energy, $\epsilon$,  is highly sensitive to the temperature. Indeed, $\epsilon$ can be approximated by $\epsilon \propto \rho T^\nu$ where the exponent $\nu$ depends on the nuclear reaction chain: $\nu \approx 4$ for the $pp$ chain and $\nu \approx 15$  for the CNO chain \citep[see, {e.g.},][]{HansenKawaler1994}. $\epsilon$ increases with increasing temperature such that $\delta \epsilon$ is always positive at maximum compression. As a consequence, the  $\epsilon$-mechanism, which acts in the burning region, is always a driving mechanism. It was originally proposed by \citet{Eddington1926} as the main driving agent for Cepheid pulsators.

An apparent difficulty is that the $pp$ and CNO chains have a very long time-scale (evolution time-scale) much longer than any  modal periods. 
Nevertheless, some intermediate reactions have a time-scale of a few hours, which matches the period of some  gravity or acoustic modes. However, acoustic modes have in general very small amplitude in the burning regions (generally located in the core or deep in the interior of the stars). 
Following \citet{Cox74}, we will establish here the criteria that must be verified in order to have an efficient $\epsilon$-mechanism for acoustic modes. 

The  $\epsilon$-mechanism operates if it counterbalances the damping, which is in general due to radiation (see Sect.~\ref{radiative_damping}). In other words, one must have
\eqn{
\frac{\rm \epsilon~mechanism}{\rm radiative~damping} \gtrsim 1 \, .
}
Using the  equation of mass conservation, \eq{conservation},  it can easily be shown that \citep[see][]{Cox74}
\eqn{
\frac{\rm \epsilon~mechanism}{\rm radiative~damping}  \propto \frac{ \left ( \delta \rho/ \rho\right )_c^2}{ \left (  \delta \rho/ \rho \right )_s^2} \approx \frac{\xi_c^2}{\xi_s^2}\, ,
}
where the subscript $c$ refers to the core and $s$ to the surface, and $\xi$ is the mode displacement. 
As a rough order of magnitude, one can show that \citep[see][Vol.~2, Chap.~27]{Cox68}
\eqn{
\frac{\xi_c}{\xi_s} \approx \frac{\rho_c}{\left < \rho \right >} \, ,
}
where $\left < \rho \right >$ is the mean density.  Accordingly, we have 
\eqn{
\frac{\rm \epsilon~mechanism}{\rm radiative~damping}  \propto \frac{\xi_c^2}{\xi_s^2}   \approx  \frac{\rho_c^2}{\left < \rho \right >^2} \, .
\label{ratio-emechanism-damping}
}
In view of \eq{ratio-emechanism-damping} excitation of acoustic modes by the $\epsilon$-mechanism can only be efficient for either compact objects or very massive stars. For instance, a fully radiative star can be very roughly described by a polytrope of index  $n=3$ and for such a polytrope  we have ${\xi_c^2}/ {\xi_s^2} \sim 1/400$.  
On the other hand fully convective stars can be roughly described by a polytrope of index  $n=1.5$, which  leads to  ${\xi_c^2}/ {\xi_s^2} \sim 1$. Such fully convective objects are more  compact than fully radiative stars and the $\epsilon$-mechanism can in principle operate. 
In very massive stars radiation pressure generally dominates over the gas pressure. In that case $\Gamma_1 \rightarrow 4/3$ and we can show that $\xi$ slowly varies inside the stars \citep[see][Vol.~2, Chap.~27]{Cox68} such that the ratio ${\xi_c^2}/ {\xi_s^2} $ is also of the order of unity.  Accordingly, excitation by the $\epsilon$-mechanism may also operate in very massive stars. It was actually early shown by \citet{Ledoux41} that this excitation is only possible for a stellar mass above $M \simeq 100~M_\odot$.   This threshold was modified to a higher value, 121$M_\odot$ by \cite{Stothers92} with the new opacity table released by \cite{Rogers92}.

Concerning the fully convective stars, \cite{Palla05} found that the radial fundamental mode is excited by the $\varepsilon$-mechanism due to the central $^2$D burning in brown dwarfs. \cite{RL12, RL14} showed that low-degree low-order $g$-modes are also excited by the non-equilibrium $^3$He burning. Although there is up to now no  bona fide detection of a pulsation signal caused by this mechanism, many authors have made observational efforts for brown dwarfs \citep[{e.g.}][]{Marconi07, Baran11, Cody14, RL15}. The $^2$D burning also excites low-degree low-order $g$-modes in the pre-main sequence stage of a $1.5M_\odot$ star \citep{Lenain06}. In these cases, the high value of $\rho_c/\langle\rho\rangle$ activates the $\varepsilon$-mechanism as discussed above.

Such situation appears also in metal-poor low-mass main-sequence stars. \cite{Sonoi12} found that the instability of the low-degree low-order $g$-modes due to the non-equilibrium $^3$He burning is induced in a wider range of stellar mass as the metallicity decreases. In this case, the ratio $\rho_c/\langle\rho\rangle$ decreases with decreasing the metallicity, since the star becomes compact as the opacity decreases. In addition, less contribution of the CNO-cycle makes the convective core smaller, and hence helps gravity waves propagate in the central region where the nuclear burning is taking place. This  $g$-mode instability due to $^3$He burning was originally discussed in connection with the solar neutrino problem, and nonadiabatic analyses were carried out for solar-like main-sequence stars with the solar metallicity \citep{Dilke72, Boury73, CD74, Boury75, Shibahashi75, Noels76}. 

\cite{Shibahashi76} suggested the possible excitation of $g$-modes due to the $\varepsilon$-mechanism at the H-burning shell in post-main sequence massive stars. Recently, \cite{Moravveji12} proposed that excitation of a $g$-mode appearing in a B supergiant, Rigel, observed by MOST could be explained by this mechanism. Theoretical works have suggested that pre-white dwarfs also have a possibility to exhibit pulsations excited by the $\varepsilon$-mechanism at the He-burning shell \citep{Kawaler86, Gautschy97}. An observed pre-white dwarf VV47 was also found to exhibit short pulsation periods ($\sim$ 130--300 s). \cite{GP06} and \cite{Corsico09} speculated that such pulsations could be excited by the $\varepsilon$-mechanism. On the other hand, \cite{Maeda14} found the excitation at the H-burning shell in models with relatively thick H envelopes. 

\subsubsection{The $\gamma$-mechanism}

The regions of partial ionisation are characterised by lower values of the adiabatic exponents. This is explained by the  increase of the number of degrees of freedoms  in that region. As a consequence, most of the (adiabatic) compression goes into ionisation energy rather than into kinetic energy of thermal motion \citep[see {e.g.}][volume 1]{Cox68}.  As an illustration, we have plotted in Fig.~\ref{gamma3-1} the quantity $\left (\Gamma_3 -1 \right )$ as a function of  $\log_{10} (T)$ for an A-type main-sequence stellar model. The regions of partial ionisation of HeII, HeI and H are clearly characterised by lower values of $\left (\Gamma_3 -1 \right )$. 
As an immediate consequence, adiabatic variations of temperature are lower at maximum compression, since 
\eqn{
\left ( \frac{\delta T}{T} \right )_{\rm ad} = \left (\Gamma_3 -1 \right ) \, \frac{\delta \rho}{\rho}  \, .
}
It then follows from the diffusion approximation, \eq{diffusion_approx}, that $\delta L_{\rm r}$ is locally decreased during compression in the region of partial inonisation. As said by \citet{Cox80}, radiation is locally ``dammed up''. 
As a consequence, $ \deriv{}{m} \delta L_{\rm r} < 0$ (resp. $ \deriv{}{m} \delta L > 0$) in the inner (resp. outer) limit of the partial inonisation regions.  
Accordingly, the inner (resp. outer) limit is potentially a driving (damping resp.) region. 
The effectiveness of the driving by the $\gamma$-mechanism will ultimately  depends on the location of the transition region with respect to the location of the regions of partial inonisation. 
Historically, \citet{Eddington41}  suggested that the driving of Cepheid pulsation is finally caused by the partial ionisation of hydrogen and not, as initially believed, by the $\epsilon$ mechanism (see Sect.~\ref{epsilon-mechanism}).  It was, however, latter shown by \citet{Zhevakin53} that this driving mainly occurs in the region of  He$^ {+}~\rightarrow$~He$^{++}$ dissociation.

 \begin{figure}
        \begin{center}
        \resizebox{10cm}{!}{\includegraphics  {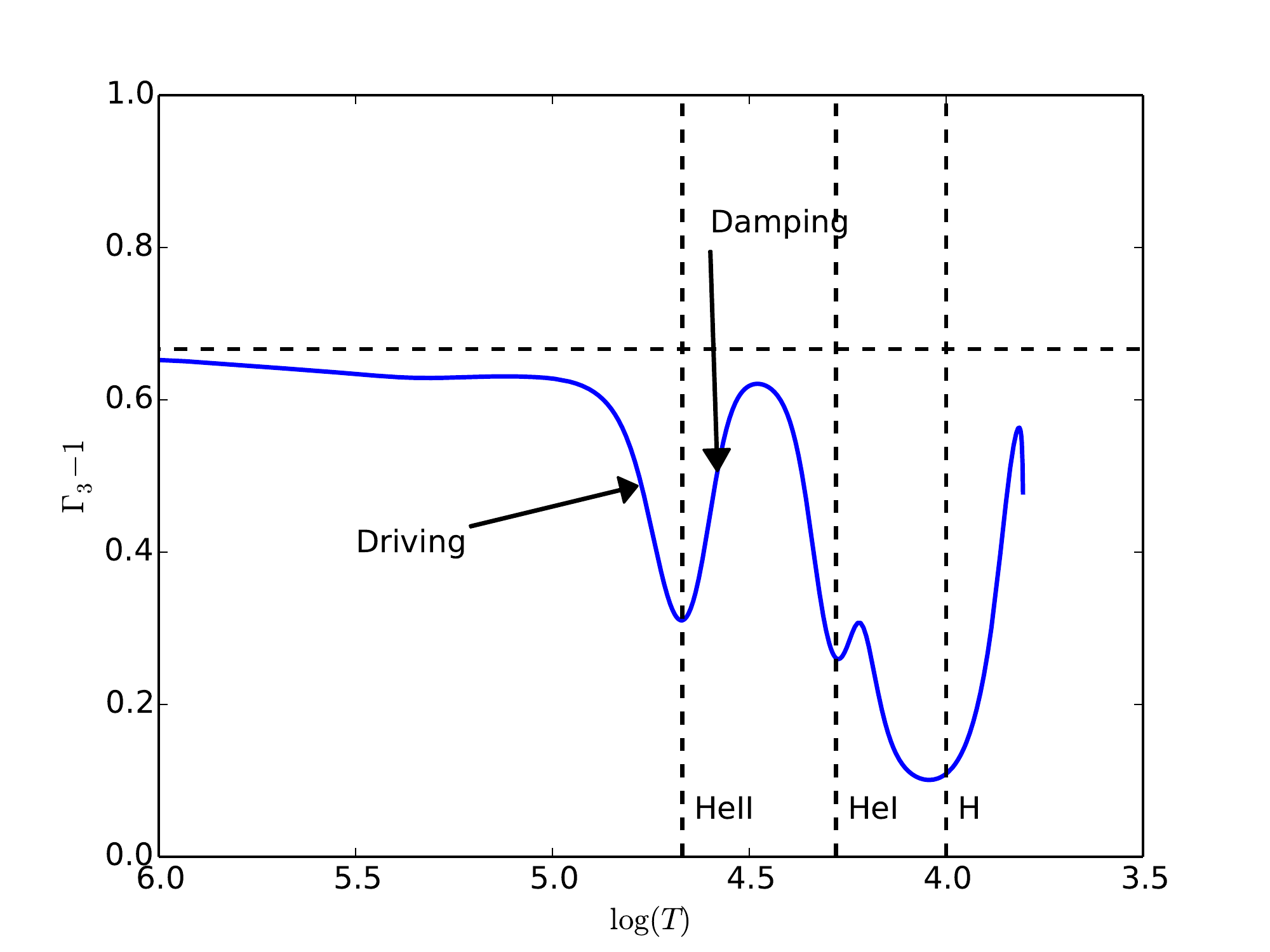}}
        \end{center}    
        \caption{$(\Gamma_3 -1)$ as a function of $\log_{10} (T)$ for an A-type main-sequence stellar model. The vertical dashed lines represent from left to right the regions where  HeII, HeI and H are ionised by an amount of 50\,\%.}
        \label{gamma3-1}
        \end{figure}

\subsubsection{The $\kappa$-mechanism}

\label{kappa_mechanism}

The variation of opacity induced by the oscillation, $\delta \kappa$, plays a role  on the mode stability through  the second term in the RHS of \eq{delta_r}. This term can potentially drive the mode when $\deriv{} {m} \left ( \delta \kappa / \kappa  \right ) > 0$. As we will see, this can happen  due to an ``abnormal'' behaviour of the stellar opacities with the temperature  in some particular regions of the star. 
To highlight this,  we decompose the Lagrangian variation of opacity, $\delta \kappa$ as\footnote{As noticed by \citet{Dupret02}, the choice of $ {\delta P}/{P}$, instead of ${\delta T }/{T}$ and ${\delta \rho}/{\rho}$,  is motivated by the fact that it is always a smooth eigenfunction, not much affected by the opacity bumps, partial ionisation zones, and convection zones.  }
\eqn{
\frac{\delta \kappa}{\kappa} = {\cal \kappa}_T \frac{\delta T }{T} + {\cal \kappa}_\rho \frac{\delta \rho}{\rho} ={\cal \kappa}_P  \, \frac{\delta P}{P} \, ,
\label{delta_kappa}
}   
where we have defined
\eqn{
{\cal \kappa}_T  = \left ( \deriv{\ln \kappa}{\ln T} \right )_\rho \, , \, {\cal \kappa}_\rho = \left ( \deriv{\ln \kappa}{\ln \rho} \right )_T \, ,
}
and 
\eqn{
{\cal \kappa}_P  = \left( \frac{\left ( \Gamma_3 -1 \right) {\cal \kappa}_T  +  {\cal \kappa}_\rho  }{\Gamma_1} \right ) \, .
\label{kappa_P}
}
In ``normal'' conditions, the opacity $\kappa$ decreases under compression ($\kappa_P <0$). Furthermore, in most part of the star, $\kappa$ varies slowly such that $\deriv{{\cal \kappa}_P} {m}  \approx 0 $. Finally, at maximum compression, $\delta P/P$ increases outward. Therefore, in ``normal'' conditions, $\deriv{} {m} \left ( \delta \kappa / \kappa  \right ) <0$ and  the variation of opacity  mainly contributes to damping. Indeed, the medium  transparency  generally decreases during compression leading to a heat leakage, and hence to the mode damping.

Nevertheless, in regions of partial ionisation, the opacity presents some ``bumps''. The existence of theses ``bumps'' are illustrated in Fig.  \ref{kappa_vs_T_rho} for the hydrogen, helium and iron group elements. In the inner part of the  partial ionisation regions, $\mathcal{K}_P$  (Eq.~\ref{kappa_P}) increases sharply outwards. This is illustrated in Fig.~\ref{fig_kappa_P} for an  A-type main-sequence stellar model. 
According to \eq{delta_kappa} and since at maximum compression $\delta P/P$ increases outward,  $\deriv{}{m} \left (  \frac{\delta \kappa}{\kappa} \right )$ is then positive and high in those regions. These regions are then potentially driving regions. However, the driving will be effective only if the inner limit of these opacity  bumps coincides with the transition region.  
This driving mechanism, which is commonly named as the $\kappa$-mechanism,  is further enhanced by the $\gamma$-mechanism.  The two mechanisms are actually linked together, such that strictly speaking one should refer to the $\kappa/\gamma$-mechanism.  
To finish, it is important to note that this  mechanism is responsible for the excitation of the majority of the self-excited oscillations (Cepheids, $\delta$ Scuti, $\beta$ Ceph, ...).  For the classical pulsators (Cepheids, RR~Lyrae, $\delta$-Scuti) this driving mechanism is located in the region of  He$^{+}~\rightarrow$~He$^{++}$ dissociation, while for other pulsators (e.g $\beta$ Ceph, SPB, DOV and DAV) it takes place in the partial ionisation regions of other chemical species.

 \begin{figure}
        \begin{center}
        	        \resizebox{10cm}{!}{\includegraphics  {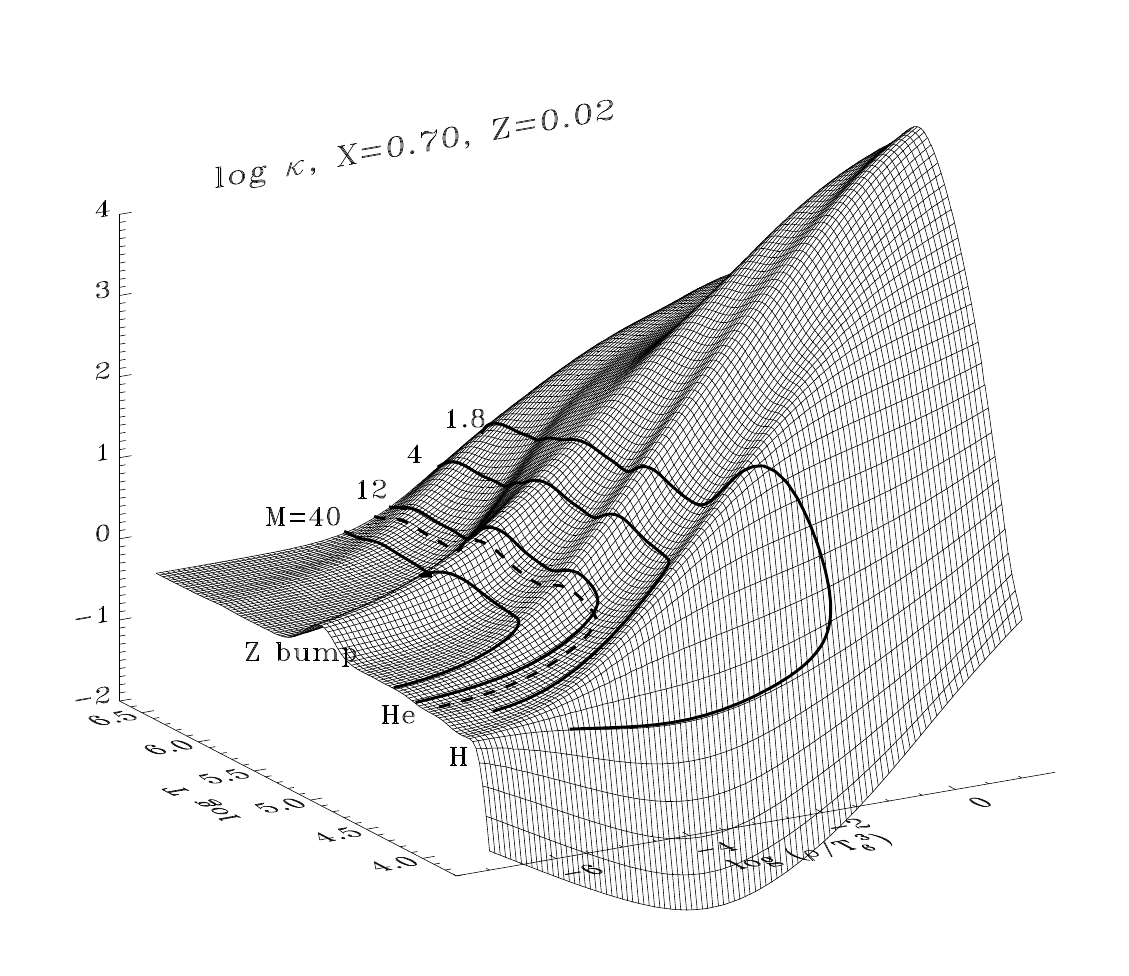}}
        \end{center}    
        \caption{Opacity versus $\log_{10}(T)$ and $\log_{10}(\rho/T_6^3)$ where $T_6 = T/10^6$.  The thick lines correspond to  Zero-Age Main Sequence models between $M= 1.8~M_\odot$ and $M = 40~M_\odot$, the dashed line to a $12~M_\odot$ model on the Terminal-Age Main Sequence. Figure reproduced from \citet{Pamyatnykh99}. Opacity data are from the OPAL project \citep{Iglesias96}.}
        \label{kappa_vs_T_rho}
        \end{figure}

 \begin{figure}
        \begin{center}
        \resizebox{10cm}{!}{\includegraphics  {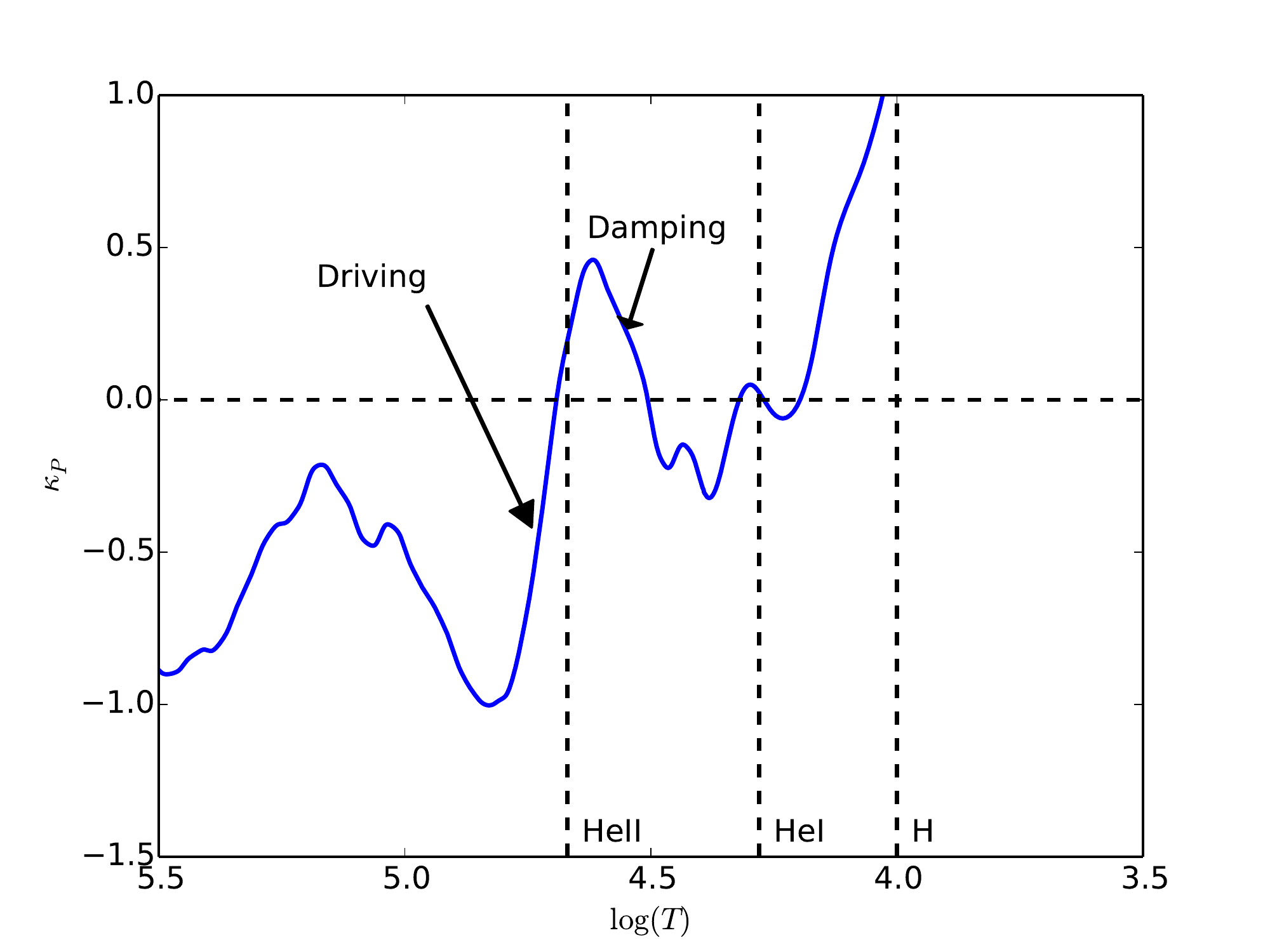}}
        \end{center}    
        \caption{Coefficient $\kappa_P$ as a function of $\log_{10}(T)$ for an A-type main-sequence stellar model. The vertical dashed lines have the same meaning as  in Fig.~\ref{gamma3-1}.}
        \label{fig_kappa_P}
        \end{figure}

\subsubsection{Convective flux blocking mechanism}

In the presence of convection, the Lagrangian variation of the total luminosity includes the convective component, {\it i.e.}  $ \delta L = \delta L_r + \delta L_c$. 
We remind that the luminosity is related to the flux of energy as $L = 4 \pi \, r^2 \, F_{\rm tot}$ where $F_{\rm tot}~ =~F_{\rm r}~+~ F_{\rm c}$ is the total flux, $ F_{\rm r} $  the radiative flux, and $ F_{\rm c} $ the convective flux\footnote{We have neglected other contributions to the flux, such as the kinetic energy, which are not relevant in that context.}. 
Above  the bottom of the upper convective zone (BCZ hereafter), $F_{\rm r}$ decreases sharply outwards as a consequence of the rapid increases of the convective flux. This  is illustrated in Fig.~\ref{Frad_Fconv_tauc} for a  $M=1.40~M_\odot$ main-sequence stellar model.  

The gradient of the radiative component of the luminosity, $\deriv{}{m} \delta L_{\rm r}$, can be decomposed as
\eqn{
\deriv{\delta L_{\rm r}}{m}  = L_{\rm r} \deriv{}{m} \left ( \frac{\delta L_{\rm r} } {L_{\rm r}}  \right ) + \deriv{L_{\rm r}}{m}  \, \frac{\delta L_{\rm r}}{L_{\rm r}} \, .
\label{delta_Lrad_decomposed}
}
It can be shown that the first term in the RHS of \eq{delta_Lrad_decomposed} is negligible w.r.t. the second term.  Accordingly, \eq{delta_Lrad_decomposed} simplifies to
\eqn{
\deriv{\delta L_{\rm r}}{m}  \simeq  4 \pi r^2  \, \left ( \deriv{F_{\rm r} }{m} \right ) \, \frac{\delta L_{\rm r}}{L_{\rm r}}  \, .
}
The diffusion approximation \eq{diffusion_approx} applies in this region, accordingly, 
the term  ${\delta L_{\rm r}}/{L_{\rm r}}$ is given by \eq{delta_r} and it can be shown that  $\delta L_{\rm r} > 0 $ during the compression. Since $\deriv{F_{\rm r} }{m} <0$  in the vicinity of the upper limit of the BCZ, we have  $ \deriv{}{m} \left (  \delta L_{\rm r} \right ) <0$ in that region. As a consequence, the sharp decrease of $F_{\rm r}$    drives the modes for which the transition region coincides with the BCZ. 

 Nevertheless since $ \delta L = \delta L_r + \delta L_c$, the corresponding  sharp increase of  $F_{\rm c}$  counterbalances in  principle this driving. However, in the vicinity of the  BCZ the convective time-scale is much longer than the modal period. This is depicted in  Fig.~\ref{Frad_Fconv_tauc} where we have plotted the product $\omega \tau_c$ where $\tau_{\rm c}$ is the convective time-scale and  $\omega = 2 \pi/\Pi$ for a period $\Pi=0.2$~day representative of the gravity modes  in $\gamma$-Doradus stars. 
In this region $\omega  \tau_{\rm c} \gg 1$, that is convection is too slow to effectively counterbalance the destabilisation effect of the radiative flux, leading to an effective driving by the radiative flux. This driving mechanism is named the \emph{convective flux blocking} mechanism because the radiative flux variation entering into the convective zone has no time to be transported by convection and is injected into the mode. 

 \begin{figure}
        \begin{center}
        \resizebox{10cm}{!}{\includegraphics  {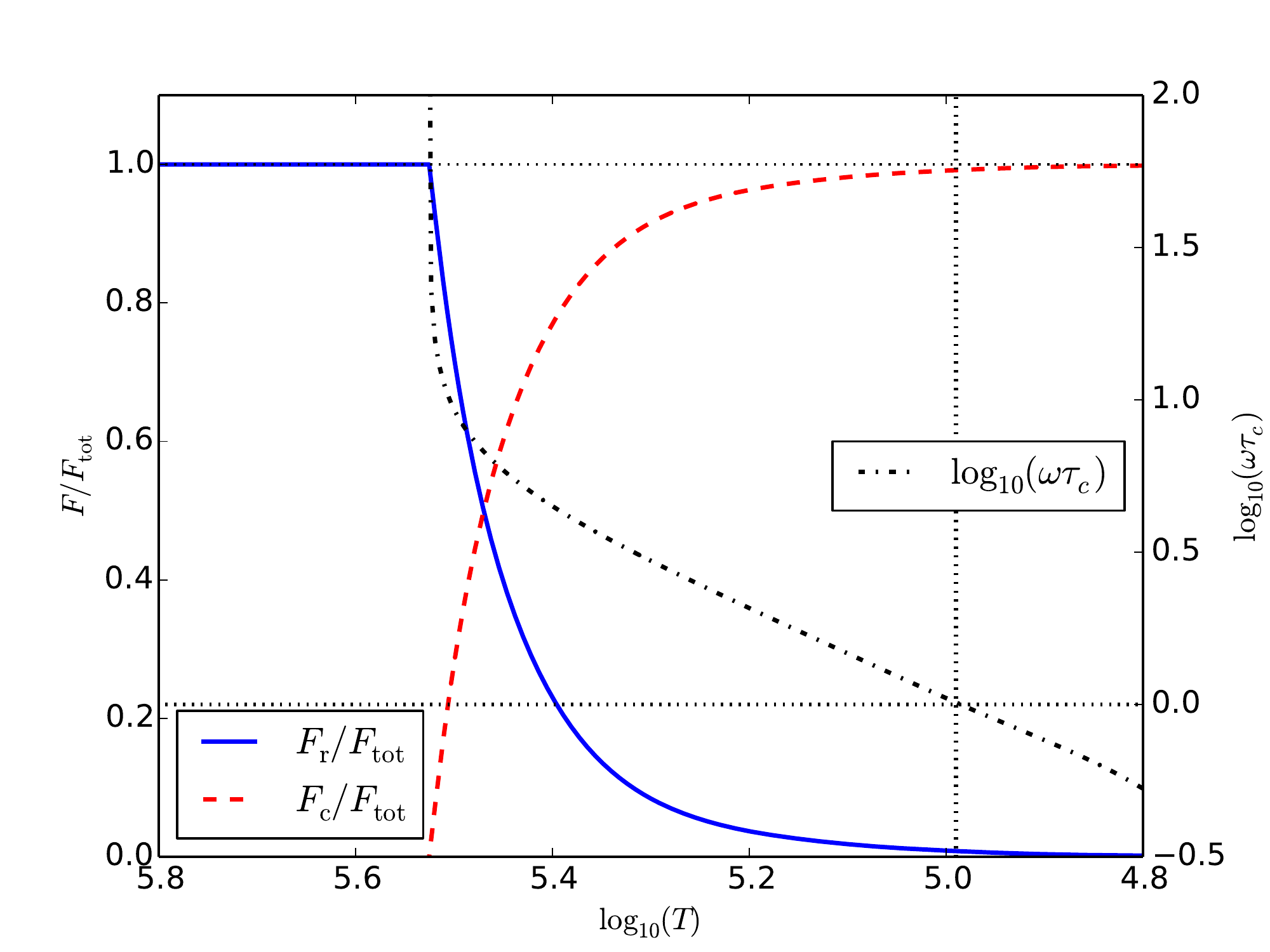}}
        \end{center}    
        \caption{{\bf Left y-axis}: Normalised radiative flux $F_{\rm rad}/F_{\rm tot}$ (blue solid line), and convective flux $F_{\rm c}/F_{\rm tot}$ (red dashed line) as a function of $\log_{10} (T)$ for a $M=1.40~M_\odot$ main-sequence model. {\bf Right y-axis:}  $\log_{10} (\omega \tau_c)$ (dot-dashed line) as  a function of $\log_{10} (T)$  where $\tau_c$ is the convective time-scale,  $\omega = 2 \pi/\Pi$ and $\Pi=0.2$~day. The vertical dotted line shows the layer where $\omega \tau_c = 1$. }
        \label{Frad_Fconv_tauc}
        \end{figure}

This mechanism was first suggested by \citet{Pesnell87} as a possible driving mechanism in pulsating white dwarfs. It was finally shown later to operate in the $\gamma$-Doradus stars \citep{Guzik00,Warner03}. 
Indeed, for these objects, the transition region of low radial order gravity modes coincides with the BCZ.
There is, however, a difficulty. As seen in   Fig.~\ref{Frad_Fconv_tauc}, the convective time-scale $\tau_{\rm c}$ becomes rapidly shorter than the modal period, so that convection can no  longer be considered as a passive actor (``frozen convection''). Therefore, the region where the assumption of ``frozen convection'' is valid is very tiny and a Time-Dependent Convection treatment (hereafter TDC, which will be addressed in Sect.~\ref{mode_damping}) must be considered. 
\citet{Dupret04} calculations based on TDC finally confirmed the effectiveness of the convective flux blocking mechanism \citep[see also][]{Dupret05}. 

Theoretical calculations by \citet{Dupret04} also successfully reproduced the observed instability strip of the  $\gamma$-Doradus stars,  at least as it was known before the observations made by \emph{Kepler}. Indeed, recently the \emph{Kepler} satellite detected a large number of hybrid $\delta$~Scuti - $\gamma$~Doradus pulsators lying on the left side of the blue  edge of $\gamma$~Doradus \citep{Balona11,Balona14}, as shown in Fig.~\ref{balona_2014_fig3}. As it is clearly seen, a large fraction of these pulsators are located at hotter temperature w.r.t. the temperature of the theoretical blue edge of the $\gamma$~Doradus stars. This is then not yet clear why,  $g$~modes in those stars   (characteristics of the $\gamma$~Doradus pulsators),  are  excited   \citep[see the discussion in][]{Balona14}. 

 \begin{figure}
        \begin{center}
        \resizebox{0.8\hsize}{!}{\includegraphics  {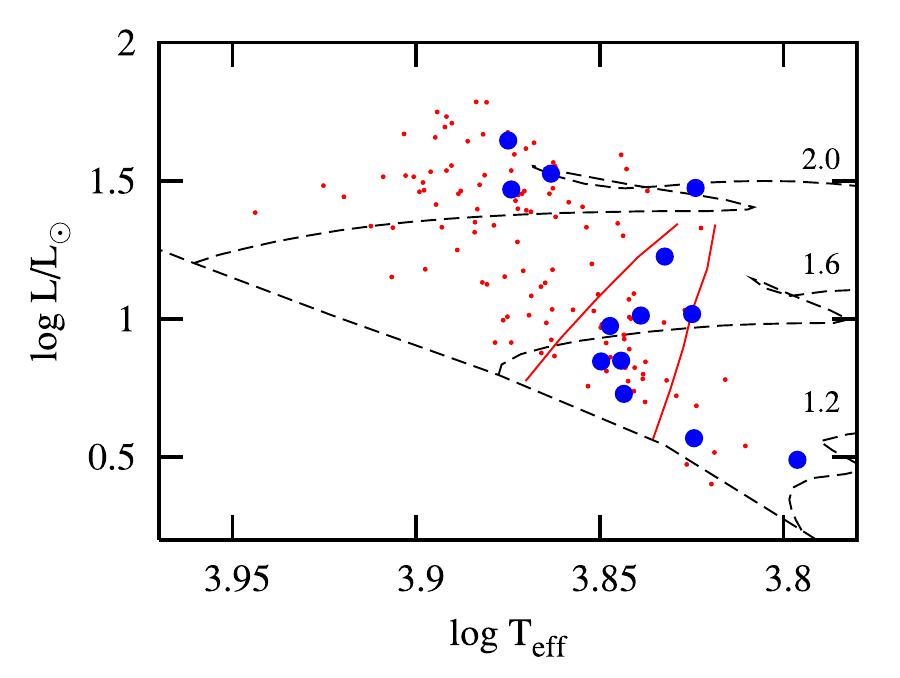}}
        \end{center}    
        \caption{Location in the HR diagram of the  hybrid $\delta$ Scuti - $\gamma$ Doradus pulsators detected by the {\it Kepler} satellite.  The small red filled circles correspond to those pulsators  that would be classified as ``pure'' $\delta$~Scuti stars from the ground while the large blue filled circles are those with large amplitude that would be classified as ``hybrids''. The two solid lines are the blue and red theoretical edges of the $\gamma$~Doradus instability strip computed by \citet{Dupret04}.  Figure from \citet{Balona14}. }
        \label{balona_2014_fig3}
        \end{figure}

\subsubsection{Convective driving}

This driving mechanism occurs in the convective regions and is caused  by  a modulation of the convective flux by the mode. It was originally proposed by \citet{Brickhill83} to explain the gravity-modes detected in the ZZ Ceti stars (A-type white dwarf puslator, also named DAV, see Fig.~\ref{pulsation-HR} and Table~\ref{tab-pulsating-stars}). To highlight this  mechanism, we follow the didactic approach proposed by \citet{Saio13}. Note that an extensive review about the various type of pulsating  white dwarf stars can be found in \citet{Winget08}.

A-type white dwarf stars have shallow upper convective envelopes, where energy is predominantly transported by convection (efficient convection, $F_{\rm c} \gg F_{\rm rad}$).
Furthermore, the transition region of the gravity modes observed in  ZZ Ceti stars coincides  with their convective zone \citep{Winget82}. 
%Accordingly, only the modulation of the convective luminosity, $\delta L_c$, can potentially drive the mode. \textcolor{red}{KEVIN: ce n'est pas si trivial car meme si le flux radiatif est faible sa perturbation peu être importante. Je propose donc: "
In such a situation, it has been shown that the modulation of the convective luminosity, $\delta L_c$, can potentially drive the mode. 
Following \citet{Saio13} we now establish the expression for $\delta L_{\rm c}$  on the basis of   mixing-length theory (MLT hereafter).
The  MLT yields the following relation for the convective flux \citep[see {e.g.}][]{Bohm89,Cox68}
\eqn{
F_{\rm c} \propto \alpha^2 \,  \left ( \frac{\nablaad}{\rho} \right ) ^{1/2}\, \left (  \frac{P T}{g}  \right )^{3/2} \, \left ( - \deriv{S}{r} \right )^{3/2} \, ,
\label{F_c}
} 
where $\alpha$ is the mixing-length parameter, $g$ the gravity, $ \nablaad =  \left ( \deriv{\ln T}{\ln P} \right )_s $ the adiabatic gradient, and $S$ the entropy. 
The convective turnover time $\tau_c$ turns to be much shorter than the periods of the gravity modes. Accordingly, convection instantaneously adjusts to pulsation so as to maintain the entropy gradient (which is nearly isentropic, {\it i.e.} $dS/dr \approx 0$). As a consequence $\delta \left ( \deriv{S}{r}   \right ) \approx 0$ during the pulsation cycle.
Since $L_{\rm c} = 4 \, \pi \, r^2 \, F_{\rm c}$,  we establish with the help of \eq{F_c} the following expression for $\delta L_{\rm c} $
\eqn{
\delta L_{\rm c} \approx L_{\rm c} \, f_{\rm c} \, \frac{\delta P }{P} \, ,
\label{delta_L_c}
}
with
\eqn{
f_{\rm c} =  \frac{1}{2} \left [  3 \left ( 1 + \nablaad \right )  - \frac{1}{\Gamma_1} + \left ( \deriv{\ln \nablaad}{\ln P} \right )_T + \nablaad \left (  \deriv{\ln \nablaad} {\ln T} \right )_P \right ]
\label{f_c}
}
Substituting  \eq{delta_L_c} into the expression of \eq{growth_rate3} for the mode growth rate, gives
\eqn{
\gamma = - \frac{1}{2 \omega^2 \, I }  \int_0^M \, \left ( \frac{\Gamma_3-1}{\Gamma_1} \right )  \,  \left | \frac{\delta P}{P} \right |^2 \, \left (   \deriv{f_{\rm c}}{m}  \right )     \, {\rm d}m\, .
 \label{growth_rate_convective_driving}
}
To establish \eq{growth_rate_convective_driving}, we have used the fact that ${\delta P}/{P}$ is  nearly constant in the convective zone. On the other hand, the quantity $f_{\rm c}$ given by \eq{f_c} decreases outwards as depicted in Fig.~\ref{convective_driving}. Accordingly, $\gamma >0$, so that the mode is unstable (effectively excited).  The decrease of $f_{\rm c}$ is directly linked with the partial ionisation of hydrogen: energy is absorbed  during  compression  in the region of partial ionisation and released to the modes during expansion. The convective  mechanism operated then in very similar way than the $\gamma$ mechanism. 
Note that a more complete approach was proposed by \citet{Goldreich99}.
Furthermore, theoretical calculations  by \citet{vangrootel12}, based on a TDC treatment, confirm the origin of this driving. However, while the authors successfully reproduce the observed blue edge of the ZZ ceti stars, they predict a much cooler red edge.

 \begin{figure}
        \begin{center}
        \resizebox{\hsize}{!}{\includegraphics  {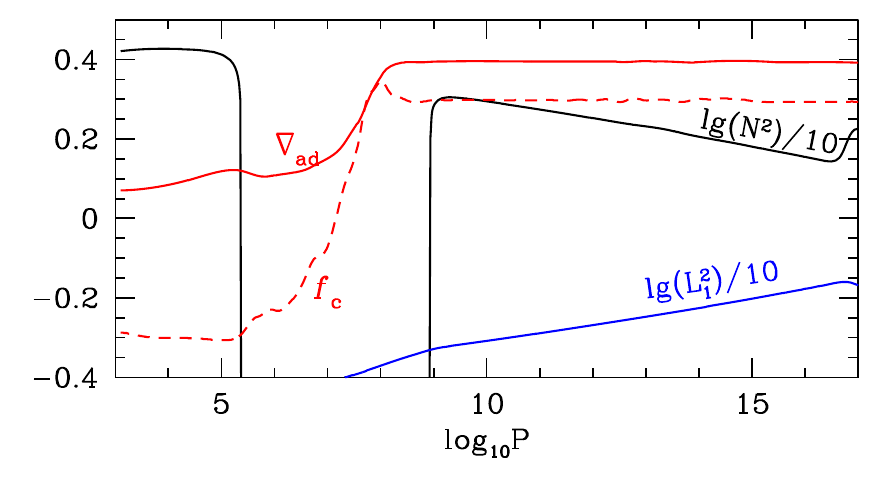}}
        \end{center}    
        \caption{Adiabatic gradient $\nablaad$ (red solid line) and the quantity $f_{\rm c}$ (red dashed line) given by \eq{f_c}    as a function of $\log_{10} P$ for an A-type dwarf stellar model with $M=0.60~M_\odot$ and $\teff= 11 700$~K. Figure reproduced from \citet{Saio13}.}
        \label{convective_driving}
        \end{figure}

\subsection{Instability strips}

As it is clearly seen in Fig.~\ref{pulsation-HR}, pulsating stars do not exist everywhere in the HR diagram but inside in characteristic strips, named ``instability strips''.  For instance,  Cepheids, RR~Lyrae and $\delta$ Scuti stars lie in the same strip, which is named the classical instability strip. As we will show now, the existence of these characteristic strips is  directly linked with the coincidence of the transition region (TR hereafter) with a partial  ionisation region (IR hereafter) of a given chemical element. Following \citet{Cox80}, we establish under which condition the TR and the IR coincide.  
We hence define the ratio $\phi = \tauth / \Pi $,  where the thermal time-scale is given by \eq{tau_th4}. Accordingly, we have
\eqn{
\phi \approx \frac{ c_v \, T \, \Delta M} {  \Pi \, L} \, ,
\label{phi}
}
where $\Delta M$  is the mass of a  given shell. 
The hydrostatic equation gives the relation $P \propto M \, \Delta M / R^4$. We also use  the period-density  relation  $\Pi \propto M^{1/2} \ R^{-3/2}$. Finally, we assume a polytrope  {\it i.e.}  $P \propto T^{n+1}$ and adopt the index $n=3$ corresponding to a fully radiative star. Combining these three relations into \eq{phi} yields to the following expression
\eqn{
\phi \propto \frac{R^{5/2} \, T^5} {M^{1/2} \, L} \, .
\label{phi_2}
}
The TR corresponds by definition to the layer where $\phi =  1$.
According to \eq{phi_2}, at fixed $M$ and $L$, the temperature at the TR scales then as $T_{\rm TR} \propto R^{-1/2}$. As we have seen in Sect.~\ref{driving_criteria}, driving by the $\kappa (/\gamma)$-mechanism is efficient only when the TR and the IR coincides, that is when  $T_{\rm TR}  \approx T_{\rm IR}$ where $T_{\rm IR} $ is the temperature of a given ionisation region ({e.g.} H, He$^{+}$, He$^{++}$, ... etc). 
This situation is illustrated in Fig.~\ref{IS_blue_edge_1}.

\begin{figure}
        \begin{center}
        \resizebox{\hsize}{!}{\includegraphics  {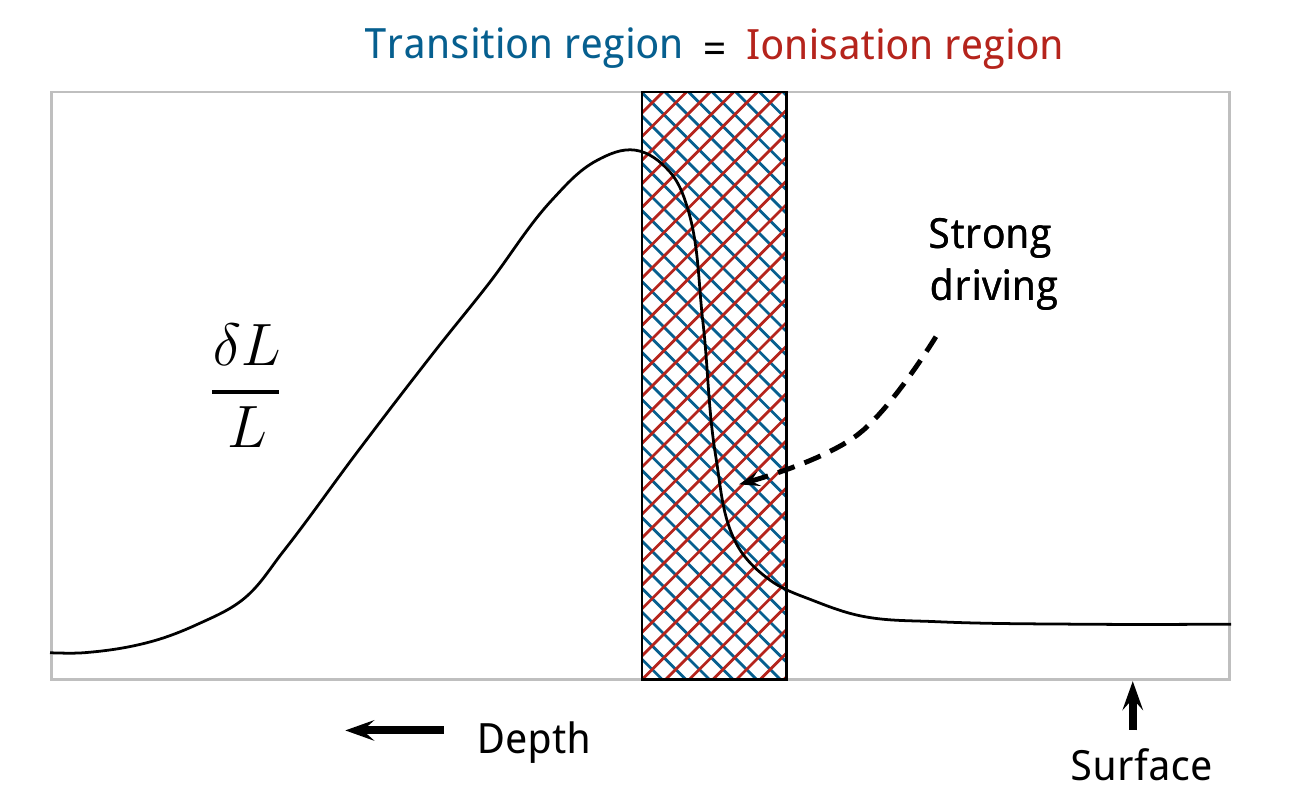}}
        \end{center}    
        \caption{Relative luminosity perturbation $\delta L/L$ as a function of depth. This  corresponds to the situation when the Transition Region (TR) and the Ionisation Region (IR) coincide.  Figure adapted from a figure published in Gastine's PhD thesis \citep{Gastine09}.}
        \label{IS_blue_edge_1}
        \end{figure}

At fixed $M$ and $L$, the TR coincides with the IR  when the star radius $R$ reaches a given  critical radius $R_{\rm crit}$  or equivalently at a given effective temperature since $L \propto R^2 \, \teff^4$. 
A less evolved star (\emph{i.e.} a hotter star) will have   $R < R_ {\rm crit}$ and since $R \propto T_{\rm TR}^{-1/2}$, we have necessarily $T_{\rm TR} > T_{\rm IR}$. In such a case the TR lies \emph{below} the IR which  lies  in the strongly non-adiabatic regions. This is illustrated in Fig.~\ref{IS_blue_edge_2}. In this situation the driving by the $\kappa$-mechanism is inefficient and is dominated by the strong damping occurring in the vicinity of the TR.
All the modes are stable. The star is  too hot and lies outside the instability strip, on the left side of the \emph{blue} edge (see the case A shown in Fig.~\ref{pulsation-HR-IS})

\begin{figure}
        \begin{center}
        \resizebox{\hsize}{!}{\includegraphics  {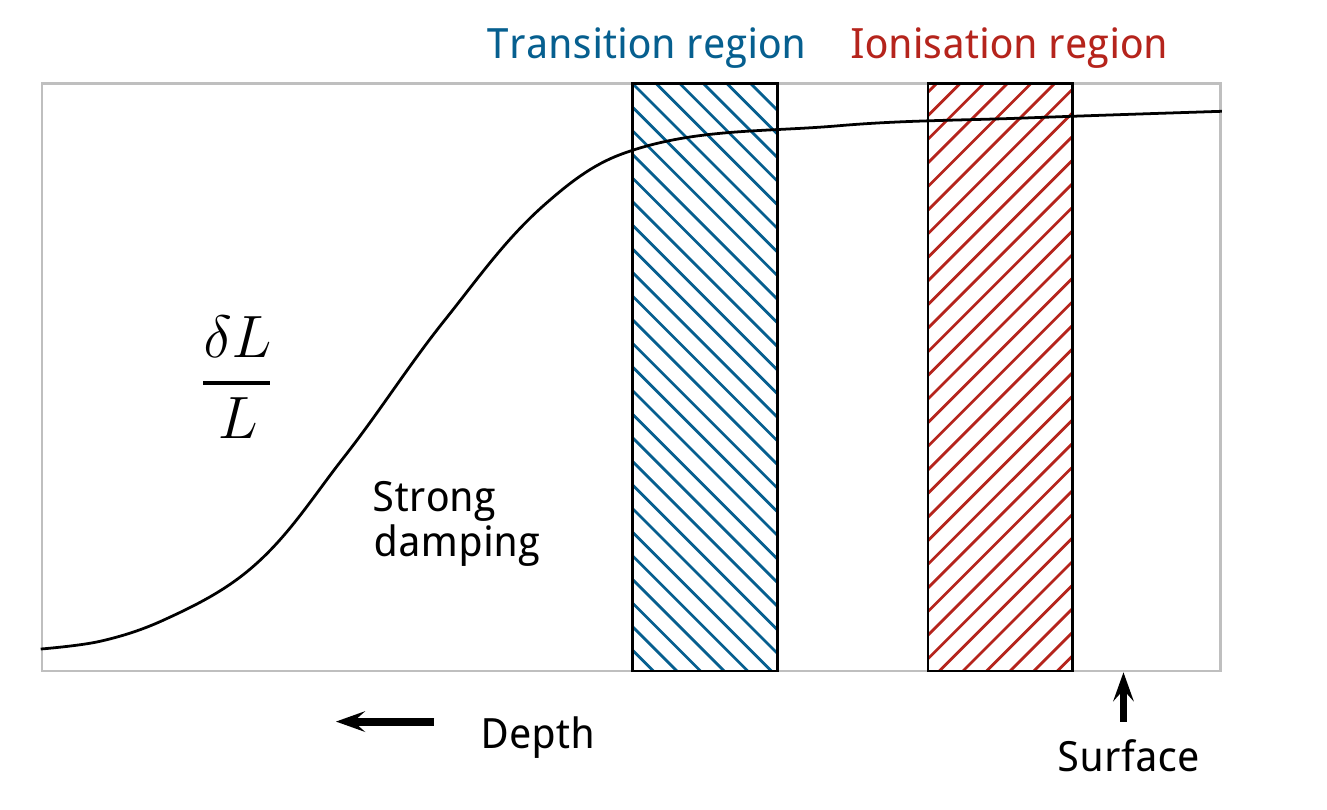}}
        \end{center}    
        \caption{Same as Fig.~\ref{IS_blue_edge_1} in the situation when the TR lies \emph{below} the IR.   Figure adapted from a figure published in Gastine's PhD thesis \citep{Gastine09}.}
        \label{IS_blue_edge_2}
        \end{figure}

On the opposite, a more evolved star ({\it i.e.} a cooler star), will have   $R > R_ {\rm crit}$ and accordingly  $T_{\rm TR} < T_{\rm IR}$. As illustrated in Fig.~\ref{IS_blue_edge_3}, 
the TR is  \emph{above} the IR, which lies in the quasi-adiabatic region. The driving is inefficient and counterbalanced by a dominant damping occuring just above the IR. 
The modes are stable. The star is  too cool and lies outside the instability strip, on the right side of the \emph{red} edge (see the case C shown in   Fig.~\ref{pulsation-HR-IS}). 

Finally, for $R= R_{\rm crit}$, the  driving by the $\kappa$-mechanism will be sufficiently efficient to counterbalance the damping. In that case the star will show one or several unstable modes. The star lies in the instability strip associated with the partial ionisation region of a considered element,  {e.g.}  the 'classical' instability strip that is associated with the He$^{+}~\rightarrow~$He$^{++}$ dissociation (see the case B shown in Fig.~\ref{pulsation-HR-IS}). 
Note that each ion ({e.g.} H$^+$, He$^+$, He$^{++}$, iron-group ions) is generally associated to a given instability strip. This is the main reason why pulsating stars lie along well defined vertical strips in the HR diagram (see Fig.~\ref{pulsation-HR}).

\begin{figure}
        \begin{center}
        \resizebox{\hsize}{!}{\includegraphics  {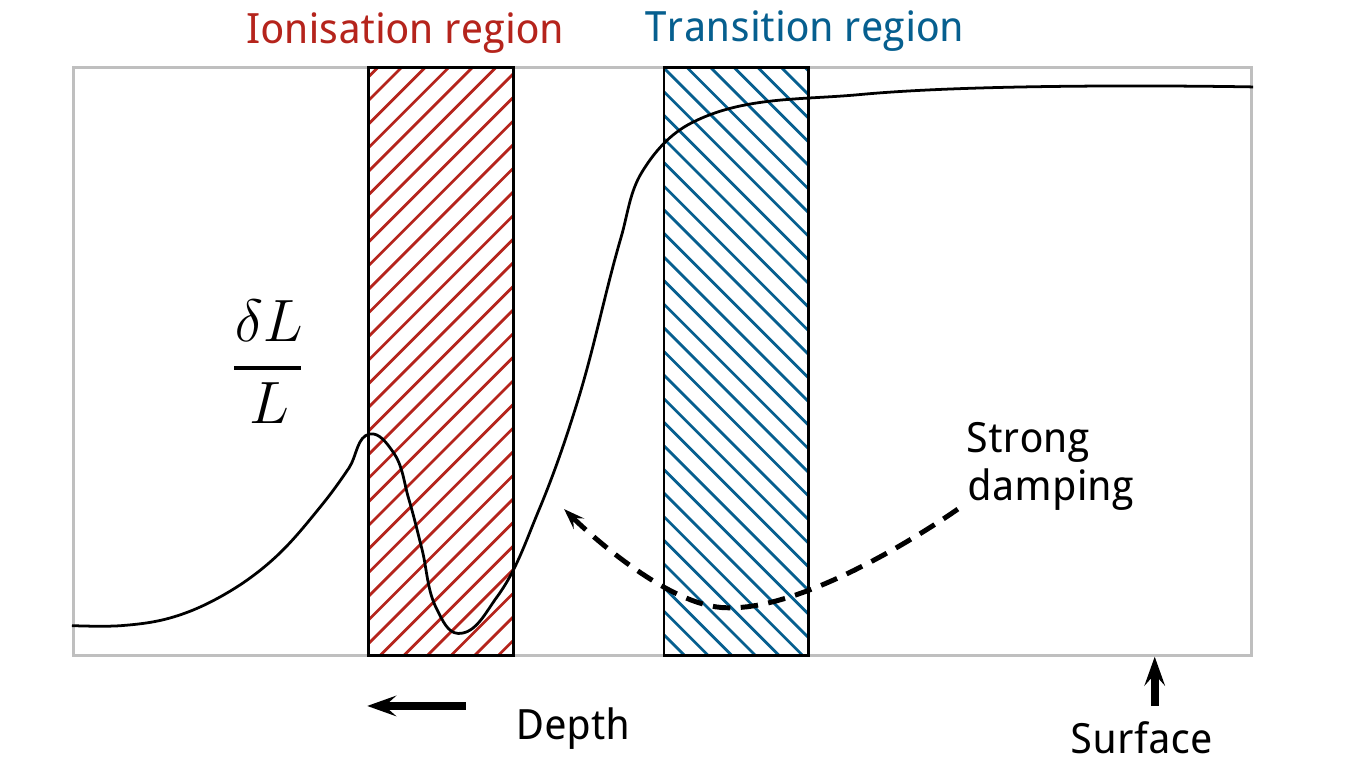}}
        \end{center}    
        \caption{Same as Fig.~\ref{IS_blue_edge_1} in the situation when the TR lies \emph{above} the IR.   Figure adapted from a figure published in Gastine's PhD thesis \citep{Gastine09}.}
        \label{IS_blue_edge_3}
        \end{figure}

\begin{figure}
        \begin{center}
        \resizebox{0.8\hsize}{!}{\includegraphics  {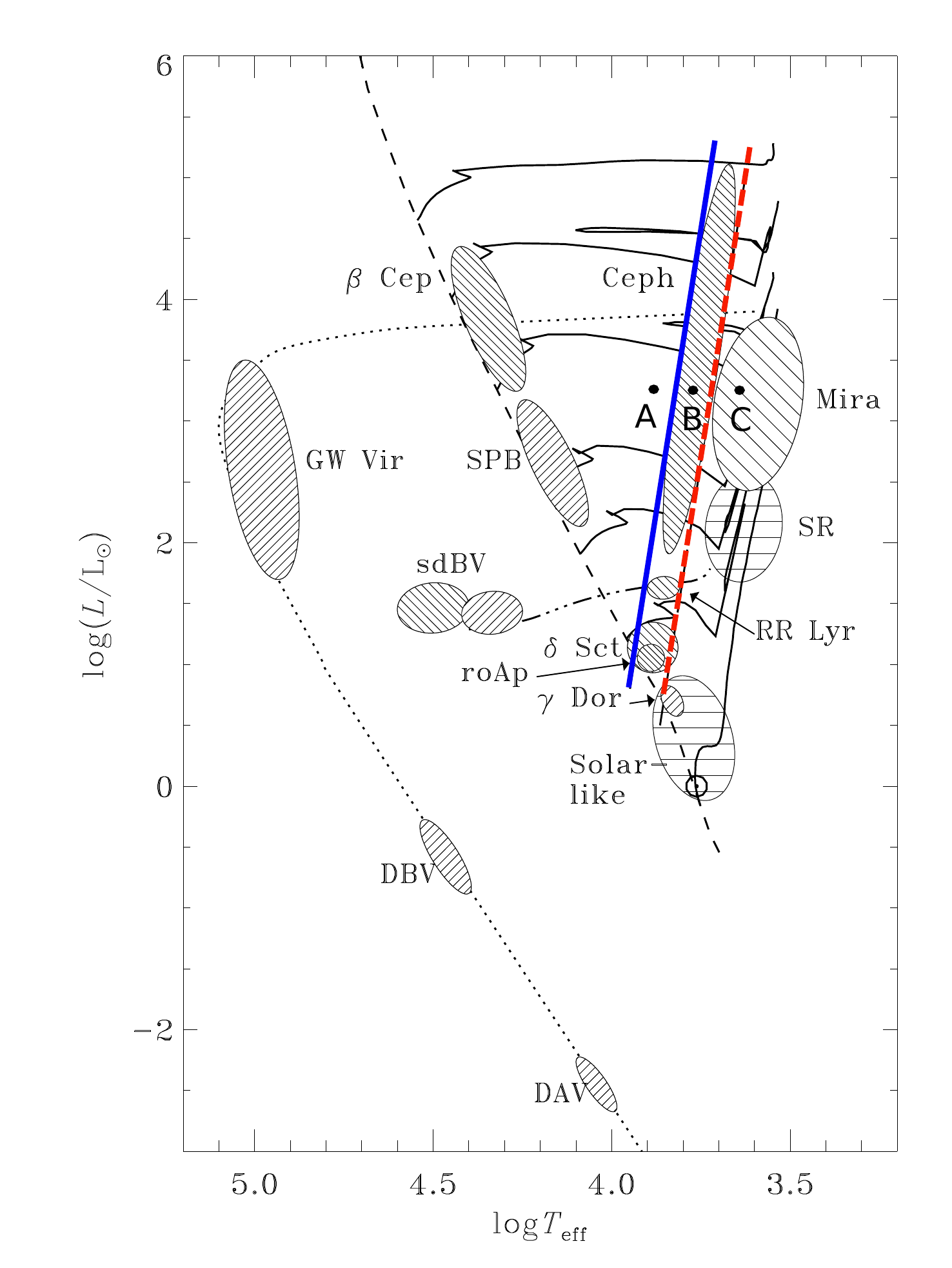}}
        \end{center}    
        \caption{Location of different classes of pulsating stars in the HR diagram. The blue solid  line represents the blue edge of the  classical instability strip (which is associated with the He~$\rightarrow$~He++ dissociation) and the red dashed  line its red edge. The position A (resp. C) corresponds to a star of given $L$ and $M$ for which $R < R_ {\rm crit}$ (resp. $R > R_ {\rm crit}$).  The position B corresponds to the case of a star lying in instability strip, and  for which  $R \approx R_ {\rm crit}$. Figure adapted from the figure generated by Christensen-Dalsgaard.   }        
\label{pulsation-HR-IS}
        \end{figure}

The existence of the blue and red edges are qualitatively well explained by the scaling relation of \eq{phi_2}. Fully non-adiabatic calculations in general quantitatively explain the observed  blue edge of the classical instability strip.  However, until only about 15 years ago, it was impossible  to predict the red edge of the $\delta$ Scuti instability strip, the theoretical red edge being much cooler than the observed one \citep[see {e.g.}][]{Pamyatnykh00}. This is actually because convection is in general treated as a passive process. However, near the observed red edge the convective time-scale $\tauc$ turns out to be of same order as the thermal time-scale $\tauth$ such that convection can no longer be considered as ``frozen''. 

This is illustrated in Fig.~\ref{conv-time-scale} where we have compared $\tauc$ with $\tauth$ for two stellar models. 
For the model  lying within the instability strip  of the $\delta$ Scuti stars, $\tauc$  is  much larger than $\tauth$ in the region of partial ionisation of He$^{++}$ (where the $\kappa$ mechanism mainly operates). For the hotter model located near the observed red edge,  $\tauc$  becomes smaller than $\tauth$. In that case, ``frozen convection'' is no longer a valid approximation and  TDC treatment must be considered (see Sect.~\ref{mode_damping}). 
Several authors have considered various TDC treatments  \citep{Houdek00,Xiong01,Dupret04} and finally successfully reproduced the observed red edge of the $\delta$ Scuti stars.  
For instance, calculations performed by \citet[]{Dupret04} \citep[see also][]{Dupret05} match the observed red edges associated with both radial and non-radial modes, provided, however, that the solar calibrated mixing-length parameter ($\alpha=1.8$) is adopted. This  result is, however, not consistent with results from 3D hydrodynamical models \citep{Ludwig99,Trampedach11,Trampedach14}, since the latter  predict that the  mixing-length decreases with increasing $ \teff$ ($\delta$ Scuti stars  are A-type stars hence significantly hotter than the Sun).

\begin{figure}
        \begin{center}
        \resizebox{0.8\hsize}{!}{\includegraphics  {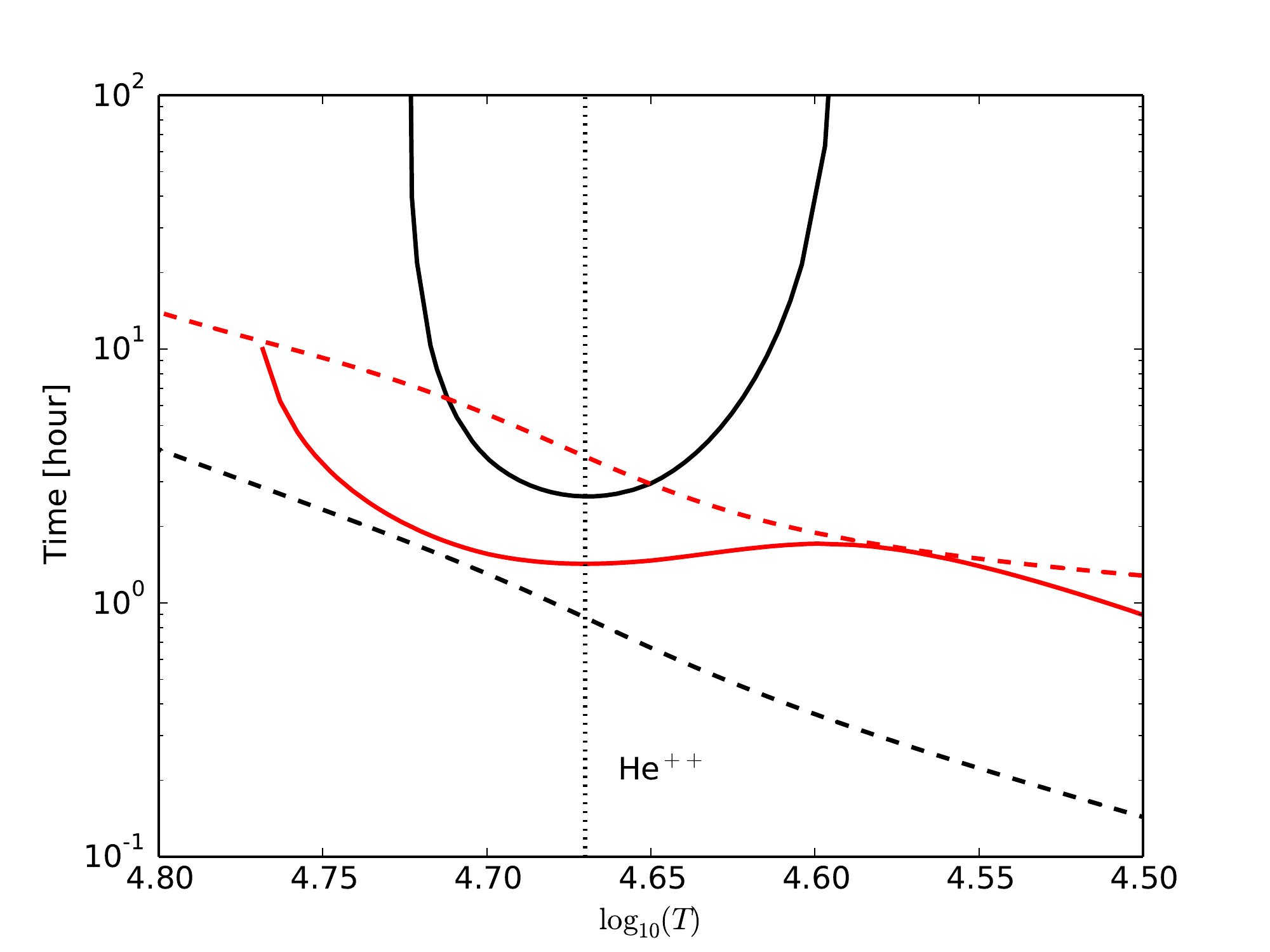}}
        \end{center}    
        \caption{Thermal and convective time-scales as a function of $\log_{10} (T)$. The solid lines correspond to the convective time-scale, $\tauc$, and the dashed lines to the thermal time-scale, $\tauth$. The red color is for a stellar model with $\teff$ = 6~950~K, which lies near the observed red edge  and the black line for a model with  $\teff$ = 7~900~K, which is close to the blue edge and within the instability strip. The vertical dotted line represents  the location of the partial ionisation of He$^{++}$.   }        
\label{conv-time-scale}
        \end{figure}

\subsection{$\kappa$-mechanism and micro-physics}

\subsubsection{$\kappa$-mechanism and opacity}

Prior to the 90's, it was not possible to explain the existence of the acoustic modes detected so far in the $\beta$ Cephei pulsators \citep[for an early review see][]{Osaki86}. \citet{Simon82} suggested that this problem can be solved with an increase by a factor 2-3 of the opacity of the heavy elements (iron-group elements, Fe, Ni, Cr and Mn). Figure~\ref{Simon_1982} taken from \citet{Simon82} compares the opacity that was available at that time (Los Alamos data) with opacity augmented by a factor 2-3 for the heavy elements  in the temperature range between 10$^5$ and 10$^6$~K, {\it i.e.} near the ``bump'' of the iron-group ions (``Z-bump''), which is located around $\log_{10} T \sim 5.3$. 
Such an increase of the opacity of the iron-group elements could explain why   $\beta$ Cephei stars do pulsate. 
In the 90's, new opacities including a large number of bound-bound transitions from the iron-group elements, have  been released by the OPAL project  \citep{Rogers92}.
They result in an enhancement of the opacity near the  ``Z-bump'', which finally permits   an effective driving by the $\kappa$ mechanism  of the acoustic modes in  $\beta$ Cephei stars \citep{Cox92,Kiriakidis92,Moskalik92}. This new opacity table  explained also the existence of the SPB pulsators \citep{Gautschy93,Dziembowski93} and by the way resolved also the problem of period ratios for double-mode Cepheids \citep{Moskalik92b}. 

Opacities from the OPAL project now  explains most of the observed  $\beta$ Cephei and SPB stars. However, the discovery of B-type pulsators in low-metallicity environments as well as the existence of unpredicted hybrid SPB-$\beta$ Cephei pulsators have more recently attracted some attention. In that respect, it was shown that the opacity from the Opacity Project \citep[][OP hereafter]{Seaton96} together with the new solar chemical mixture by \citet[AGS095 hereafter]{Asplund05b} better explain the instability strip of metal-poor SPB and $\beta$ Cephei stars \citep{Miglio07,Pamyatnykh07}. While the theoretical calculations based on the opacity from the OP and the AGS05   chemical mixture predicted unstable $p$ and $g$ modes down to Z=0.005 and Z = 0.01 respectively,  numerous  $\beta$ Cephei and SPB pulsators are detected in the Small Magellanic Cloud (Z=0.0027) and the Large  Magellanic Cloud (Z=0.0046).  According to  \citet{Salmon12}, this discrepancy could be solved if the  Ni opacity peak is increased by 50\,\%.

\begin{figure}
        \begin{center}
        \resizebox{0.8\hsize}{!}{\includegraphics  {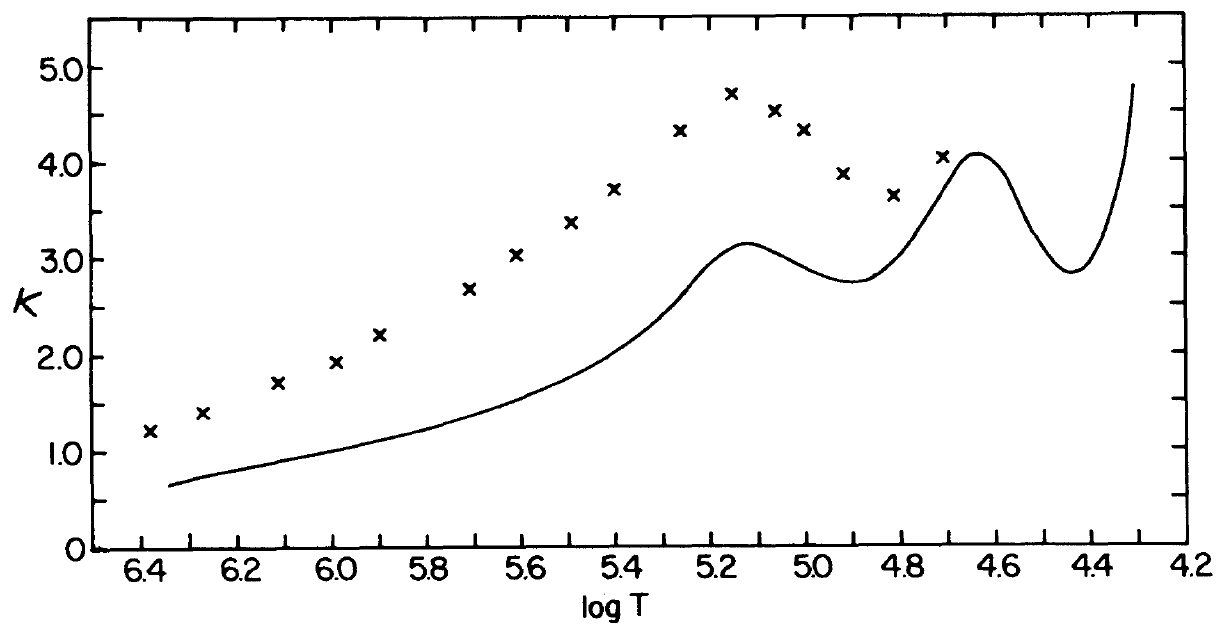}}
        \end{center}    
        \caption{Opacity as a function of $\log_{10} (T)$. Solid line: opacity prior the 90's ('old' opacity). Crosses:  modified  opacity  corresponding to the 'old' opacity augmented by a factor 2-3 for the heavy elements in the temperature range between 10$^5$ and 10$^6$~K. Figure reproduced from \citet{Simon82}.   }        
\label{Simon_1982}
        \end{figure}

\subsubsection{$\kappa$-mechanism and microscopic diffusion}

Hot subdwarf B stars (sdB)   are core He burning stars that have lost most of their H envelope. They lie along the Extreme Horizontal Branch (EHB) but their formation is not well known \citep[for a review see][]{Heber09}. 
Some of these sdB are found inside the instability strip of the iron-group ions and do pulsate. They are named sdB pulsators (see Fig.~\ref{pulsation-HR} and Table~\ref{tab-pulsating-stars}) and were predicted by \citet{Charpinet96} before their first detection \citep[for a review see][]{Charpinet01}.
Indeed, while for stellar models with solar metal abundance and homogenous composition the mode damping mechanism dominates over the $\kappa$ mechanism, it was  shown by \citet{Charpinet96}  that, for homogenous  models with enhanced metal abundance,  the $\kappa$ mechanism  takes over the damping.  
However, the relative large overall metal abundance required for the mode to be unstable is not realistic and a local enhancement by some microscopic diffusion in the vicinity of the ``Z-bump'' region  must be considered in order to explain the driving. 
Microscopic diffusion results in general from a balance between gravitational settling (\emph{heavy elements sink faster}) and radiative levitation (\emph{photons communicate momentum}). Inclusion of these microscopic diffusion processes boosts the metal in the ``Z-bump'',  making  the driving possible  \citep[for more details see][]{Charpinet01}. 

% \begin{figure}
%         \begin{center}
%         \resizebox{0.8\hsize}{!}{\includegraphics  {charpinet_2001_fig4}}
%         \end{center}    
%         \caption{Logarithm of the iron abundance as a function of $\log_{10} (q)$ where $q = m /M$ and $M$ is total mass, for a set of sdB models computed by including  gravitational settling and radiative levitation. The red vertical line represents the location of the ``Z-bump''. The horizontal dotted line indicates the solar iron abundance. Figure adapted from  \citet{Charpinet01}. \textcolor{red}{KEVIN: question d'eric,figure non-citée dans le texte?}}        
% \label{Charpinet_2001_fig4}
%         \end{figure}

\subsection{Radiative damping and blue supergiants}

\label{radiative_damping}

In the diffusion approximation, variation of the radiative component of the luminosity induced by a mode, $\delta L_{\rm r}$, is given by  \eq{delta_r}. 
The first term in the  RHS of \eq{delta_r} is responsible for damping since  at maximum compression $\left ( \deriv{r} {\ln T} \, \deriv{}{r}  \right ) \, \left ( \frac{\delta T}{T} \right )$ is always positive. This radiative damping contributes to mode stabilisation  and can then prevents the existence of unstable modes.  

Pulsations in SPB stars correspond to $g$~modes that become unstable because the $\kappa$ mechanism  in the  ``Z-bump'' region is strong enough to overcome the radiative damping. 
Until the space-borne ultra-high photometry missions, $g$ modes in more evolved B-type stars, were not expected to be excited because of  the strong  damping occuring in the radiative core \citep{Gautschy93,Dziembowski93}.  
This is the main reason why the region of the HR diagram located above the SPB region is almost free of g-mode pulsators (see Fig.~\ref{SPB-betaCep-regions}).

\begin{figure}
        \begin{center}
        \resizebox{0.8\hsize}{!}{\includegraphics  {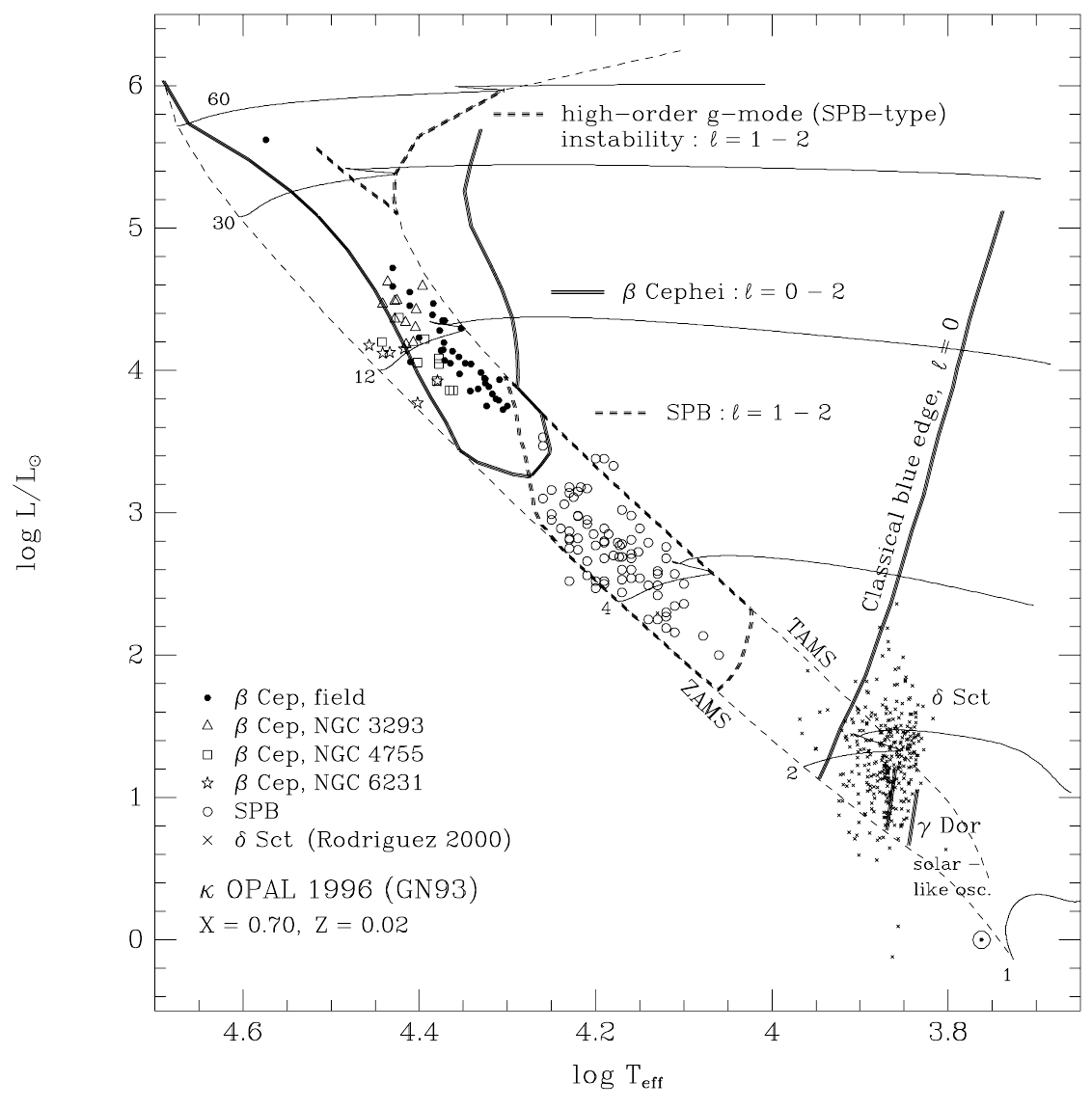}}
        \end{center}    
        \caption{Location of the $\beta$ Cephei and SPB pusators in the HR diagram. Figure reproduced from \citet{Pamyatnykh02}  }        
\label{SPB-betaCep-regions}
        \end{figure}

To highlight the existence of this strong damping in supergiant B-type stars, we perform a local analysis and decompose the relative temperature fluctuation  as $\delta T / T \propto e^{-i  k_r . r}$ where $ k_r$ is the local radial wave-number. This permits us to write the first term in the  RHS of \eq{delta_r}  as
\eqn{
\left (  \deriv{r} {\ln T} \, \deriv{}{r}  \right ) \, \left ( \frac{\delta T}{T} \right ) \approx k_r \, r \, \left (   \deriv{\ln r} {\ln T}\right ) \left ( \frac{\delta T}{T}   \right ) \; .
\label{radiative-damping-local-analysis}
}
It is clear from  \eq{radiative-damping-local-analysis} that the higher the wavenumber ({\it i.e.} the shorter the wavelength), the stronger the radiative damping. 

For $g$ modes, the dispersion equation  yields $k_r \approx \sqrt{ \ell \left (\ell +1 \right )}  {N_{\rm  b}}/{\left (\omega \,r \right ) }$ where $N_{\rm b}$ is the \BV\ frequency. In post-main  sequence stars, $N_b$ reaches very high values because of the sudden contraction of the core when the star leaves the main-sequence. 
Accordingly, $N_{\rm b} \gg \omega$ and as a consequence $k_r \, r \gg  1 $. 
The $g$ modes  in those post-main sequence stars have thus very short wavelengths in the inner layers  and  are therefore substantially damped preventing them from being unstable. 

Despite this expected strong damping,  light-variations with period of few days were  detected in a limited number of blue supergiants \citep[see {e.g.}][and references therein]{Waelkens98,Lefever07}. 
The clear detection by the MOST satellite of a larger number of gravity modes   in the blue supergiant HD~163899  has, however, motivated new and intense theoretical studies to explain the existence of such modes \citep{Saio06}.
%% This detection was unexpected because --~~ as mentioned above ~--  gravity modes are expected to be strongly damped  in the radiative region of such evolved B-type stars. 
According to  \citet{Saio06}, the existence of these gravity modes could be due to the presence of an Intermediate Convection Zone (ICZ hereafter) associated with the hydrogen burning shell. Indeed, some $g$ modes can be partially reflected in this ICZ preventing them from entering into the  radiative core.  
The existence of an ICZ in a representative blue supergiant model is illustrated in Fig.~\ref{saio_2006_fig7}. Figure~\ref {godart_2009_fig4} presents the radial displacement of two $g$ modes computed for a supergiant model.  One of these modes crosses the ICZ whereas the other is reflected on the ICZ.   
The two modes have approximately the same amplitudes in the envelope, however, within the radiative core, the first one has both large amplitudes  and a short-wavelength  whereas the second one has much smaller amplitude. As a consequence the first one is strongly damped and remains stable whereas for the second one the damping remains small enough such that the $\kappa$ mechanism, which as for the SPB pulsator takes place in the iron group opacity bump, is able to destabilise the mode. 

% While the first one is strongly damped, The latter is then weakly damped and , and is then wekaly. 

\begin{figure}
        \begin{center}
        \resizebox{0.8\hsize}{!}{\includegraphics  {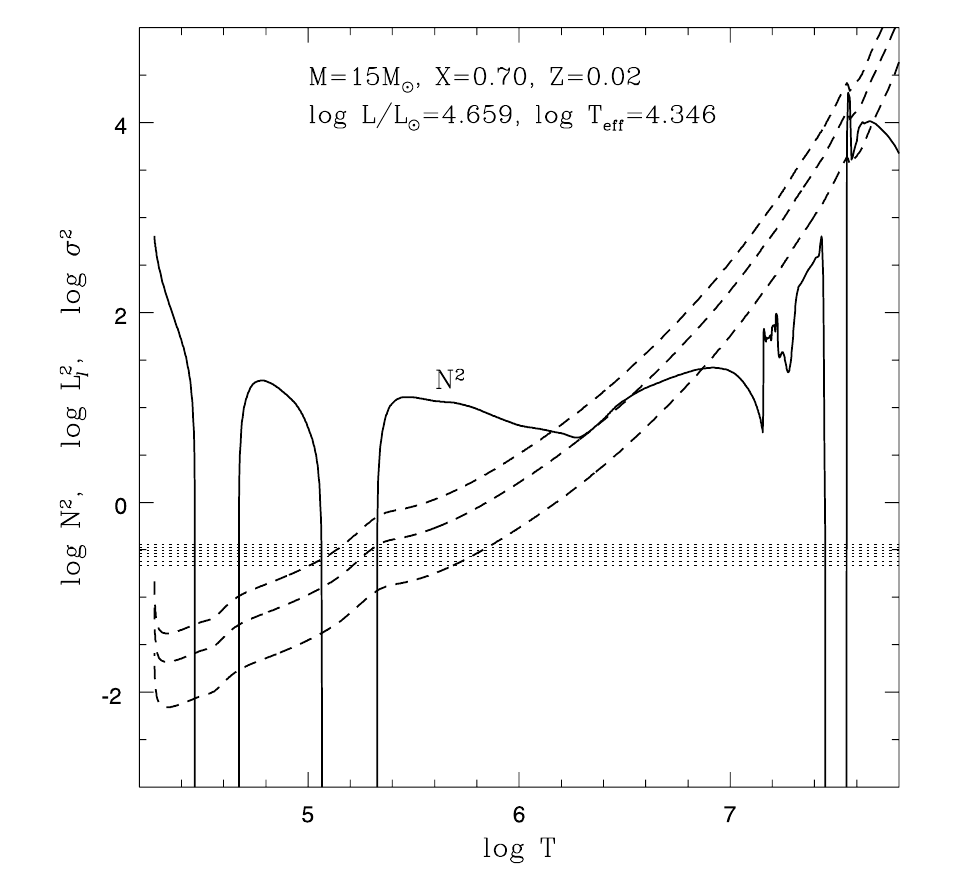}}
        \end{center}    
        \caption{Square of the \BV\ frequency $N_b^2$ (solid line), and the square of Lamb frequency $L_\ell^2$ for $\ell = 1,2,3$ (dashed lines) as functions of $\log_{10} T$ for a 15~$M_\odot$ stellar model. Horizontal dotted lines indicate the angular frequencies ($\sigma$) of the g-modes  $\ell =2$ found to be unstable in the considered blue supergiant model. All the quantities in this figure are normalized by $G \, M \, R^{-3}$.  The Intermediate Convection Zone (ICZ) is located in this model around  $\log_{10} T \sim 7.5$. Figure from \citet{Saio06}.}         
\label{saio_2006_fig7}
        \end{figure}

\begin{figure}
        \begin{center}
        \resizebox{0.8\hsize}{!}{\includegraphics  {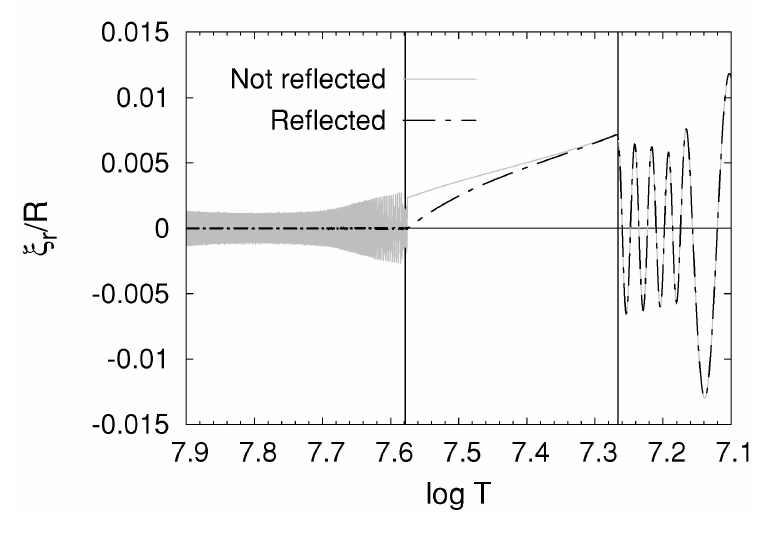}}
        \end{center}    
        \caption{Radial displacement eigenfunction versus $\log_{10} T$ in the vicinity of  the  Intermediate Convection Zone (ICZ) of  a blue supergiant model. The ICZ is shown by vertical solid lines. The grey line represents a $g$ mode that crosses  the ICZ and enters into the radiative core and the  black dashed one represents  a $g$ mode reflected on the ICZ.  Figure from \citet{Godart09}.}        
\label{godart_2009_fig4}
        \end{figure}

As shown by \citet{Godart09} and \citet{Lebreton09}, the existence of such ICZ strongly depends on the strength of the mass loss, the amount of overshooting, and the adopted convection criteria ({e.g.} Ledoux versus Schwarzschild convection criterion). Finally, as an alternative interpretation, \citet{Daszynska13} suggested that the $g$ modes are rather partially reflected at the chemical composition gradient surrounding the radiative helium core. As for the ICZ, this partial reflection prevents them from being strongly damped below in the radiative core.

%   in JWKB\footnote{Refers to Jeffreys, Wentzel, Kramers and Brillouin} approximation.

% in which it has large amplitude, whereas 
% that case, the amplitude in the core is small. The amplitudes in the envelope
% are roughly the same. The behaviour of the eigenfunctions in the centre is
% an artefact coming from our use of the asymptotic approximation (equation
% 10, where N and r come to zero).

\subsection{Strange modes}

\label{strange_modes}

Strange modes were originally found by the numerical study of \cite{Wood76}, who analyzed high luminosity helium stars. At that time, they were not yet  called  ``strange modes''; \cite{Cox80b} named them as such in the study of pulsations in hydrogen deficit carbon stars. After that, many authors have been working on analyses of strange modes in helium stars \citep{Saio84, Saio88, Gautschy90, Gautschy95b, Saio95}, Wolf-Rayet stars \citep{Glatzel93_WR, Kiriakidis96}, massive stars \citep{Gautschy92, Glatzel93_HD, Glatzel93_massive, Kiriakidis93, Glatzel96, Saio98, Saio09, Saio11, Saio13b, Godart10, Godart11, Sonoi14PASJ}, ... etc. In particular, \cite{Sonoi14PASJ} carried out the nonadiabatic analysis with time-dependent convection for massive stars. They found that convection certainly weakens the excitation of strange modes, although the instability still remains.    

In modal diagrams, strange modes show different behaviors from ordinary modes appearing in most pulsating stars. The growth or damping timescale is extremely short and comparable to their oscillation periods. Then, the instability of strange modes might lead to such nonlinear phenomena as mass loss, and might be influential in the stellar evolution. Although nonlinear analyses have been carried out, we have not yet obtained a definitive conclusion \citep{Dorfi00, Chernigovski04, Grott05, Lovekin14}. On the other hand, there are observational candidates for pulsations related to strange modes. Pulsations in a luminous B star, HD50064 \citep{Aerts10}, and in $\alpha$ Cygni variables \citep{Gautschy90, Saio13b} could correspond to strange modes according to their periods.

\subsubsection{Strange modes with and without an adiabatic counterpart}

Previous theoretical studies have found that there are two types of strange modes, with or without a corresponding adiabatic solution, or an adiabatic counterpart. Strange modes with adiabatic counterparts appear due to a narrow acoustic cavity in outer layers. By carrying out the local analysis \citep[{e.g.}][]{Saio98}, we can derive the lowest frequency for the wave propagation of radial pulsations, which writes in a dimensionless form
\begin{eqnarray}
  \omega_c=\frac{c_s}{2H_p}\sqrt{\frac{R^3}{GM}},
\end{eqnarray}
where $\sqrt{R^3/(GM)}$ is the dynamical timescale. In massive main-sequence stars, the opacity bump due to ionization of Fe group elements is formed, and induces convection. Hence, the density gradient with respect to the pressure becomes less steep.  Roughly speaking, the critical frequency $\omega_c$ is proportional to $\sqrt{P/\rho}$. Then, the less steep gradient of  density makes a cavity on the profile of the critical frequency (see Fig. \ref{strange:fig:omega_c}). As the stellar mass increases, the Fe opacity bump becomes more conspicous, and hence the cavity becomes deeper. The acoustic waves are then trapped, and the eigenmode is confined in the cavity. As a result,  the $\kappa$-mechanism due to the Fe bump can efficiently excite the mode. Their growth timescale is extremely short and comparable to their pulsation periods.

  \begin{figure}
        \begin{center}
        \resizebox{\hsize}{!}{\includegraphics  {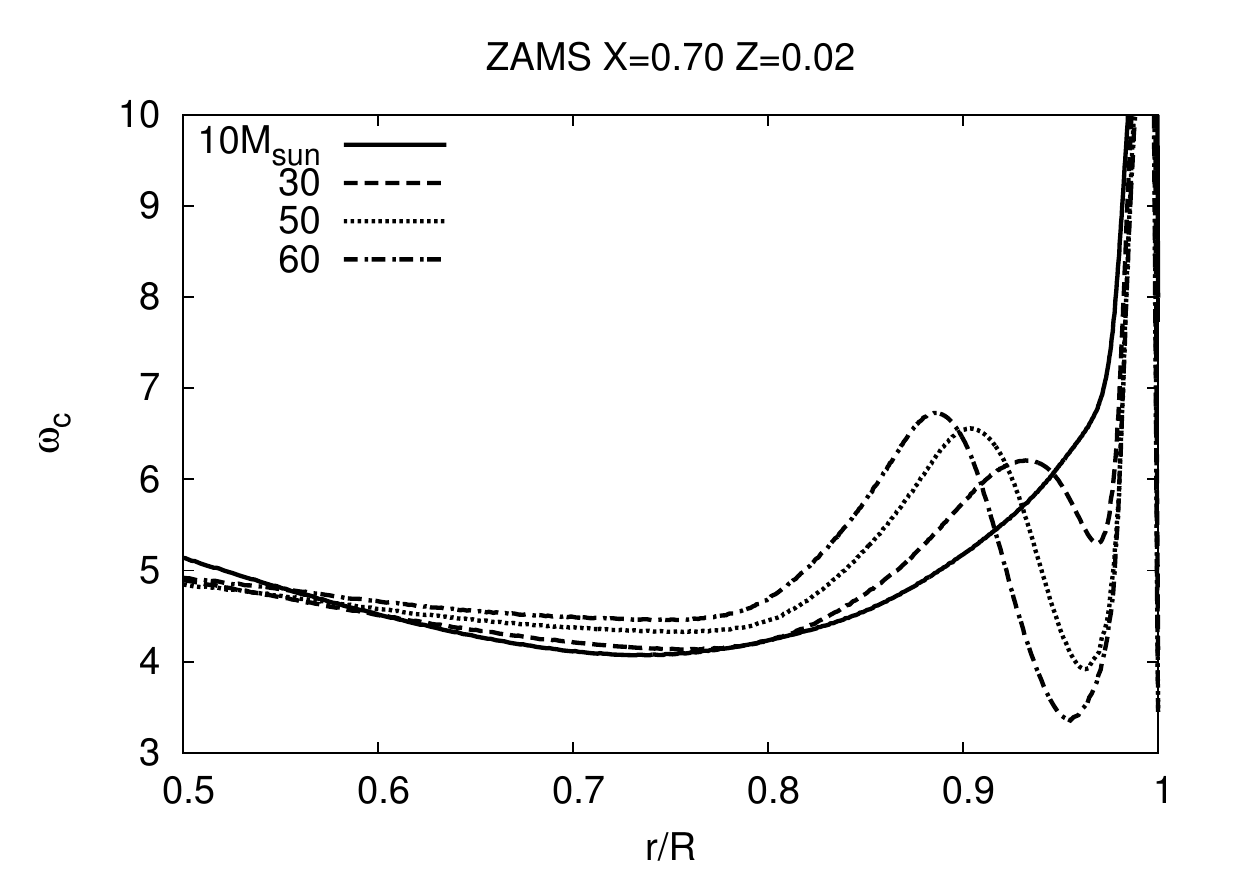}}
        \end{center}    
        \caption{Profiles of the nondimensional critical frequency $\omega_c(=c_s/2H_p \, \sqrt{\frac{R^3}{GM}})$ for radial pulsations in the Zero Age Main Seqnence models with $M=10,\,30,\,50$ and $60M_{\odot}$.}
        \label{strange:fig:omega_c}
        \end{figure}

On the other hand, strange modes without adiabatic counterparts are related to extreme nonadiabaticity in outer layers of very luminous stars. \cite{Wood76} pointed out that there is no one-to-one correspondence between solutions by adiabatic and nonadiabatic analyses of high luminosity helium stars. \cite{Shibahashi81} found that strange modes appear in cases of $L/M\gsim 10^4L_\odot/M_\odot$ by a systematic analysis of models with different $L/M$ ratios. \cite{Gautschy90} found strange modes in the nonadiabatic reversible (NAR) approximation. 
In this  approximation, the heat capacity and hence the thermal timescale are set to zero in the linearized energy equation. In this situation, the radial gradient of the relative luminosity perturbation becomes zero, namely,
\begin{eqnarray}
  \frac{d}{dr}\frac{\delta L}{L}=0,
\end{eqnarray}
if we neglect the term of the nuclear energy generation. Besides, the eigenfrequency and the eigenfunctions should be real, or their complex conjugates should be also eigen solutions if they are complex.
In the NAR approximation, the classical $\kappa$-mechanism can no longer work,  hence an alternative physical explanation of the mechanism for the instability has been needed. It has been called as ``strange-mode instability.'' \cite{Saio98} proposed that the restoring force may be radiation pressure gradient. With a local analysis adopting the plane parallel approximation, they derived an approximate relation 
\begin{eqnarray}
  \delta P\propto i\kappa_\rho\kappa F_{\rm rad}\frac{\delta\rho}{\rho},
\end{eqnarray}
where $F_{\rm rad}$ is the radiative flux and $\kappa_\rho\equiv(\partial\ln\kappa/\partial\ln T)_\rho$. This relation implies that a large phase lag between the pressure and the density perturbations may lead to strong instability. That was also indicated by the local analysis in \cite{Glatzel94}.

\subsubsection{Numerical examples}

    \begin{figure}
        \begin{center}
        \resizebox{\hsize}{!}{\includegraphics  {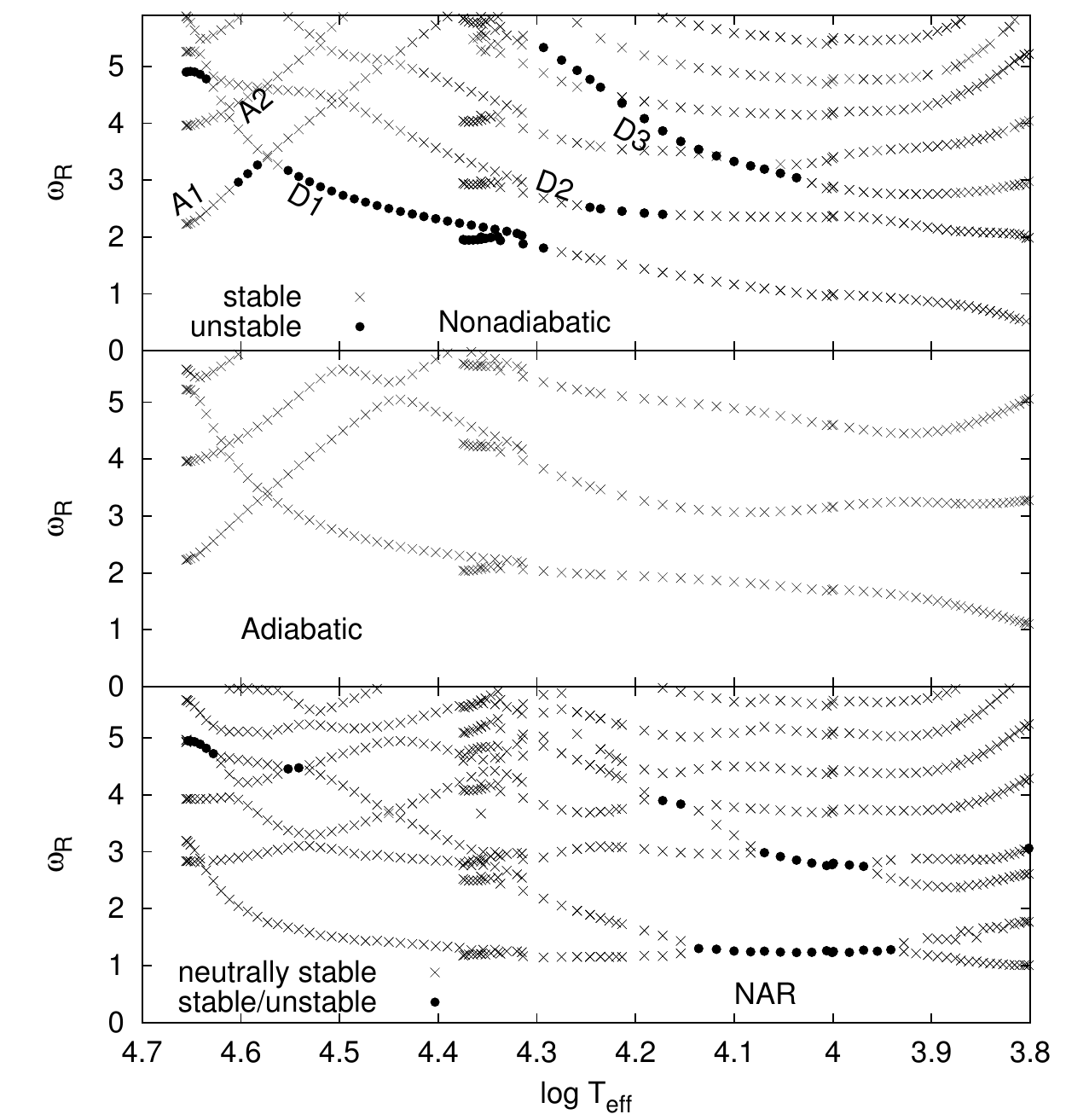}}
        \end{center}    
        \caption{Modal diagrams of radial modes obtained by the nonadiabatic (top panel) and the adiabatic analyses (middle panel) and by the NAR approximation (bottom panel) for $50M_{\odot}$ evolutionary models with $X=0.70$ and $Z=0.02$. The horizontal axis is the effective temperature, which is an indicator of stellar evolution, and the vertical axis is the real part of eigenfrequency normalized by the dynamical timescale, $\sqrt{R^3/(GM)}$.}
        \label{strange:fig:modal}
        \end{figure}

 \begin{figure}
        \begin{center}
        \resizebox{\hsize}{!}{\includegraphics  {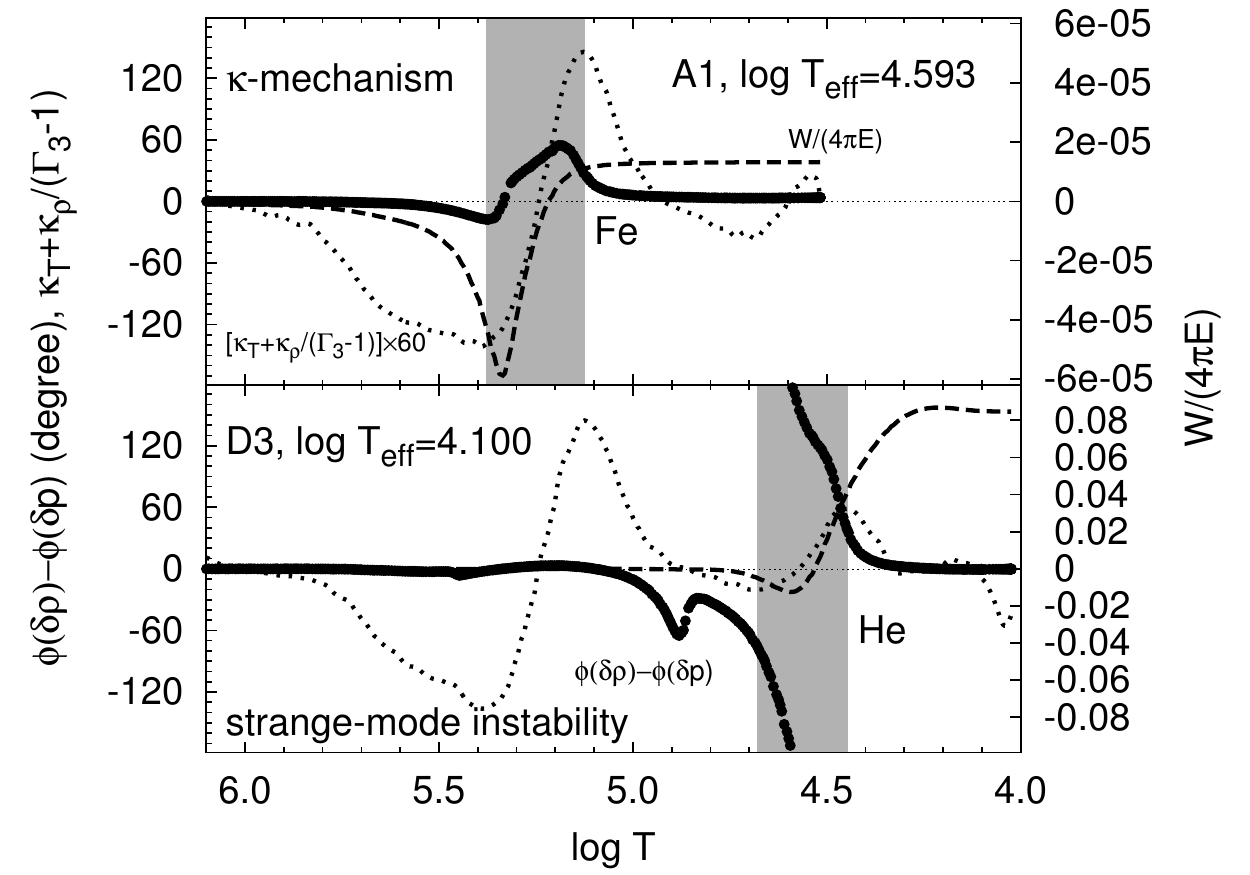}}
        \end{center}    
        \caption{Profiles of strange modes with and without adiabatic counterparts on the A1 and D3 sequences in the top panel of Fig.~\ref{strange:fig:modal}, respectively. The one with an adiabatic counterpart (A1) is excited by the $\kappa$-mechanism around the Fe opacity bump, while  the one  without an adiabatic counterpart (D3) is excited  by the strange-mode instability around and above the He opacity bump. Each panel shows the phase lag between the density and the pressure perturbations in degree (dots), work integral normalized with the kinetic energy of the mode (dashed line), and the quantity related to the opacity derivatives, $[\kappa_T+\kappa_\rho/(\Gamma_3-1)]$ (dotted line). Roughly speaking, the $\kappa$-mechanism should take place in regions with $d[\kappa_T+\kappa_\rho/(\Gamma_3-1)]/d\log T<0$, which are hatched in the panels. The horizontal axis is the logarithmic temperature coordinate.}
        \label{strange:fig:work}
        \end{figure}

Figure \ref{strange:fig:modal} shows radial modes obtained by the nonadiabatic (the top panel) and the adiabatic analyses (the middle panel) and by the NAR approximation (the bottom panel) for evolutionary models of $50~M_{\odot}$ with $X=0.70$ and $Z=0.02$. As we can see, there are sequences ascending and descending with the decrease in the effective temperature. The ascending ones such as the A1 and A2 sequences correspond to ordinary modes, which have the adiabatic counterparts as shown in the middle panel. The descending ones such as D1, D2 and D3, on the other hand, are composed of strange modes. While some of them have the adiabatic counterparts like the D1 sequence, we can find ones which do not appear in the adiabatic case (D2 and D3). 
On the other hand, such sequences are reproduced by the  NAR approximation as shown in the bottom panel.  
% But the NAR approximation seems to realize the features of such sequences as shown in the bottom panel. 
Eigenmodes having complex conjugates appear when two sequences of neutrally stable modes join together. This issue is discussed in detail by \cite{Gautschy90}.

Among the ascending sequences in the top panel of Fig. \ref{strange:fig:modal}, the A1 sequence has unstable modes. They are excited by the $\kappa$-mechanism, and the imaginary part of the eigenfrequency is much smaller than the real part. For modes in the descending sequences, on the other hand, the imaginary part is comparable to the real part. The unstable modes in the D1 sequence, which has the adiabatic counterpart, are excited by the $\kappa$-mechanism, while the strange-mode instability takes place for the ones in the D3 sequence. For ones in the D2 sequence, the $\kappa$-mechanism and the strange-mode instability work together.

Figure \ref{strange:fig:work} shows profiles of an ordinary mode on the A1 sequence and a strange mode on the D3 sequence. For the ordinary mode, as shown in the top panel, the excitation zone, where the work integral increases outward, satisfies the analytically derived condition for the occurrence of the $\kappa$-mechanism  (Eq. 26.14 in \citet{Unno89}, which was derived in a different way from that in  Sect.~\ref{kappa_mechanism}).

 In the top and the bottom panels of Fig. \ref{strange:fig:work}, the hatched zones correspond to the excitation zones by the $\kappa$~mechanism due to the Fe and the He bumps, respectively. For the strange-mode instability, on the other hand, the work integral indicates that the excitation takes place also outside the zone satisfying  $d[\kappa_T+\kappa_\rho/(\Gamma_3-1)]/d\log T<0$  as shown in the bottom panel. 
Although we no longer have the exact periodicity due to the high growth/damping rates, we use the work integral for convenience of knowing the driving/damping regions. Note that the zero or low heat capacity is not a problem in adopting the total work integral \citep[see][]{Glatzel94}. Figure \ref{strange:fig:work} also shows the phase lag between the density and the pressure perturbations. Since the $\kappa$-mechanism is close to adiabatic, the phase lag is not so large. But it is much larger in the strange-mode instability as predicted by the local analyses in the previous studies. Indeed, it increases to 180 degrees in the excitation zone.

\section{Stable and stochastically excited oscillations}
\label{stochastically_excited}

Up to now, we have considered self-excited oscillations (\emph{i.e.} oscillations that result from an instability). We now shift to an other class of pulsating stars whose amplitudes result from a balance between an external driving and damping. This class of oscillations are called \emph{solar-like oscillations} and exhibit very low amplitudes so that they are difficult to observe. It explains why they have been observed for a relatively short time for the Sun (see Sect.~\ref{forewords}) and even shorter for other stars \citep[see][for a comprehensive review]{Chaplin2013}. Nevertheless, in term of seismic diagnostic on the stellar interiors, solar-like oscillations have so far provided  much more information than self-excited oscillations.

Anticipating on the following, we note that such oscillations are driven and damped in the uppermost layers of low-mass stars and more precisely in the super-adiabatic region. With the advent of the space-borne missions CoRoT and \emph{Kepler}, such oscillations have been detected for thousands of stars from the main-sequence to the red-giant phases. Therefore, those observations are currently used to infer the interior properties of the low-mass stars as a function of their evolution \citep[e.g.][]{Mosser2014}. It is thus of prime importance to understand the physical mechanisms responsible for the mode driving and damping. In the following, we will  explain what is the current knowledge on those issues.  

\subsection{Forewords on solar-like oscillations}
\label{forewords}

The mechanism of acoustic noise generation by turbulence is a longstanding problem in fluid mechanics \citep[see][for details]{Lighthill78}. The discovery of solar five-minute oscillations by \cite{Leighton62} and \cite{Evans62}, reinforced by their interpretation in terms of normal modes by \cite{Ulrich70} and \cite{Leibacher71}, made the issue more concrete since these pulsations are excited and damped by turbulent convection. 

Indeed, a solar-like normal mode can be considered as a damped and excited oscillator so that the eigendisplacement $\delta r$ follows 
\begin{align}
\label{eq:oscillator}
\deriv{^2 \delta r}{t^2} + 2\pi \Gamma \, \deriv{ \delta r}{t} + (2\pi)^2 \nu_0 \, \delta r = F(t) \, , 
\end{align}
where $\Gamma$ is the linewidth, $\nu_0$  is the eigenfrequency, and $F(t)$ is the forcing term. The Fourier transform of \eq{eq:oscillator} is thus
\begin{align}
\label{eq:oscillator:fourier}
\widehat{\delta r} = \frac{\widehat{F}(\nu)}{(2\pi)^2 \left[ \nu_0^2 - \nu^2 + i \nu \Gamma \right]} \, ,
\end{align}
where the symbols $\widehat{\delta r}$ and $\widehat{F}$ are the Fourier transforms of the eigen-displacement and forcing term respectively, and $\nu$ is the frequency. In the power spectrum, for $\nu \approx \nu_0$, \eq{eq:oscillator:fourier} can be approximated by a Lorentizan function such as \citep[{e.g.}][]{Baudin05,Appourchaux14}
\begin{align}
\label{eq:oscillator:fourier2}
\left\vert \widehat{\delta r} \right\vert^2 = \frac{H}{1+x^2} \,, \quad {\rm with}\quad x=2(\nu-\nu_0)/\Gamma \, , 
\end{align}
and $H$ stands for the mode height. 

\begin{figure}
\begin{center}
\includegraphics[height=8cm,width=10cm]{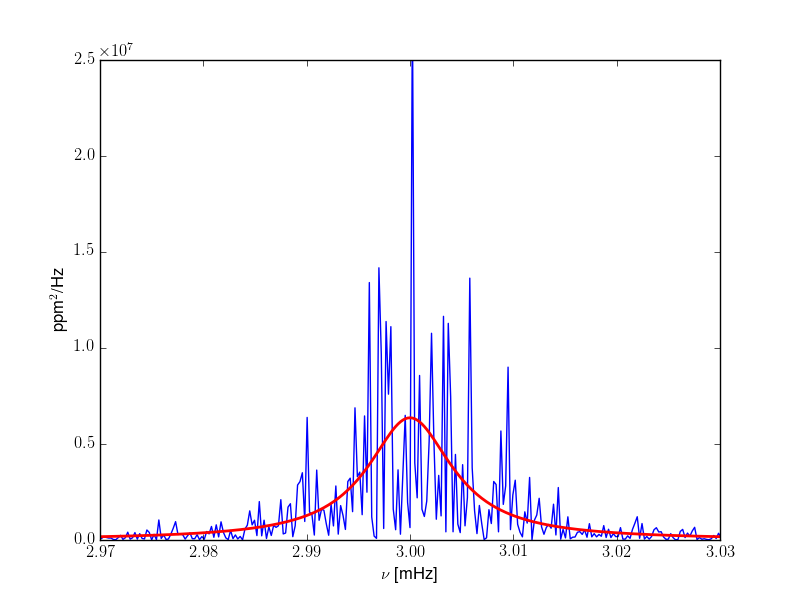}
\caption{Stochastically excited mode at the frequency $\nu_0 = 3$ mHz. The red line corresponds to a Lorentzian schematic profile as described by \eq{eq:oscillator:fourier2}.}
\label{profile_schematic}
\end{center}
\end{figure}

Therefore, as depicted by \eq{eq:oscillator:fourier2}, a solar-like mode exhibits a Lorentzian profile in the Fourier power spectrum. This is a major observational characteristic of such a solar-like mode, compared to an unstable mode that is a sinusoidal function in the time-series (and thus a square sinus cardinal in the power spectrum). Note, however, that \eq{eq:oscillator:fourier2} supposes an observation time duration longer the than mode lifetime defined as $\tau = 1/\eta$. 
Another observational characteristic is their stochastic nature. Indeed, as shown by Fig.~\ref{profile_schematic}, a solar-like mode is characterized, in the Fourier power spectrum, by a \emph{speckle-like Lorentzian} profile that is the result of both the convective driving and damping. 
Finally, modes in solar-like pulsators are mainly driven and damped in the uppermost part of the convective region, \ie, in the superadiabatic region and just below it --- a region near the photosphere and at the transition between the convective and radiative atmosphere. 
In these layers, convection becomes inefficient and convective velocities increase rapidly over a relatively small radial scale to sustain the convective flux. As a result, in this region the convective time-scale reaches a minimum (which is of the order of 5~min for the Sun), while the kinetic flux is maximal. Given the fact that the efficiency of the driving crucially depends on the magnitude of the  kinetic flux and on the convective time-scale (see Sect.~\ref{driving}), acoustic modes with periods of the order of a few minutes can be efficiently excited in the uppermost part of the convective region.

%Concerning the thermal structure of the uppermost layers, the thermal time-scale (also called the thermal relaxation time scale) is also of the order of 5 mins. This is of the utmost importance since it defines how the star's structure reacts to any thermal perturbation and will, as we will see later, define the strength of thermal normal-mode leakages (see Sect.~\ref{damping}).  

%In the Sun \textcolor{red}{(generalized au solar-like)}, several millions of modes (stochastically excited by turbulent convection as well as damped) have been detected. The maximum amplitude of the excitations is found at a frequency of about $3$ mHz with typical values of amplitudes of $2.5$ part per million (ppm) (see \cite{Michel09} for details). In terms of mode surface oscillation velocities, the observed maximum amplitude is about $20$ cm.s$^{-1}$, which immediately allows us to deduce that solar (and solar-like) oscillations are small perturbations of turbulent convection; thus justifying the widely-used and commonly-accepted linear approximation to model them. More interestingly, the frequency of the peak amplitude (around $3$ mHz) corresponds to a time-scale of about $5$ mins. For mode damping the frequencies are found to be of the order of several $\mu$Hz, corresponding to life-times of several days. 

In the following, after exhibiting the relations between mode energy and the observables in Sect.~\ref{mode_energy}, we will focus on the physical mechanisms responsible for both mode damping (Sect.~\ref{mode_damping}) and mode driving (Sect.~\ref{mode_driving}).

\subsection{Seismic constraints: relation between mode energetics and observables}
\label{mode_energy}

%In this section, our aim is to exhibit the relation between mode energy and observables (\emph{i.e.} the mode height, $H$, and the mode linewidth, $\Gamma$). More precisely, we will show how those observables are related to the driving and damping process.

\subsubsection{Relation between mode energy, mode driving, and mode damping}
\label{steady_state}

To this end, we first formally admit that the mode energy follows an equation of the type
\begin{align}
\label{E_osc_dt}
\deriv{E_{\rm osc}}{t} = \mathcal{P} + \mathcal{D}
\end{align}
where $\mathcal{P} $ stands for the driving and more precisely for the amount of energy injected 
per unit of time into a mode by an arbitrary source, $\mathcal{D}$ stands for the damping. 
To go further, let us now dwell  on the mode damping by first recalling that the mode total energy (potential plus kinetic) 
is  
\begin{align}
E_{\rm osc} (t)  =    \int \rho_0 \,    \vec v_{\rm osc}^2 (\vec r,t)  \, {\rm d}^3 \vec r \, , 
\label{E_osc_3}
\end{align}
where $\vec v_{\rm osc}$ is the mode velocity at the position $\vec r$ and the instant $t$, 
and $\rho_0$ is the mean density. 

Mode damping occurs over a time-scale much longer than  the time-scale associated with the driving.
Accordingly, damping and driving occurs on two different characteristic time-scales and thus can be decoupled 
in time. In addition, we assume a constant and linear damping such that, over a time scale much larger than 
the characteristic time-scale of the driving, one gets
\begin{align}
  \deriv{ \vec v_{\rm osc} (t) }  {t}  = - \eta \; \vec v_{\rm osc} (t) \, , 
\label{damping}
\end{align}
where $\eta$ is the (constant) damping rate.  The time derivative in \eq{damping} is
performed over a time scale much larger than  the characteristic time
over which the driving occurs.  Consequently, using \eq{damping}, the time derivative of \eq{E_osc_3} is injected into \eq{E_osc_dt} so that
\begin{align}
\label{balance} 
\deriv{E_{\rm osc}}{t} (t)  =  {\cal P} - 2\ \eta \, E_{\rm osc} (t) \;.
\end{align}

Solar-like oscillations are known to be stable over time. 
 As a consequence, their energy cannot grow on a time scale much 
longer than the time scales associated with the damping  and driving process. 
Accordingly, averaging \eq{balance}  over a long time scale gives 
\begin{align}
\label{balance_2}
\overline{\deriv{E_{\rm osc}}{t} (t) } = 0 \quad \Longrightarrow  \quad \overline{ E}_{\rm osc}    =   \frac{\overline{ {\cal P} }}{2 \eta} \, , 
\end{align}
where the notation $\overline{()}$ refers to a time average. We then clearly see from \eq{balance_2} that a steady state implies that \emph{the mode 
energy is controlled by the balance between driving and damping.} 

%Then, the major difficulties are to model the processes that are at the origin of the driving and the damping.  
%For ease of  notation, we will  drop from now on the symbol $\overline{()}$ from $E_{\rm osc}$ and ${\cal P}$. 

\subsubsection{Relation between mode energy, mode amplitude, and mode mean-squared surface velocity} 
\label{mode_eosc_amplitude}

It is worth emphasizing the relation between the mode energy ($E_{\rm osc}$) and the mode amplitude (hereafter denoted by $A$). 
To this end, the mode displacement, $\delta \vec{r}_{\rm osc}$, can be written in terms of the adiabatic eigen-displacement $\vec \xi$,  
 and the instantaneous amplitude $A(t)$
\begin{align}
 \delta \vec r_{\rm osc}  \equiv {1 \over 2} \, \left ( A(t) \,  \vec {\xi} (\vec r) \,
 e^{-i \omega_{\rm osc} t } +cc \right ) \, ,
\label{delta_osc_2}
\end{align}
where $cc$ stands for the complex conjugate, $\omega_{\rm osc}$ is the mode eigenpulsation, and $A(t)$ is the instantaneous
amplitude resulting from both the driving and the damping mechanisms. Note that, since the normalisation of $\xi$ is arbitrary, the actual \emph{intrinsic} 
mode amplitude is fixed by the term $A(t)$, which remains to be determined.

%The  mode velocity, $\vec v_{\rm osc}$ , is then given by:
%\eqn{
%\vec v_{\rm osc}\, (\vec r,t) = \deriv{\delta \vec{r}_{\rm osc}}{t}  =    {1\over 2} 
%(  -i \omega_{\rm osc} \, A(t) \,  \vec {\xi} (\vec r) \,  e^{-i \omega_{\rm osc} t } +cc)
%\label{At}
%}
%where cc means complex conjugate. Note that we have neglected in
%\eq{At} the time derivative of $A$.  This is justified since the mode
%period ($2\pi/\omega_{\rm osc}$)  is in general much shorter than the mode
%lifetime ($\sim 1/\eta$)

It is then possible to write the mode energy by using \eq{E_osc_3} and the time derivative of 
\eq{delta_osc_2}\footnote{Note that the time derivative of $A$ is neglected since the mode
period ($2\pi/\omega_{\rm osc}$) is in general much shorter than the mode lifetime ($\sim 1/\eta$)} 
so that
%From Eqs.~(\ref{At}) and (\ref{E_osc}), we derive the expression for
the mean mode energy reads 
\begin{align}
{E}_{\rm osc} =  \frac{1}{2} \, \overline{ \vert A \vert ^2 }  \,  I  \, {\omega_{\rm osc}}^2 \, ,
\label{E_osc_4}
\end{align}
where $\overline{ \mid A \mid ^2 }$ is the mean squared amplitude, and $I$ is christened mode inertia
\begin{align}
 I \equiv   \int_0^{M} \vec \xi^\ast \cdot \vec \xi \; {\rm d}m \, .
\label{inertia}
\end{align}

The mean squared velocity can also be expressed as a function of the mean squared amplitude by using the time-derivative of 
\eq{delta_osc_2}, so that 
\begin{align}
\label{v_s_sq}
\vec{v}_s^2 (r_h) =  \, \frac{1}{2}  \overline{ \vert A \vert^2} \, \omega_{\rm osc}^2 \, \left( \vert \xi_{\rm r} (r_h) \vert^2 + \ell (\ell+1) \vert \xi_{\rm h} (r_h) \vert^2 \right)\, , 
 \end{align}
where $r_h$ is the radius at which the mode velocity is measured, $\ell$ is the mode angular degree, $\xi_r$ and $\xi_h$ are the radial and horizontal components of the eigenfunction, respectively. Note that to derive \eq{v_s_sq} the eigenfunction has been projected onto the spherical harmonics and integration over the solid angle has been performed. 

Using \eq{E_osc_4} and \eq{v_s_sq}, the relation between the mode energy and the mean squared surface velocity is given by 
\begin{align}
{E}_{\rm osc} = {\cal M}  \, \vec{v}_s^2 \, , 
\label{E_osc_5}
\end{align}
where ${\cal M}$ is the mode mass, related to the inertia by
\begin{align}
 {\cal M} (r_h)  \equiv  { I \over {   \mid \xi_{\rm r}  (r_h) \mid^2 }} \, .
\end{align}
It should be noticed, that although ${\cal M}$ and $v_s$ depend on the
choice for the radius $r_h$, ${E}_{\rm osc}$ is by definition
intrinsic to the mode  and hence is independent of 
$r_h$.

\subsubsection{Relation between mode mean-squared surface velocity, mode height, and mode linewidth} 
\label{mode_vosc_gamma_H}

Using  Eqs.~(\ref{balance_2}), and (\ref{E_osc_5}), we derive 
\begin{align}
\vec{v}_s^2 (r_h , \omega_{\rm osc}) = \frac{ \mathcal P}{2 \, \pi \, \mathcal M \, \Gamma} 
\label{v_s}
\end{align}
where $\Gamma=\eta/\pi$ is the mode linewidth. From \eq{v_s}, one again sees that the mode surface velocity is the result of the balance between excitation ${\cal P}$ and damping. However, it also depends on the mode mass ${\cal M}$: for a given driving (${\cal P}$) and damping ($\Gamma$), the larger the mode mass (or the mode inertia), the smaller the mode velocity. 

When the frequency resolution and the signal-to-noise are high enough,
it is possible to resolve the mode profile and then to measure \emph{both}
$\Gamma$ and the mode height $H$ in the power spectral density. In that case $v_s$ is given by the relation \citep[see {e.g.}][]{Baudin05} 
\eqna{
v_s^2 (r_h , \omega_{\rm osc})  & = \pi \,C_{\rm obs} \, H \, \Gamma 
\label{v_s_2} 
}
where the constant $C_{\rm obs}$ takes the observational technique and
geometrical effects into account \citep[see][]{Baudin05}.
%The integration of power spectral density is performed over $] -\infty , + 
%  \infty [$  to take both the negative and the positive side of the
%    spectrum into account. 
From \eq{v_s} and (\ref{v_s_2}), one can then infer from the observations
the mode excitation rate ${\cal P}$ as 
\eqn{
{\cal P}   (\omega)  =  2
\pi^2 \, {\cal M} \, C_{\rm obs} \,H \, \Gamma ^2  \; .
\label{pow_obs}
}
Provided that we can measure $\Gamma$ and $H$, it is then possible to constrain ${\cal P}$. However, we point out that
the derivation of ${\cal P}$ from the observations is also based on models since  ${\cal M}$ is required. 
Furthermore, there is a strong anti-correlation between $H$ and $\Gamma$ \citep[see {e.g.}][]{Chaplin98,Chaplin08}, which can introduce
important bias. This anti-correlation vanishes when considering the squared mode amplitude, $v_s^2$, since $v_s^2 \propto  H \, \Gamma$ (see \eq{v_s_2}).  However, ${\cal P}$ still depends on $\Gamma$, which is strongly anti-correlated with $H$. 

As an alternative, it is possible to compare theoretical results and observational mode heights ($H$) 
as proposed by \cite{Chaplin05},  according to the relation 
\eqn{
 H = { {\cal P } \over {2 \pi^2 \, {\cal M} \, C_{\rm obs} \,\Gamma ^2 \, .
   } } 
\label{H}
}
However, as  pointed-out by \cite{Belkacem06b},  $H$ strongly depends on the observation technique.
The quantity $ C_{\rm obs} \,H $ is less dependent on the observational data  but still depends on the instrument since different instruments probe different layers of the atmosphere (see below). Therefore, it is difficult to compare values of $H \,C_{\rm obs} $ coming from different instruments.

\subsection{Damping of solar-like oscillations}
\label{mode_damping}

The relation between the mode energy (or mode amplitude, mean-squared surface velocity) and the observables being now clarified, it is now worth dwelling on the physical mechanisms responsible for mode damping.  

\subsubsection{Setting the stage}
\label{mode_damping_historical}

From a historical point of view, our knowledge about the underlying physics of mode damping does not follows a linear path. The first non-adiabatic  calculations \citep[{e.g.}][]{Ando75} were unable to reach a clear conclusion and the issue of mode stability was still pending. However, important  observational efforts made it indisputable that solar $p$-modes are stable due to observational evidence of Lorentzian mode profiles \cite[{e.g.}][]{Toutain1992}. It thus emphasised the need of an extra physical ingredient to stabilise solar $p$-modes. 

In this framework, \cite{Goldreich91} proposed that the shear due to Reynolds stresses, modeled by an eddy-viscosity, is of the same order of magnitude as the non-adiabatic component of the perturbation of gas pressure. \cite{Gough80} and \cite{Balmforth92a} found that the damping is dominated by the modulation of turbulent pressure, while \cite{MAD05}, \cite{MAD06c}, and \cite{Belkacem2012} also include the perturbation of the dissipation rate of kinetic energy into heat that acts to compensate the perturbation of turbulent pressure. Therefore, there is still no clear picture concerning the physics of mode damping. It is however likely that the main contributions have been identified, but their relative contribution as well as the possible mutual cancellations are still an issue.  

It seems that the main shortcut available in modeling mode-damping rates is the way in which turbulent convection is described. One major deficiency of these formalisms is that they use the mixing-length theory. It thus reduces the whole of the turbulent cascade to a single length-scale. While this can be an acceptable assumption for modeling the convective background, the perturbation of the mixing-length cannot account for the relation between oscillations and the turbulent cascade.  \cite{Xiong00} proposed an alternative approach using a Reynolds stress formalism \citep[{e.g.}][]{Canuto92} to model convection and, using a perturbation method, computed mode damping rates. However, in their analysis some modes are found unstable, contrary to the observational evidence.  

%%\textcolor{green}{Petite discussion sur les derniers résultats obs de kepler et corot + sous-geantes et geantes radiatif}

\subsubsection{Coupling between convection and oscillation}
\label{mode_damping_coupling}

\begin{figure}[t]
\begin{center}
\includegraphics[height=8cm,width=10cm]{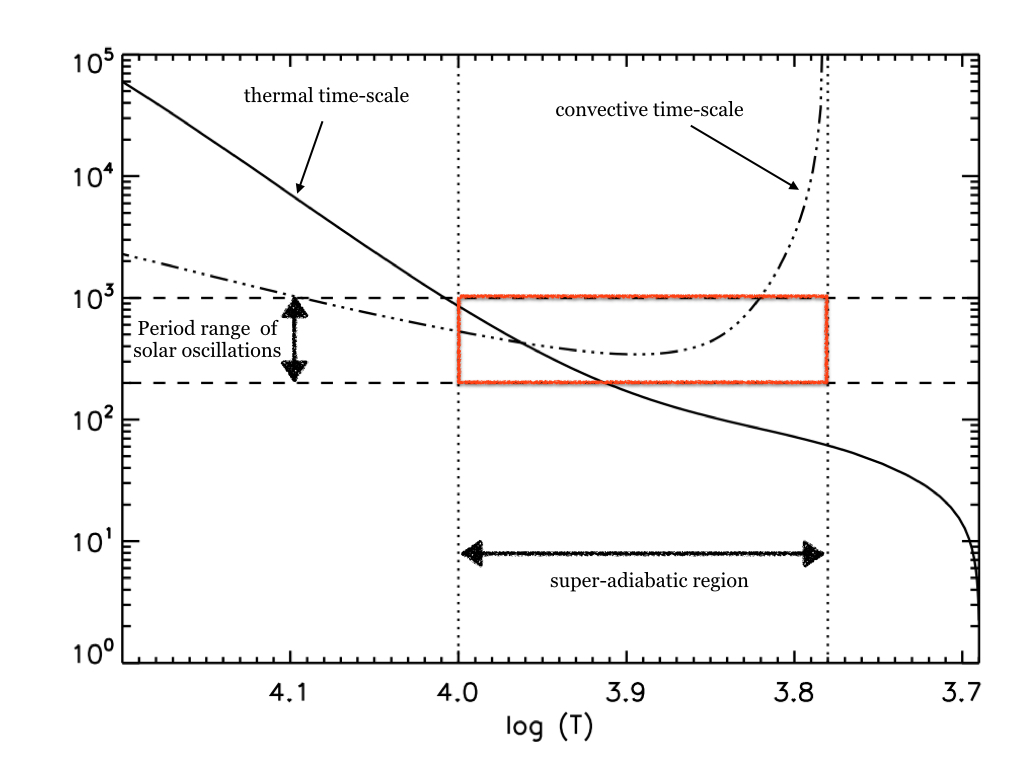}
\caption{Characteristic time-scales as a function of the logarithm of the temperature for a solar model. The solid line corresponds to the thermal time-scale computed as described in Sect.~\ref{timescales} and the dashed dotted line corresponds to the convective time-scale. The red rectangle delimits the location of the triple resonance. }
\label{profile_timescales}
\end{center}
\end{figure}

As already explained in Sect.~\ref{mode_damping_historical}, the calculation of mode-damping rates is a difficult and still unsettled problem for solar-like stars. The assumption of adiabatic pulsation must be abandoned, resulting in a higher-order problem to solve. Indeed, a measure of the degree of non-adiabaticity can be obtained by comparing the thermal time-scale $\tau_{\rm th}$ as already introduced in Sect.~\ref{timescales}, to the modal period. If both time-scales are of the same order of magnitude, the adiabatic assumption is no longer valid and one must consider the full non-adiabatic equations to account for the energy exchange between the oscillations and the background. Last but not least, convection must be also considered as a leading factor.  If one compares the convective time-scale to the thermal time-scale and modal period, one finds for solar-like pulsators in the super-adiabatic regions
\begin{align}
P_{\rm osc} \simeq \tau_{\rm th} \simeq \tau_{\rm conv} \, , 
\end{align}
This relation, even if not accurate, will be named \emph{the triple resonance}. This is illustrated for the Sun in Fig.~\ref{profile_timescales}. Consequently, the computation of mode damping rates requires to account for the full non-adiabatic equations together with turbulent convection. To the authors knowledge, only a few codes are currently able to do this in the framework of present description of convection in stars \citep[see for instance the review by ][]{Houdek08}. 

From a conceptual point view, the computation of mode damping rates can be derived as displayed by Fig.~\ref{sketch}. The first step is to separate the full hydrodynamical equations into two sets of equations related to the mean structure and to the convective fluctuations. Both systems of equations are related to each other. For instance, turbulent convection induces an additional contribution to the pressure named \emph{turbulent pressure}. This additional pressure thus modifies the hydrostatic equilibrium and therefore the stratification. 
The main difficulty does not lie in the modeling of the mean equations but rather in the modeling of turbulent convection. Indeed, a realistic modeling of turbulent convection is still challenging except if one uses 3D hydrodynamical simulations. 

For 1D approaches, several types of models have been proposed  based on the mixing length theory \citep[{e.g.}][]{Gough77,Unno67} as well as on a Reynolds stress approaches \citep[{e.g.}][]{Xiong89,Canuto92}. Based on these models, one needs to develop a formalism to compute non-adiabatic equations including the coupling with convection. Indeed, one of the key elements is the coupling between convection and oscillation, which requires a time-dependent treatment of turbulent convection. In addition, one must be able to determine the perturbation induced by the oscillations on convection and subsequent feedback of perturbed convection on the oscillations (see Fig.~\ref{sketch}). To our knowledge, there are mainly three types of formalisms able, up to now, to compute non-adiabatic oscillations including the coupling with convection for low-mass stars. The first one has been used for instance by \cite{Gough80,Balmforth92a,Houdek99} and is based the mixing-length formalism proposed by \cite{Gough77}. The second is derived from the \cite{Unno67,Unno77} convective model and was extended by  \cite{Gabriel96} and \cite{MAD05}. The last formalism is based on the Reynolds stress model of convection by \cite{Xiong89} \citep[{e.g.}][]{Xiong00}. 

In these lecture notes, our objective is not to provide a full account for the different treatments of time-dependent convection since it would require full review on the subject. In the following, we essentially base our discussion on the formalism developed by \cite{MAD05} and mainly discuss the potential physical mechanisms responsible for mode damping and the available observational contraints at our disposal. 

\begin{figure}[t]
\begin{center}
\includegraphics[height=8cm,width=10cm]{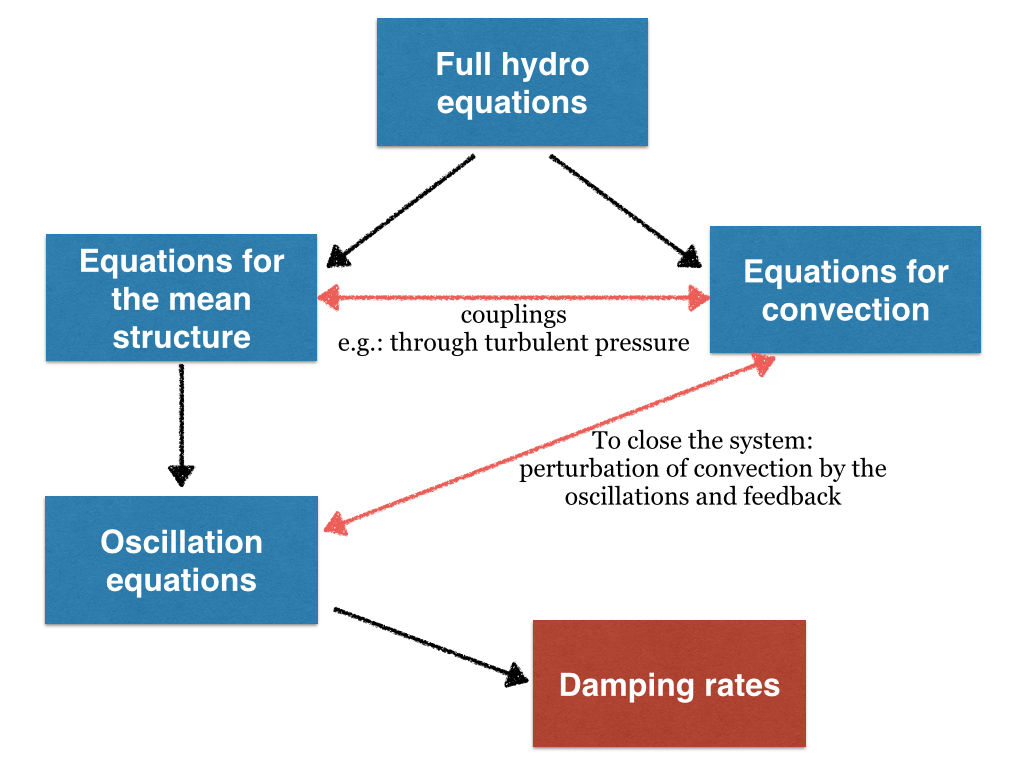}
\caption{Schematic sketch displaying the derivation of the mode damping rates from the hydrodynamical equations. The black arrows illustrate derivation processes while the double red arrows illustrate a coupling.}
\label{sketch}
\end{center}
\end{figure}

\subsubsection{Main contributions to the mode damping rates}
\label{mode_damping_contributions}

%In this section, our ambition is not to provide an comprehensive view of the non-adiabatic problem but rather to discuss the different physical processes at %work. Consequently, it is worthwhile to explicitly derive those contributions.

Let us start with the fluid equations. Neglecting viscous effects, rotation, and magnetic field, the governing equations reads
\begin{align}
\label{damp_01}
\frac{\partial \rho}{\partial t} + \vec \nabla \cdot \left( \rho \vec v \right) &= 0 \, , \\ 
\label{damp_02}
\frac{\partial v}{\partial t} + \vec v \cdot \vec \nabla \vec v + \nabla \psi + \frac{1}{\rho} \vec \nabla P&=0 \, , \\
\Delta \psi - 4\pi G \rho &= 0 \, , \\
\label{damp_03}
T \frac{{\rm d} S}{{\rm d}t} - \epsilon - \frac{1}{\rho} \vec \nabla \cdot \vec F &= 0\, , 
\end{align}
where $\rho$ is the density, $\vec v$ the fluid velocity, $\psi$ the gravitational potential, $G$ the gravitational constant, $T$ the temperature, $S$ the specific entropy, $P$ the total pressure (\ie, the gas, turbulent, and radiative pressure), $\epsilon$ the rate of energy generation, and $\vec F$ the energy flux. Note that one must also include equations to describe the radiative and the convective flux, as well as the equation of state. 

To go further, we perturb the set of Eqs.~(\ref{damp_01})--(\ref{damp_03}). To this end, several additional assumptions are required. First, we suppose that the background is at rest, \ie, any dynamical processes such as convection are ignored except through the inclusion of the convective flux in the energy equation or turbulent pressure in the momentum conservation equation. 
%This is indeed an inconsistent procedure since a perturbation from the mean structure can be related to the oscillatory or convective motion. 
%However, it makes the problem tractable since retaining it prevents us from solving 
%the problems of convection and oscillation simultaneously. %Moreover, such a simplification permits us to separate the issues of mode damping and driving (as already mentioned in Sect.~\ref{intro_part2}).  
To perform the perturbation procedure, it is useful to use the Lagrangian variation projected onto spherical harmonics, such that any quantity $y(\vec r,t)$ reads
\begin{align}
\delta y(\vec r,t) = \delta y(\vec r) \, Y_\ell^m(\theta,\phi) \, e^{i\sigma t} \, , 
\end{align} 
where $ Y_\ell^m$ is the spherical harmonic with degree $\ell$ and azimuthal number $m$.
Note that the frequency $\sigma=\omega_{\rm osc} + i \, \eta$ is a complex quantity, where $\omega_{\rm osc}$ is the modal frequency and $\eta$ the damping rate. To avoid superfluous complexity, we will restrict our discussion to radial modes and radiative pressure will be neglected.  Finally, the perturbed equations governing the problem are found to be \citep[see][for a detailed derivation]{MAD05}
\begin{eqnarray}
\label{damp_04}
\frac{\delta \rho}{\rho} + \frac{1}{r^2} \frac{\partial}{\partial r} \left( r^2\, \xi_r \right)  &=0 \, , \\ 
\label{damp_05}
  \sigma^2 \xi_r -\frac{{\rm d} \delta \psi}{{\rm d}r} - \frac{1}{\rho} \frac{{\rm d}\delta P}{{\rm d}r} - g \frac{\delta \rho}{\rho} &= 0 \, , \\
\label{damp_06}
\sigma^2 r \, \xi_h - \delta \psi - \frac{\delta P}{\rho} &= 0 \, , \\
\label{damp_07}
  i \sigma T \, \delta S + \frac{{\rm d}\delta L}{{\rm d}m} 
 + \delta \Big[ \beta_{\rm g} \otimes \vec \nabla \vec V + \vec V \cdot \vec \nabla P_g \Big] &= 0\, , 
\label{eqs01}
\end{eqnarray}
where $\otimes$ stands for the tensorial product, $\delta$ is the Lagrangian perturbation associated with the oscillation, $\xi_r, \xi_h$ are the radial and horizontal eigenfunctions of the displacement, $\vec V$ is the convective velocity, and $\beta_{\rm g}$ is the non-diagonal part of the gas pressure tensor. 
To derive Eqs.~(\ref{damp_04})--(\ref{damp_07}) it was assumed that turbulence is isotropic, and $\epsilon$ is null. We recall that one also has to add an expression for the perturbation of the radiative and convective fluxes as well as the perturbed equation of state. 

%$\delta \rho$ is the perturbation of density,  $P_g$ the gas pressure, $\xi_r, \xi_h$ the radial and horizontal eigenfunctions, $\delta \psi$ is the perturbation of the gravitational potential, $\delta P$ the perturbation of total pressure,  $\delta S$ the perturbation of specific entropy, $\delta L$ the perturbation of luminosity (including both radiative and convective luminosity), $\vec V$ the convective velocity, and $\beta_{\rm g}$  the non-diagonal part of the gas pressure tensor. 
%To derive Eqs.~(\ref{damp_04})--(\ref{damp_07}) it was assumed that turbulence is isotropic, and $\epsilon$ is null. We recall that one also has to add an expression for the perturbation of the radiative and convective fluxes as well as the perturbed equation of state. 

Hence, it is possible to write down the integral expression of mode-damping rates by combining \eq{damp_05} (multiplied by $\xi_r^\star$) with \eq{damp_06}, then integrating over the star mass. This gives, finally,
\begin{align}
\label{damp_final}
\eta = \frac{1}{2\omega_{\rm osc} I} \int_{0}^{M}  \mathcal{I}m \left[ \frac{\delta \rho}{\rho}^* \frac{\delta P}{\rho} \right] {\rm d}m \, , 
\end{align}
where the star denotes the complex conjugate, $\mathcal{I}m$ the imaginary part, and $I$ the mode inertia. 

Equation~(\ref{damp_final}) is the integral expression of the damping rates. To get more insight into the physical mechanisms at work, it is useful to rewrite  \eq{damp_final} by first noting that the perturbation of total pressure is the sum of gas and turbulent pressure ($\delta P = \delta P_{\rm turb} + \delta P_g$). In addition, we use the thermodynamic relation
\begin{align}
\label{damp_thermo}
\frac{\delta P_g}{P_g} &= P_T \frac{\delta S}{c_v} + \Gamma_1 \frac{\delta \rho}{\rho} \, , 
\end{align}
where 
\begin{align}
P_T = (\Gamma_3 - 1) \frac{c_v \rho T}{P} \quad  {\rm and} \quad
 \left(\Gamma_3 - 1\right) = \left(\frac{\partial \ln T}{\partial \ln \rho}\right)_s \, , 
\end{align}
then Eq.~(\ref{damp_final}) together with \eq{damp_thermo} allows us to express the damping rate in a more explicit form 
\begin{equation}
\label{damp_08}
\eta = \frac{1}{2 \, \omega_{\rm osc} I} \int_{0}^{M} \mathcal{I}m \left[
 \left(\frac{\delta \rho}{\rho}^* T \delta S\right) \left(\Gamma_3 - 1\right) 
+ \left(\frac{\delta \rho}{\rho}^* \frac{\delta P_{\rm turb}}{\rho}\right) \right] \textrm{d}m \, .
\end{equation}
One can then go a step further by inserting \eq{damp_07} (where the last term of this equation is named $\delta \epsilon_2$ for short) into \eq{damp_08}, to obtain 
\begin{align}
\label{final}
\eta &= \underbrace{\frac{1}{2 \, \omega_{\rm osc} I} \int_{0}^{M} \mathcal{I}m \left(\frac{\delta \rho}{\rho}^* \frac{\delta P_{\rm turb}}{\rho}\right) \textrm{d}m }_{\rm turbulent \; pressure \; contribution}
+ \underbrace{\frac{1}{2 \, \omega_{\rm osc}^2 I} \int_{0}^{M} \mathcal{R}e \left[(\Gamma_3-1) \frac{\delta \rho}{\rho}^* \deriv{\delta L_c}{m} \right] \textrm{d}m}_{\rm convective\;  flux  \; contribution} \nonumber \\
&+ \underbrace{\frac{1}{2 \, \omega_{\rm osc}^2 I} \int_{0}^{M} \mathcal{R}e \left[(\Gamma_3-1) \frac{\delta \rho}{\rho}^* \deriv{\delta L_r}{m} \right] \textrm{d}m}_{\rm radiative\;  flux  \; contribution}
- \underbrace{\frac{1}{2 \, \omega_{\rm osc}^2 I} \int_{0}^{M} \mathcal{R}e \left[(\Gamma_3-1) \frac{\delta \rho}{\rho}^* \delta \epsilon_2 \right] \textrm{d}m}_{\rm dissipation\; of \; kinetic \; energy \; contribution} \, ,
\end{align}
where $\mathcal{R}e$ refers to the real part. 

The first term of \eq{final} is the contribution of turbulent pressure, which originates from the perturbation of the mean part of the Reynolds stress tensor. The oscillation loses part of its energy by performing a work $\delta P_{\rm turb} {\rm d}V$, where the variation of volume ${\rm d}V$ induced by the oscillation are related to the mode compressibility $\vec \nabla \cdot \vec \xi = - \delta \rho / \rho$.  These losses of energy are mainly controlled by the phase differences between $\delta \rho$ and $\delta P_{\rm turb}$. The second term of \eq{final} is the damping associated with the perturbation of the convective heat flux. This contribution is certainly the more complex to evaluate, since it strongly depends on how the convection and oscillations are coupled, and consequently it depends on the dynamic modeling of convection. The third contribution to the damping rates is related to the perturbation of the radiative flux. It contains two dominant terms: the opacity effect that is responsible for the instability of the self-excited modes but negligible in solar-type stars (see Sect.~\ref{kappa_mechanism}), and a contribution related to temperature fluctuations $\delta T$. Finally, the last contribution of \eq{final} is the contribution to the damping associated with the perturbation of the dissipation rate of turbulent kinetic energy into heat. This contribution was introduced by \cite{Ledoux58} and more recently by \cite{MAD05}; it partly compensates the effect of turbulent pressure and, in the limit of a fully ionised gas in which radiative pressure can be ignored, the sum vanishes. 
Note that \eq{final} contains what is nowadays considered as the dominant contributions, but additional possible sources of damping have been investigated and discussed \citep[see][]{Houdek99,Belkacem2011LNP}.

\subsection{Driving of solar-like oscillations}
\label{mode_driving}

Let us now consider the problem of mode driving. A first attempt to model mode driving and to explain the observed solar five minute oscillations was carried out by \cite{Unno62} and was followed by \cite{Stein67}. The latter generalised the approach of \cite{Lighthill52} to a stratified atmosphere, and concluded that the Reynolds stresses should be the major source of acoustic wave. Except for a transient debate on the relative contributions of the Reynolds stresses and the non-adiabatic part of gas pressure (or entropy term), this conclusion is still favoured today \citep[see for details][]{Samadi11}. A noticeable leap forward has been  made by \citet{GK77b}. Despite an under-estimation of the observed amplitudes as shown by \cite{Osaki90}, the work of \cite{GK77b} still constitutes the foundation of the current formalisms for modeling mode driving. Since these pioneering works, different improved models have been developed \citep{Dolginov84,Balmforth92c,GMK94,Samadi00I,Chaplin05,Samadi02II,Belkacem06b,Belkacem08,Belkacem2010}. These approaches differ from each other in the way  either the turbulent convection or the excitation processes are described.  

\subsubsection{Modeling mode driving}
\label{driving}

Let us start from the perturbed momentum and continuity equations
\eqna{
\derivp{\rho \vec v  } {t}  + \vec \nabla \cdot ( \rho \vec v \vec v ) + 
\vec \nabla P_1 - \rho_ 1 \vec g_0   = 0
\label{perturbed_momentum_eqn}
\\
\derivp{ \rho_1} {t} + \vec \nabla . (  \rho \vec v) = 0 \; ,
\label{perturbed_continuity_eqn}
}
where $P$, $\rho$, $\vec v$ and $\vec g$ denote, respectively, the gas pressure, 
density, velocity and gravity. In Eqs.~(\ref{perturbed_momentum_eqn}) and (\ref{perturbed_continuity_eqn}) the subscript $1$ denotes Eulerian fluctuations and the subscript $0$, equilibrium quantities. These equations must be supplemented by an Eulerian description of the perturbed equation of state:
\eqna{
P_1 = c_s^2 \rho_1 +    \alpha_s  s_1 + {\cal R}(\rho_1,s_1) \, , 
\label{perturbed_state_eqn}
}
where  $s$ is the entropy, $\displaystyle{\alpha_s =\left ( \partial P_0  /\partial s_0  \right )_\rho}$,  $\displaystyle{c_s ^ 2 = \Gamma_1 \,  P_0/\rho_0}$ is the average sound speed and $\displaystyle{\Gamma_1 = \left(  \partial \ln P_0 / \partial \ln \rho_0 \right )_s }$ the  adiabatic
exponent. The term ${\cal R}(\rho_1,s_1)$ in the RHS  of Eq.~(\ref{perturbed_state_eqn}) represents higher-order terms, which are shown to have a negligible contribution to mode driving \citep[see][]{GK77b}.  The velocity field is decomposed into a component due to the oscillations ($\vec v_{\rm osc}$) and a component due to the turbulent motions ($\vec u$), that is $\vec v = \vec v_{\rm osc} + \vec u$. 

By combining  Eqs.~(\ref{perturbed_momentum_eqn}), (\ref{perturbed_continuity_eqn}), and (\ref{perturbed_state_eqn}), we obtain the inhomogeneous wave equation
\eqna{
\rho_0 \left ( \derivp  { ^2 } {t^2}  -  \vec L  \right ) \vec v_{\rm osc} 
+ {\vec D} \left [ \vec v_{\rm osc}\right ]  & = &  \derivp{{ \vec S}}{t} - {\vec C}
\label{inhomogeneous_wave}
}
with 
\eqna{
   {\vec S} & \equiv & \Big[ \quad \underbrace{\vec \nabla : \left (\rho_0 \, \vec u \, \vec u \right ) -  \vec \nabla : \left < \rho_0 \, \vec u \, \vec u \right >}_\text{Reynolds stress contribution}   \quad - \underbrace{\vec \nabla \, \left ( \bar{\alpha}_s \, s_t \right )}_\text{Entropy contribution} \Big]
\label{S}
}
where $s_t$ is the \emph{Eulerian} turbulent entropy fluctuation.
The terms in Eq.~(\ref{S}) are the driving sources, namely the Reynolds stress tensor and a source term due to  entropy fluctuations, respectively. The last term ${\vec C}$ in the RHS of Eq.~(\ref{inhomogeneous_wave}) involves higher-order driving terms that are found to be negligible \citep[see][]{Samadi01,GK77b}. Finally, the term ${\vec D}$ in the LHS of Eq.~(\ref{inhomogeneous_wave}) gathers terms that couple linearly the oscillation- and turbulence-induced fluctuations. 

To solve the inhomogeneous wave equation (Eq.~(\ref{inhomogeneous_wave})) we first solve the homogeneous equation (\textit{i.e.}, Eq.~(\ref{inhomogeneous_wave}) without the forcing terms on the RHS), supplemented by appropriate boundary conditions. The solutions are the  adiabatic eigendisplacement $\vec \xi$ and associated eigenfrequency  $\omega_{\rm osc}$. Then, we assume that the solution of Eq.~(\ref{inhomogeneous_wave}) takes the form
\eqn{
 \delta \vec r_{\rm osc}  \equiv {1 \over 2} \, \left( A(t) \,  \vec {\xi} (\vec r) \,
 e^{-i \omega_{\rm osc} t } +cc \right) \, , 
\label{delta_osc}
}
where $cc$ indicates complex conjugate, $\omega_{\rm osc}$ is the mode eigenfrequency, and $A(t)$ is the instantaneous amplitude resulting from both driving and damping.
Substituting Eq.~(\ref{delta_osc}) into Eq.~(\ref{inhomogeneous_wave}), multiplying by $\vec \xi^*( \vec r , t )$ and integrating over the stellar volume gives, finally, 
\eqn{
  \deriv {A} {t} + \Delta \sigma~ A   = 
 \frac{1}{2 \omega_{\rm osc}^2 I} \int \,  \vec \xi^*  \cdot \derivp{\vec S }{t} \; {\rm d}^3 x \quad  {\rm with}  \quad I \equiv   \int_0^{M}  \, \vec \xi^* \, . \, \vec \xi \; {\rm d}m \; ,
\label{dA_dt_2}
}
where the term $\Delta \sigma$ comes from the contribution of $\vec D$. The latter is replaced by the damping rate $\eta$ in order to take (a posteriori) all sources of damping into account. Indeed, the real part of $\vec D$ results in a (negligible) frequency shift, while the imaginary part contributes to the damping. 

Equation~(\ref{dA_dt_2}) is straightforwardly solved and one obtains 
\eqn{
A(t)  =  \frac {i e^{-\eta t} } {2 \omega_{\rm osc} I}  \int_{-\infty}^{t} {\rm d}t^\prime   
\int_{\cal V} {\rm d}^3 x \;  e^{(\eta+i\omega_{\rm osc}) t^\prime} \,  
 \vec{\xi}^* (\vec x).  \vec { S }  (\vec x, t^\prime) \, , 
\label{A_t} 
}
where the spatial integration is performed over the stellar volume ($\cal V$). 

In order to simplify subsequent theoretical derivations we will consider the Reynolds source term only. In addition, since the excitation arises from a turbulent medium, the sources are random so that $A(t)$ cannot been determined in a deterministic way. We thus derive an expression for the average squared $\langle \left | A \right | ^2 \rangle$, where the average is performed over a larger set of independent realisations. Furthermore, it can been shown with the help of Eq.~(\ref{balance}) that $\langle \left | A \right | ^2 \rangle$ is related to $\cal P$ and $\eta$ according to ${\cal P} = \eta \, I \, \omega_{\rm osc}^2 \, \langle \left | A \right | ^2 \rangle$. Finally, by using Eq.~(\ref{A_t}), we establish the following expression for $\cal P$ (see SG for a detailed derivation) 
\eqn{
{\cal P} = { 1 \over {8 \, I} } \,   \int_{{\cal V}}  {\rm d}^3x_0 \, \rho_0^2 \, \int_{-\infty}^{+\infty} {\rm d}^3r \; {\rm d}\tau \, e^{-i\omega_{\rm osc} \tau}  \left (\nabla_i \xi_j^* \right )_1   \, \langle  \left ( u_i \, u_j  \right )_1  \,  \left ( u_k u_m  \right )_2  \,  \rangle \,\left (\nabla_k \xi_m \right ) _2  \label{P_2} \; ,
}
where $x_0$ is the position in the star; $\vec r$ and $\tau$ are the spatial and temporal correlation lengths associated with turbulence and the subscripts 1 and 2 refer to quantities evaluated at the spatial and temporal positions $[ \vec x_0-\frac{\vec r}{2}, - \frac{\tau} {2}]$  and $ [\vec x_0+\frac{\vec r}{2}, \frac{\tau} {2}]$.

We further assume that the eigendisplacement is spatially decoupled from the source terms. In other words, $\vec \xi$ varies on a scale larger than the characteristic scale of turbulence. This permits us to reduce Eq.~(\ref{P_2}) to
\eqn{
{\cal P} = { 1 \over {8 \, I} } \,   \int_{{\cal V}}  {\rm d}^3x_0 \, \rho_0^2 \,  \nabla_i \xi_j^* \,  \nabla_k \xi_m    \,  \int_{-\infty}^{+\infty} {\rm d}^3r \, {\rm d}\tau \, 
e^{-i\omega_{\rm osc} \tau} \left <    \left ( u_i \, u_j  \right )_1  \,  \left ( u_k u_m  \right )_2   \right >       \; .
\label{P_3}
}
The second integral of Eq.~(\ref{P_3}) involves the term  $\left <    \left ( u_i \, u_j  \right )_1  \,  \left ( u_k u_m  \right )_2   \right >$  which is a two-point spatial \emph{and} temporal correlation product between $u_i\,u_j$ and $u_k\,u_m$. 
If one adopts the quasi-normal approximation (hereafter QNA), it is possible to decompose the fourth-order correlation product as follows
\eqna{
 \langle (u_i \, u_j )_1 \, (u_k  \, u_m  )_2 \rangle = & \langle (u_i \, u_j )_1 \rangle
 \, \langle  (u_k \, u_m )_2 \rangle  +   \, \langle ( u_i )_1 \, (u_m)_2\rangle \,  \langle ( u_j )_1 \,
 (u_k)_2\rangle      \nonumber  \\  &  + \, \langle ( u_i )_1 \, (u_k)_2\rangle \,  \langle ( u_j )_1 \,
 (u_m)_2\rangle  &  \; .
\label{qna}
}
Note that, strictly, the decomposition of \eq{qna} is only valid when the velocity is normally distributed 
\citep[see][for an extensive discussion]{Belkacem06a}.  
  
We now define $\phi_{i,j}$ to be the spatio-temporal Fourier transform of
$ \langle ( u_i )_1 \, (u_j)_2 \rangle$.
For an inhomogeneous, incompressible, isotropic and stationary turbulence, there is a relation between $\phi_{i,j}$ and the kinetic energy spectrum $E$, which is \citep{Batchelor70}
\eqna{
\phi_{ij}( \vec k,\omega) &  =& \frac { E( k,\omega) } { 4 \pi k^2}  \left( \delta_{ij}- \frac {k_i k_j} {k^2}  \right) \, ,
\label{phi_ij}
} 
where $k$ and $\omega$ are the wavenumber and frequency 
associated with the turbulent elements, and $\delta_{i,j}$ is the
Kronecker symbol.  Following \cite{Stein67}, for each layer we decompose 
$E(k,\omega)$ as 
\eqna{
E(k,\omega) & = & E(k) \; \chi_k(\omega) \, , 
\label{ek_chik}
}
where $E(k)$ is the time-averaged kinetic energy spectrum and
$\chi_k(\omega)$ is the frequency component of $E(k,\omega)$.
Note that $\chi_k(\omega)$ and $E(k)$ satisfy the normalisation conditions 
\begin{align}
\int_{-\infty}^{+\infty} {\rm d} \omega \,  \chi_k (\omega)  = 1 \quad {\rm and} \quad 
\int_0^{\infty}{\rm d}k\,E(k)  =  \displaystyle  {1 \over 2} \, {\langle \vec u^2 \rangle  }  =    \displaystyle {3 \over 2} \,  u_0^2 \; ,
\label{eqn:E:normalisation}
\end{align}
where we have defined the characteristic velocity $u_0^2 \equiv \langle  u_z^2 \rangle$ with $u_z$ the vertical component of the velocity field. 

Now, using  Eqs.~(\ref{qna}) to (\ref{ek_chik}), Eq.~(\ref{P_3}) can be written for \emph{radial} modes as
\eqna{
{\cal P}  & =  & { \pi^3 \over {2 \, I} } \,  \left( {16 \over 15 } \right) \, \int_0^M {\rm d} m  \, {   \rho_0 \,  u_0^3  \over
  {k_0^4  } }   \,  \left |  \deriv { \xi_{\rm r}} {r}
\right | ^2 \, \tilde{S}_R(r,\omega_{\rm osc}) \; , \label{P_4}
}
where we have defined the dimensionless source function
\begin{align}
\label{SR}
\tilde{S}_R &  = \left ( k_0^4 \, / \, u_0^3 \right ) \, \int_0^\infty {\rm d}k \,
 \frac {E^2(k,r)} { k^2} \int_{-\infty}^{+\infty}  {\rm d}\omega \; \chi_k (
\omega_{\rm osc} + \omega, r) \, \chi_k (\omega, r)  \, ,
\end{align}
and where we have introduced the characteristic wavenumber
$k_0 \equiv {2 \pi} / \Lambda $ with  $\Lambda$ is a characteristic size of the most-energetic eddies\footnote{This characteristic size can be determined from the kinetic $E(k)$ or by default using some prescriptions (for more details see Sect. 11.5.1 in \cite{Samadi11}).}.  
A similar dimensionless source function can be derived for the source term associated with the entropy fluctuations.  We also point out that the present formalism has been generalised for non-radial acoustic modes \citep{Belkacem08} as well as for gravity modes \citep{Belkacem09}.

\begin{figure}
\begin{center}
\includegraphics[height=8cm,width=11cm]{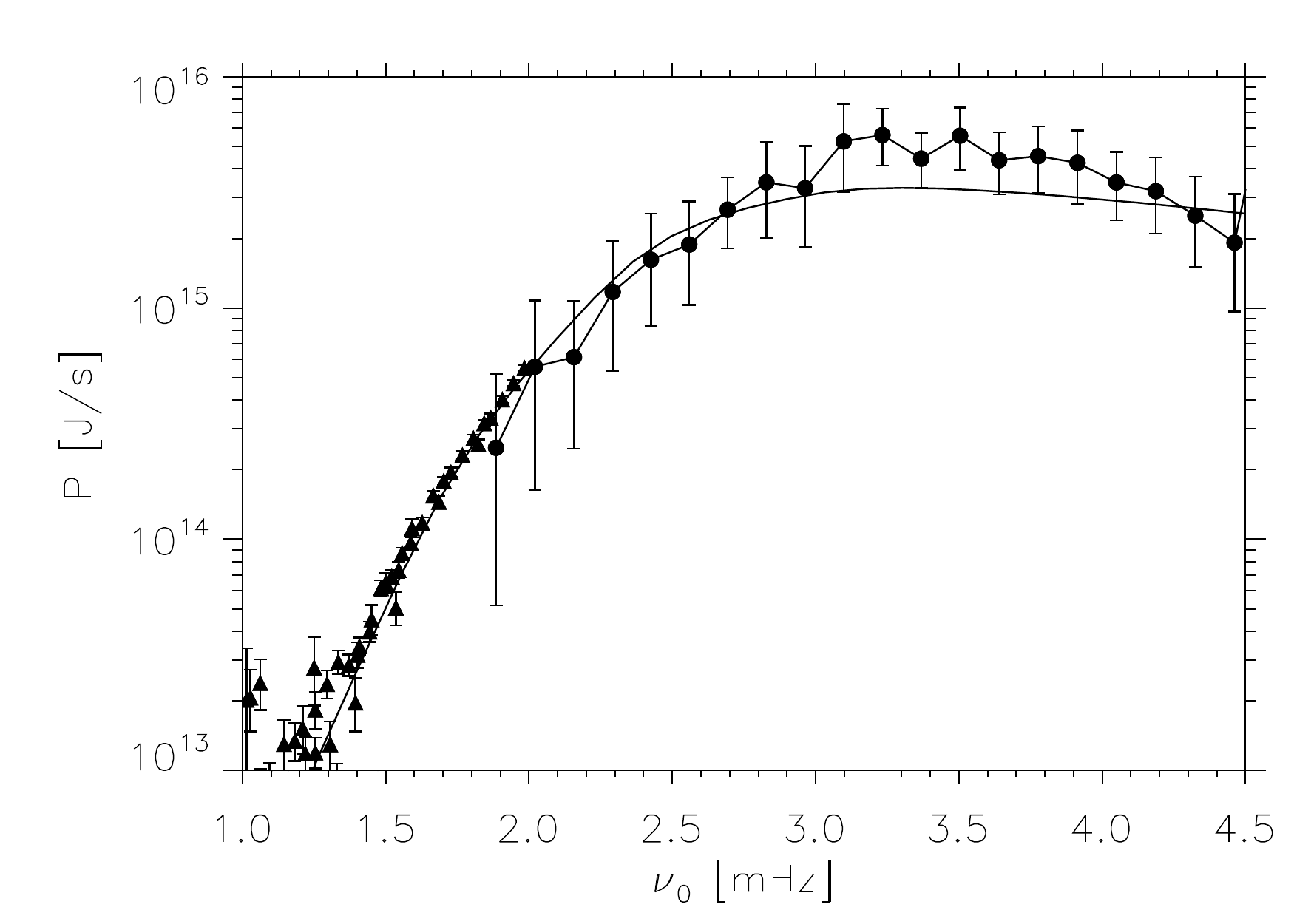}
\caption{Solar p-mode excitation rates $\cal P$ as a function of the frequency $\nu_{\rm osc} = \omega_{\rm osc}/(2\, \pi)$. The dots correspond to the observational data obtained by the GONG network, as derived by \cite{Baudin05}, and the triangles corresponds to observational data obtained by the GONG network as derived by \cite{Salabert09}  for $\ell$  = 0 to $\ell$  = 35. The solid line corresponds to theoretical $\cal P$ computed as detailed in \cite{Belkacem2010}.}
\label{Psun}
\end{center}
\end{figure}

For solar modes, the most recent and realistic calculation of $\cal P$ has been undertaken by \cite{Belkacem2010}. This calculation is compared in Fig.~\ref{Psun} with helio-seismic data. The theoretical calculation results in an overall agreement with the seismic data. We stress that this result has been obtained \emph{without} adjustments of free parameters. To explain qualitatively the variation of ${\cal P}$ with mode frequency, we rewrite Eq.~(\ref{P_4}) in a more convenient form by introducing the flux of kinetic energy ($F_{\rm kin}$). For isotropic turbulence
\eqn{
F_{\rm kin} = \langle u_z \,   E_{\rm kin} \rangle \approx   {3 \over 2} \, \rho_0 \, u_0^3 \, , 
\label{F_kin}
}
where $E_{\rm kin} \equiv (1/2) \, \rho_0 \, \vec{u}^2 $ is the specific kinetic energy. 
Substituting \eq{F_kin} into \eq{P_4} yields the relation
\eqn{ 
{\cal P} \propto {1 \over I} \int_0^M \,  \Lambda^4   \,  {F_{\rm kin}} \,  \left |  \deriv { \xi_{\rm r}} {r}
\right | ^2 \, \tilde{S}_R(r,\omega_{\rm osc}) \; {\rm d} m \; ,
\label{P_5}
}
where we have introduced the characteristic life-time $\tau_0 \equiv \Lambda / u_0$.

Equation~(\ref{P_5}) permits us to highlight the key mode-driving quantities \citep[see][]{Samadi11}.
\begin{itemize}

\item {\it The mode inertia} ($I$): The lower the mode frequency, the larger the eigendisplacement in the interior. Hence Eq.~(\ref{dA_dt_2}) implies an increase of $I$ with decreasing frequency. As a consequence, for the same amount of energy, it is more difficult to drive low-frequency modes than high-frequency ones. This is the main cause of the rapid decreases of $\cal P$ with decreasing frequency as seen in Fig.~\ref{Psun}.

\item {\it The eddy characteristic size} ($\Lambda$): As seen in Eq.~(\ref{P_5}), mode-driving scales locally as $\Lambda^4$. There is, however, no simple physical principle from which this characteristic size can be derived.  Nevertheless, this size can be derived from 3D hydrodynamical simulations. According to \cite{Samadi02I}, it varies rather slowly in the upper part of the convective zone where the driving is the most efficient. The simulations also show that, from one stellar model to another, this size scales as the pressure scale-height $H_p$ at the photosphere \citep{Freytag97,Samadi08} and  roughly scales as the ratio $T_{\rm eff} /g$, where $g$ is the surface gravity \citep[but see][]{Trampedach2013}.

\item {\it The flux of  kinetic energy} ($F_{\rm kin}$): In the framework of the mixing-length approach, it can be shown that $F_{\rm kin}$ is roughly proportional to the convective flux $F_c$, The latter increases as we go up into the convective region. In the upper part of the convective envelope (where mode-driving is most efficient), $F_c$ is almost uniform and scales as $T_{\rm eff}^4$ where $T_{\rm eff}$ is the effective temperature. Next, in the   region between the convective zone and the atmosphere,  $F_c$ (hence $F_{\rm kin}$) decreases rapidly.

\item {\it The mode compressibility} ($d\xi_{\rm r} / dr$): This quantity reaches its maximum in the transition region between the convective and radiative regions, where the temperature decreases rapidly. The maximum of the mode compressibility is also shown to increase with increasing frequency. Therefore, the mode compressibility together with the mode inertia favors high-frequency modes.

\item {\it The (dimensionless) source function at the mode frequency} ($\tilde{S}_R$): As described by Eq.~(\ref{SR}), this term depends on the shape of the kinetic energy spectrum $E(k)$ or, more precisely, on $\chi_k$, the Fourier transform of the time-correlation function of the velocity field at a given wavenumber $k$ ({e.g.}, \cite{Lesieur97}, Chap. V-10). 
% $\chi_k(\omega)$ is a Lorentzian function ($1/ \left ( 1 + (\tilde{\tau}_k \, \omega_{\rm osc} )^2 \right ) $) 
%\cite{Samadi02II,Belkacem2010,Belkacem2011proc} where the characteristic time $\tilde{\tau}_k$ scales as $\tau_0$ \cite{Samadi02II}. 
%The quantity $\chi_k(\omega)$ decreases rapidly with increasing $\omega_{\rm osc}$.  The characteristic time $\tau_0$ reaches a minimal value $\tau_0^{\rm min}$ in the super-adiabatic layers and is of the order of 5 minutes. Therefore, for modes with a frequency larger than $\nu_{\rm osc} \gtrsim 3$~mHz, the source function $\tilde{S}_R$ decreases rapidly with frequency. This latter term balances the effect of the mode compressibility and the net result is, as seen in Fig.~\ref{Psun}, a slow decrease with $\nu_{\rm osc}$ for $\nu_{\rm osc} \gtrsim 3$~mHz. 

\end{itemize}

\subsubsection{The crucial role of the eddy time-correlation, $\chi_k(\omega)$}
\label{Eddy time-correlation}

Let us first remind that, for a turbulent fluid, one defines the Eulerian eddy time-correlation function as
\begin{align}
 \left< \vec u(\vec x + \vec r,t+\tau) \cdot \vec u(\vec x,t) \right> = \int  {\cal E}(\vec k,t,\tau) \, e^{{\rm i} \vec k \cdot \vec x} \, {\rm d}^3\vec k  \, ,
\label{time-correlation}
\end{align}
where $\vec u$ is the Eulerian turbulent velocity field, $\vec x$ and $t$ the space and time position of the fluid element, $\vec k$ the wave number vector, $\tau$ the time-correlation length, and $\vec r$ the space-correlation length. 
The function $\cal E$ in the RHS of \eq{time-correlation} represents the  time-correlation function associated with an eddy of wave-number $\vec k$.

We assume an isotropic and stationary turbulence, accordingly  ${\cal E}$ is only a function of $k$ and $\tau$.
The quantity ${\cal E}(k,\tau)$ is related to the turbulent energy spectrum according to
\begin{align}
{\cal E}(k,\tau) = \frac{E(k,\tau)}{2\pi k^2} \, .
\end{align}
where $E(k,\tau)$ is the turbulent kinetic energy spectrum which temporal Fourier transform is 
\begin{align}
\label{Ekw}
E(k,\omega)   \equiv {1 \over {2\pi}} \,  \int_{-\infty}^{+\infty}  {E}(k,\tau) \, e^{{\rm i} \omega \tau} \, {\rm d} \tau \, , 
\end{align}
where $\omega$ is the eddy frequency and $E(k,\omega)$ is written as follows \cite{Stein67,Samadi00I}  
\begin{align}
\label{decomp_E}
E(k,\omega) = E(k) \, \chi_k (\omega) \quad {\rm with}\quad  \int_{-\infty}^{+\infty} \chi_k (\omega) \, {\rm d} \omega = 1
\end{align}
where $\chi_k(\omega)$ is the frequency component of $E(k,\omega)$. In other
words, $\chi_k(\omega) $  represents -~  in the frequency  domain ~-  the temporal correlation between eddies of wave-number $k$. 

The main issue is now to model the function $\chi_k(\omega)$. Most of the theoretical formulations of mode driving explicitly or implicitly assume a Gaussian function for $\chi_k(\omega)$ 
\citep[][]{GK77b,Dolginov84,GMK94,Balmforth92c,Samadi00II,Chaplin05}. However, 3D hydrodynamical simulations of the outer layers
of the Sun show that, at the length associated with the energy bearing eddies, $\chi_k$ is rather Lorentzian \citep{Samadi02II}. This result can be easily explained by the use of a stochastic model of turbulence. More precisely, if one models turbulent convection as a noisy relaxation process \citep[\emph{i.e.} a Ornstein-Uhlenbeck process, which is a sub-class of a Gaussian Markov process, see the details in][]{Belkacem2011b}, the time correlation function is an exponential function in the time-domain and thus a Lorentzian function in the Fourier domain. However, one must be cautious with the way $\chi_k$ is modeled by a Lorentzian function. Indeed, as shown by \cite{Belkacem2010}, one has to introduce a cut-off frequency in the modeling of the eddy-time correlation function to account for the effect of short-time scales. Under the sweeping approximation, which consists in assuming that the temporal correlation of the eddies, in the inertial subrange, is dominated by the advection by energy-bearing eddies, the shape of the temporal correlation function of eddies is no longer Lorentzian for high frequencies \citep[see][for details]{Belkacem2010}.

Finally, as shown by \citet{Samadi02II} and in Fig.~\ref{resultat}, calculation of the mode excitation rates based on a Gaussian $\chi_k$ results for the Sun in a significant under-estimate of the maximum of ${\cal P}$ whereas a better agreement with the observations is found when a Lorentzian $\chi_k$ is
used. A similar conclusion were reached by \citet{Samadi08} in the case of the star $\alpha$~Cen~A. 
\begin{figure}[t]
\begin{center}
\includegraphics[height=8cm,width=11cm]{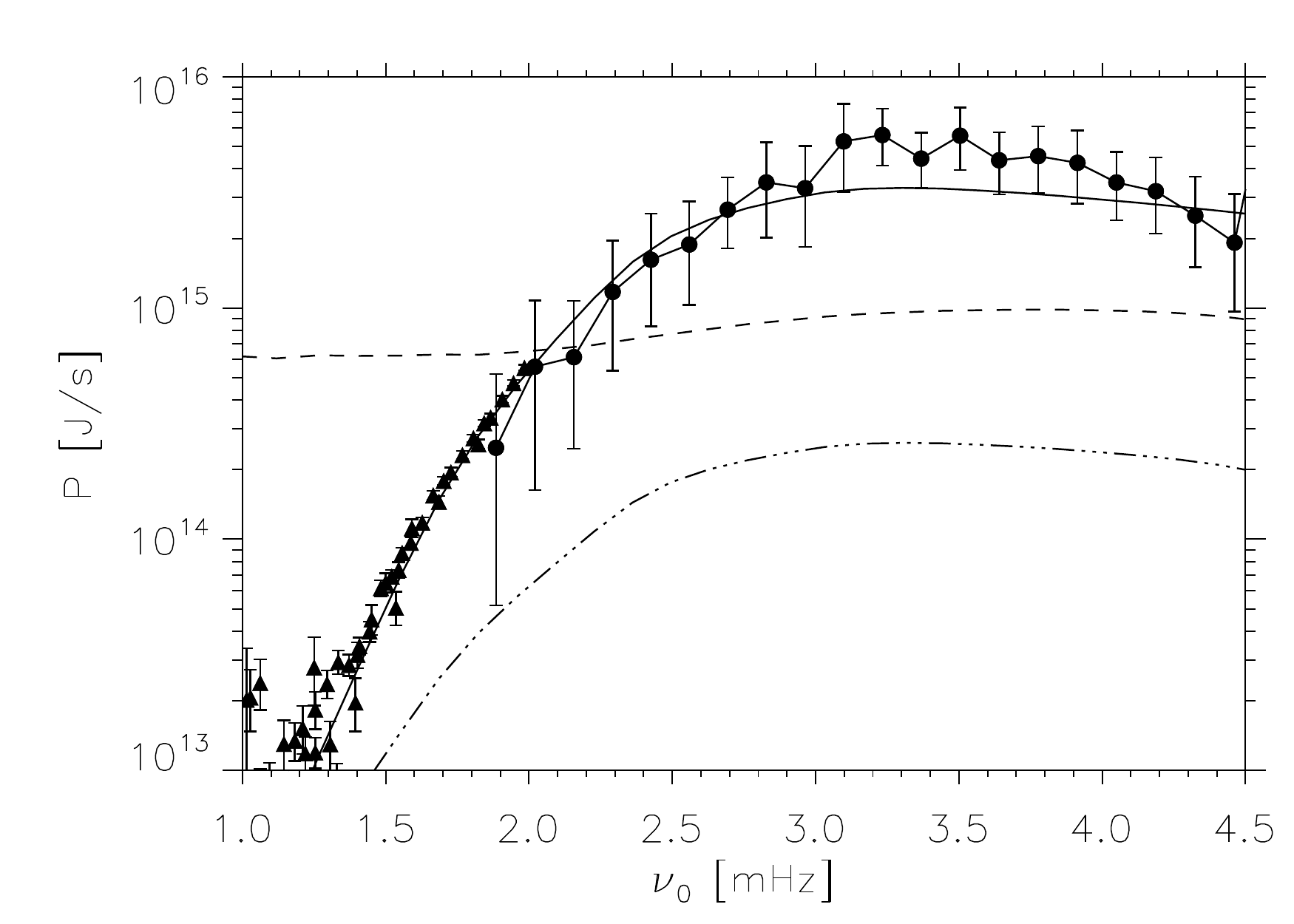}
\caption{Same as Fig.~\ref{Psun} except that the dashed line corresponds to the computation of the excitation rates using  a Lorentzien description of $\chi_k$ \emph{without} any cut-off frequency and with the MLT of convection. The triple-dot-dashed line corresponds to the same computation except that Gaussian description of $\chi_k$ is used.}
\label{resultat}
\end{center}
\end{figure}

\subsubsection{Role of the stratification and chemical composition}
\label{strat_metal}

Unfortunately, classical 1D stellar models of solar-like stars provide us with a poor modeling of the upper-most layers. Among the physical mechanisms overlooked by the 1D models,  \citet{Rosenthal99} have shown that turbulent
pressure plays an important role because it modifies the stratification. Indeed, in the super-adiabatic region, 
a model including turbulent pressure provides an additional support against gravity, hence it has a lower gas pressure
and density than a model that does not include turbulent pressure \citep[see also][]{Nordlund99b,Rosenthal99}.

In term of mode driving, taking turbulent pressure into account in a realistic way results in a much better agreement between
observed and theoretical mode frequencies of the Sun. For other stars than the Sun, the result is the same. 
Following \citet{Rosenthal99}, \citet{Samadi08} have studied the importance of turbulent pressure for the calculation of the mode excitation rates. For this purpose, they  have built two 1D models
representative of the star $\alpha$~Cen~A. One model, here referred as the ``patched'' model, has its surface
layers directly taken from a  fully compressible 3D hydrodynamical numerical  model.  A second model, here referred
as``standard'' model, has its surface layers  computed using standard physics. In particular convection is
described according to \citet{Bohm58}'s  mixing-length local theory of convection (MLT) and turbulent pressure is ignored. 
\citet{Samadi08} found that the calculations of ${\cal P}$ involving eigenfunctions computed on the 
basis of the ``patched'' global 1D model  reproduce  much better the seismic data derived for $\alpha$~Cen~A than calculations based on
the eigenfunctions computed with the ``standard'' stellar model, {\it i.e.} built with the MLT and ignoring turbulent pressure. 
This is because including turbulent pressure results in \emph{lower} mode masses ${\cal M}$ than a model ignoring turbulent
pressure. 

Another important physical ingredient for the modeling of mode driving is the metallicity. \citet{Samadi09a} have studied the role of the surface metal abundance on the efficiency of the stochastic driving. 
%For this purpose, they have computed two 3D hydrodynamical simulations representative --~ in
%effective temperature and gravity ~-- of the surface layers of HD~49933, a star which is rather metal poor compared
%to the Sun since its surface iron-to-hydrogen abundance is [Fe/H]=-0.37. 
%One 3D simulation (hereafter labeled as S0) has a solar metal abundance  and the other (hereafter labeled as S1) has [Fe/H]
%ten times smaller. For each 3D simulation they have build a ``patched'' model  in the manner of
%\citet{Samadi08}  and computed the  acoustic modes associated with the  ``patched'' model.
%As seen in Fig.~\ref{pow_hd49933}, the mode excitation rates ${\cal P}$  associated with metallicity of HD49933  are found
%to be about \emph{three times smaller} than those associated with the solar metallicity. 
Indeed, a lower surface metallicity
results in a lower opacity, and accordingly in a higher surface density.
In turn, for the same amount of energy transported by convection, 
the higher the density the smaller the convective velocities. Finally, smaller convective velocities result in a less efficient driving \citep[for details,  see][]{Samadi09a}.
This conclusion is qualitatively consistent with that  by \citet{Houdek99} who --~ on the basis of a mixing-length
approach ~-- also found that the mode amplitudes decrease with decreasing metal abundance.

\begin{figure}[t]
\begin{center}
\includegraphics[width=10cm]{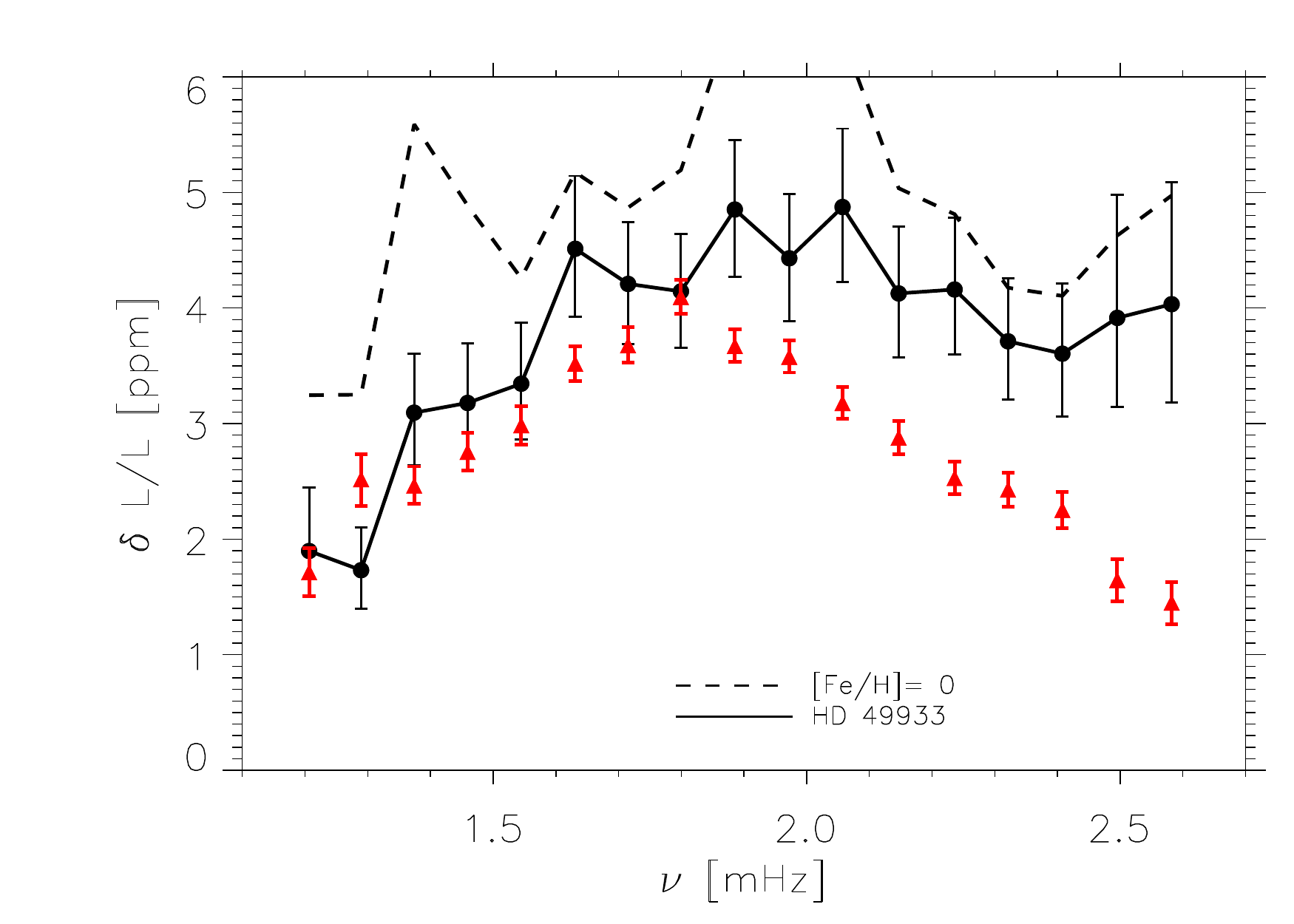}
\end{center}
\caption{Mode bolometric amplitude as a function of the mode frequency ($\nu$). The filled circles connected by the thick solid line correspond to the theoretical mode amplitudes in intensity, $\delta L/L$, derived for HD~49933 with relevant [Fe/H]. 
The thick dashed line  corresponds to the mode amplitude in intensity associated with the model with [Fe/H]$=0$.
The red triangles and associated error bars   correspond to the mode amplitudes, $(\delta L/L)_{\rm CoRoT}$, obtained from the CoRoT data \citep{Benomar09b}. These measurements have been translated into bolometric   amplitudes following \citet{Michel09}.
}
\label{pow_hd49933}
\end{figure}

Using the seismic determination of the mode linewidths measured by
CoRoT for HD~49933 \citep{Benomar09b} and the theoretical
mode excitation rates computed for the specific case of HD~49933, \citet{Samadi09b}
have derived the theoretical mode amplitudes of the acoustic
modes of HD~49933. As shown by Fig.~\ref{pow_hd49933}, except at rather high frequency ($\nu \gtrsim 1.9$~mHz), their amplitude calculations are
approximately within a 1-$\sigma$ agreement with the mode amplitudes derived
from the CoRoT data  \citep[for more details, see][]{Samadi09b}. They also show
that assuming a solar metal abundance rather than the observed 
metal abundance of the star would result in larger mode amplitudes and
hence in a larger discrepancy with the seismic data.
This illustrates the importance of taking the surface metal abundance of
the solar-like pulsators into account when modeling the mode excitation. 

\section{Scaling relations and related non-adiabatic effects}
\label{scaling}

\subsection{Introduction}

The determination of accurate stellar parameters (mass, radius, effective temperature, age, and chemical composition) is a fundamental and longstanding problem in astrophysics \citep[{e.g.},][]{Soderblom2010}. Nevertheless, such a determination is only possible by means of the use of stellar models and therefore suffers from our deficient knowledge of the physical processes taking place in stars \citep[{e.g.},][]{Goupil2011a,Goupil2011b}. 
Moreover, with the launch of CoRoT \citep{Baglin2006a,Baglin2006b,Michel2008} and {\it Kepler} \citep{Borucki2010},  solar-like oscillations have been detected in 
several hundreds of main-sequence stars and several thousands of red-giant. With such large numbers of stars it is not possible to perform  individual mode fitting of the power spectrum for each star. This is very time (and man-power) consuming so that a new method emerged through the use of seismic global parameters. The latter are typical global characteristics of the oscillation spectra such as the regularities in frequency (or period), or the frequency of the maximum amplitude. 

%The situation recently improved with the advent of space-borne asteroseismology, more precisely with the launch of CoRoT \citep{Baglin2006a,Baglin2006b,Michel2008} and {\it Kepler} \citep{Borucki2010}. Those two spacecrafts provided us with high-quality photometric data. Up to now, several hundreds of main-sequence stars with solar-like oscillations have been detected and several thousands oscillating red-giant stars, allowing for statistical analysis. With such large number of stars it is not possible to perform classical seismology, \emph{i.e.} by individual mode fitting of the power spectrum. This is very time (and man-power) consuming so that a new method emerged through the use of seismic global parameters. The latter are typical global characteristics of the oscillation spectra such that the regularities in frequency (or period), or the frequency of the maximum amplitude. 

This approach gave birth to the \emph{ensemble} asteroseismology, whose cornerstones are the relations between global seismic quantities and stellar parameters. It allows ones to infer model-independent stellar parameters as well as information on stellar structure and evolution. 
Scaling relations  between asteroseismic quantities  and stellar  parameters such as stellar mass, radius, effective temperature, and luminosity  have initially been observationally  derived  by several authors \citep[{e.g.}][]{Ulrich86,Brown91,Kjeldsen95} using ground-based data. 
CoRoT and {\it Kepler} confirmed these results by providing accurate and homogeneous measurements for a large sample of stars from main-sequence to red-giant stars \citep[{e.g.}, ][]{Mosser2010,Baudin11,Mosser2011,Mosser2011b,Mosser2012b,Mosser12a,Samadi12}.  

Among them, the relation between the frequency of the maximum height in the power spectrum ($\numax$) and the cut-off frequency ($\nuc$), the relation  between the mode line-width and the effective temperature ($T_{\rm eff}$), as well as the relation between mode amplitude and the luminosity ($L$) are particularly interesting since they all rely on the same cause: the physics of non-adiabatic oscillations. Consequently, those scaling relations give us additional constraints on those processes, still subject to many uncertainties as already mentioned in Sects.~\ref{mode_damping} and \ref{mode_driving}. 

In the following, we address the theoretical ground of these scaling relations in detail and show how they are intimately related to the non-adiabatic processes. 

\subsection{Relation between mode linewidth and effective temperature}
\label{eta_scaling}

For mode linewidths (or equivalently mode damping rates), scaling relations have been investigated only very recently. This is the result of the need for long-duration and almost-uninterrupted monitoring to resolve individual modes and to enable their precise measurements. 

\cite{Houdek99}, and later \cite{Chaplin09}, have investigated the dependence of mode-damping rates on global stellar parameters. From  ground-based measurement, \cite{Chaplin09} found that mode linewidths follow a power-law  of the form $\eta \propto T_{\rm eff}^4$ (where $T_{\rm eff}$ is the effective temperature). % and no clear tendency emerged when $\eta$ is scaled with the ratio $L/M$. 
Nevertheless, these measurements were based on short-term observations and derived from an  inhomogeneous set of analysis and instruments, resulting in a large dispersion. This was settled by \cite{Baudin11,Baudin11b}  (Fig.~\ref{observations_largeurs}, top panel) using a homogeneous sample of CoRoT data. They found that a unique power-law hardly describes the entire range of effective temperature covered by main-sequence and red-giant stars and proposed that mode linewidths of main-sequence stars follow a power-law of $T_{\rm eff}^{16 \pm 2}$, while $\eta$ in red-giant stars only weakly depends on effective temperature ($T_{\rm eff}^{-0.3 \pm 0.9}$). 
The latter result was later confirmed and extended by \emph{Kepler} observations (Fig.~\ref{observations_largeurs}, bottom panel) to main-sequence and sub-giant stars \citep{Appourchaux12,Appourchaux14}. 

\begin{figure}
\begin{center}
\includegraphics[height=10.5cm,width=7.5cm,angle=90]{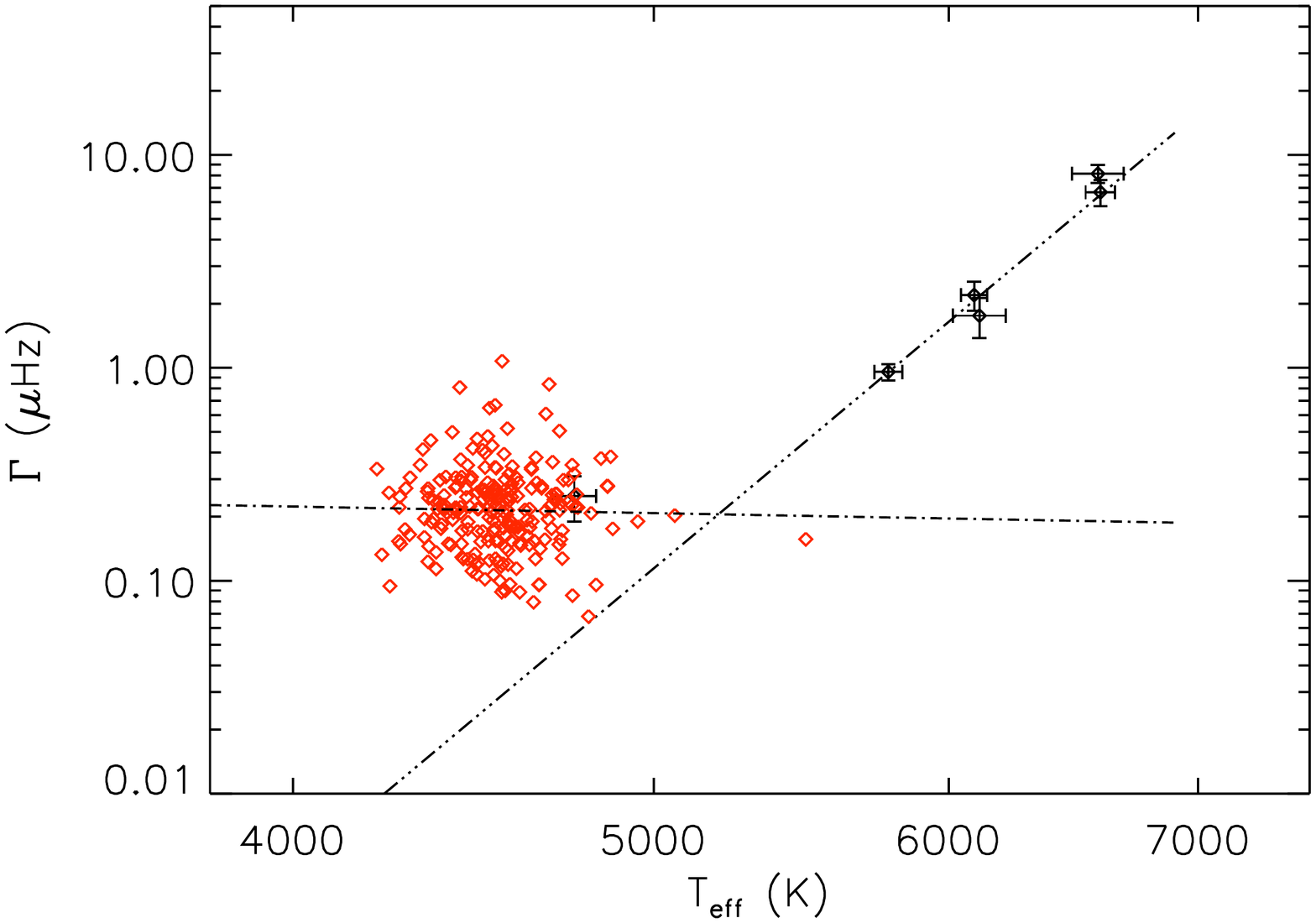}
\includegraphics[height=9.5cm,width=6.5cm,angle=90]{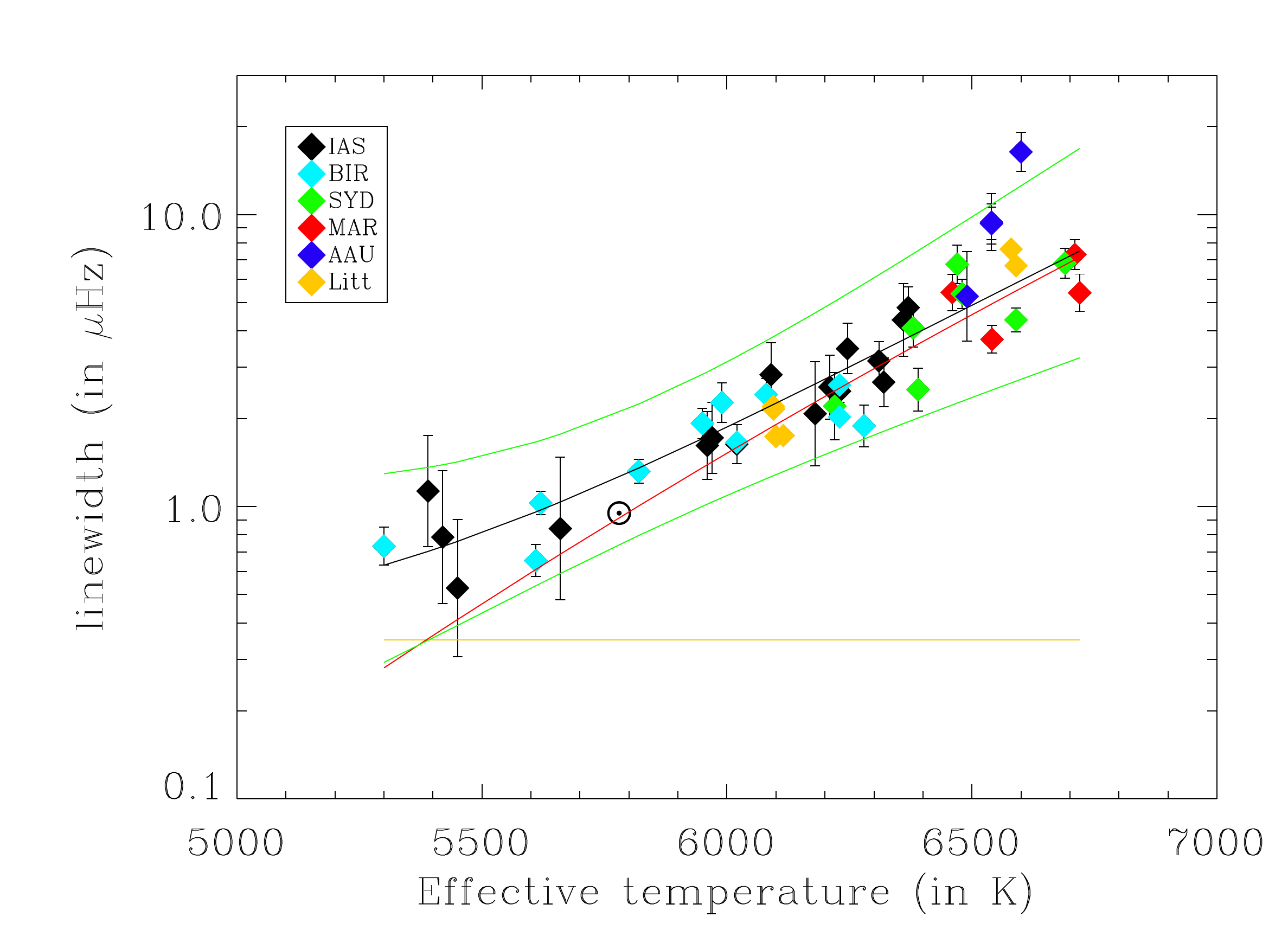}
\caption{{\it Top:} Mode linewidths versus $T_{\rm eff}$ for red giants ($T_{\rm eff} < 5000$\,K) and for main-sequence stars ($T_{\rm eff} > 5000$\,K) observed by CoRoT \citep{Baudin11}. {\it Bottom:} Average mode linewidth at maximum mode height (and their 3-$\sigma$ error bars) as a function of effective temperature, for sub-giant and main-sequence stars observed by \emph{Kepler} \citep{Appourchaux12}.}
\label{observations_largeurs}
\end{center}
\end{figure}

The theoretical work of \cite{Chaplin09}, based on the formalism developed by \cite{Balmforth92a,Houdek99} and \cite{Chaplin05}, predicted a power-law of $\eta \propto T_{\rm eff}^4$ which disagrees with CoRoT and \emph{Kepler} observations. 
In contrast, \cite{Belkacem2012}, based on the formalism of \cite{MAD05}, were able to reproduce both CoRoT and \emph{Kepler} observations. Therefore, in the following we will mainly discuss these most recent results.

\subsubsection{A probe for the damping mechanisms: \emph{an illustrative case}}
\label{amortissement_turbulent}

The relation between mode linewidths and effective temperatures is an important constraint for the modeling of damping rates. While it is difficult to settle the issue of the dominant contribution for the Sun, this relation allows us to test several mechanisms. 

A striking example concerns the contribution of turbulent viscosity to mode damping rates that had long been thought to be a dominant contribution \citep{GK77a,Goldreich91}. 
To investigate it, let us start with its integral expression, given by \cite{Ledoux58,GK77a},
\eqna{
\eta & \propto & { 1 \over {3 I }}   \, \int {\rm d} m \,
\nu_t \, \left | r \, { {\rm d} \over {{\rm d}r}} \, \left (  {\xi_{\rm r} \over r}
\right ) \right |^2
\;,
\label{eta_viscosity}
}
where $\nu_t$ is the turbulent viscosity. The simplest description of $\nu_t$ is based on the concept of eddy-viscosity, which implies $\nu_t \propto u_0 \, \Lambda$, where $u_0$ is the largest eddy velocity and $\Lambda$  the largest eddy size. Equation~(\ref{eta_viscosity}) can be simplified \citep{Samadi11} to
\eqna{
\eta & \propto & \left ( {\omega_{\rm osc} \over  c_s} \right )^2  \, \Lambda  \,
u_0 \;.
\label{eta_viscosity_2}
}
To go further, one must express the velocity $u_0$ as a function of stellar parameters. To this end, we note that the kinetic energy flux $F_{\rm kin}$ is roughly proportional to the convective flux  $F_{\rm conv}$, which can further be approximated by the total flux  ($F_{\rm conv} \approx F_{\rm tot} \propto \teff^4$). Therefore, 
\begin{equation}
u_0 \propto T^{4/3} \, \rho_s^{-1/3} \, ,
\label{approx_u0}
\end{equation}
where $\rho_s$ is the surface density. 
 
The last step is to describe the surface density. To this end, we note that the optical depth can be approximated by $\tau \approx \rho \kappa H_p$, where $\kappa$ is the mean opacity. For low-mass stars $\kappa$ is dominated by $H^{-}$ opacity, such that $\kappa \propto \rho^{1/2} \, T^9$ \citep{HansenKawaler1994}. Then, considering that in the photosphere $\tau=2/3$, this latter scaling together with \eq{approx_u0} permits us to express \eq{eta_viscosity_2} as 
\eqna{
\eta & \propto & T^{5/2} \, g^{3/2} \;.
\label{eta_viscosity_3}
}
From \eq{eta_viscosity_3}, it turns out that the damping rates related to turbulent viscosity exhibit a dependence with effective temperature that is very different from those derived from the observations (Figs.~\ref{observations_largeurs}). Therefore, this result supports that the damping from turbulent viscosity is not the dominant contribution \citep[consistently with the results by][]{Osaki90}. This result further illustrates that scaling relations of mode damping rates are a powerful tool in obtaining important constraints on the underlying physical mechanisms governing mode linewidths. Moreover, as we will show below, this provides a way to validate their modeling.   

A full computation of damping rates and a subsequent comparison with observations has been recently performed by \cite{Belkacem2012}. Among several formalisms, they used the description proposed by \cite{MAD05} which includes the time-dependent convection (TDC) treatment. 
The results of this computation are summarised in Fig.~\ref{Kepler_Corot}. It can be seen that there is an overall agreement between the theoretical computations and the CoRoT and \emph{Kepler} observations.  
This overall agreement with both CoRoT and \emph{Kepler} observations suggests that the main physical picture is well-reproduced by modeling. 

\begin{figure}
\begin{center}
\includegraphics[height=8cm,width=11cm]{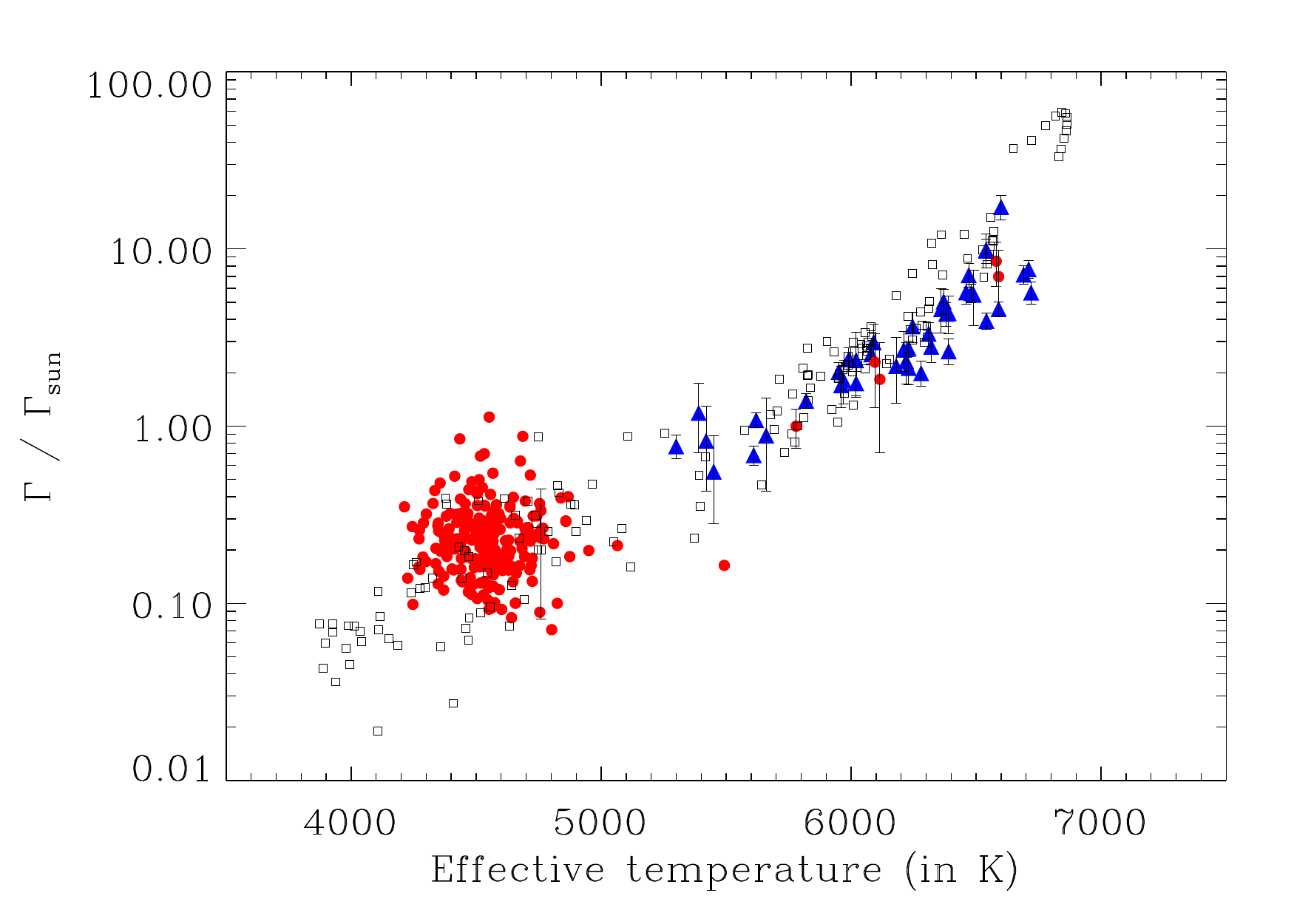}
\caption{Mode linewidths (normalised by the solar value, $\Gamma_{\rm sun} = 0.95 \, \mu$Hz) versus effective temperature. The squared symbols represent theoretical calculations computed as described in \cite{Belkacem2012}. The triangles correspond to the observations of main-sequence stars  derived by \cite{Appourchaux12} from the \emph{Kepler} data (with their 3-$\sigma$ error-bars). The dots correspond to the observations of red giants (with $T_{\rm eff} < 5200$ K) and main-sequence  (with $T_{\rm eff} > 5200$ K, with their 3-$\sigma$ error-bars) stars as derived by \cite{Baudin11,Baudin11b} from the CoRoT data. }
\label{Kepler_Corot}
\end{center}
\end{figure}

\subsubsection{Toward effective temperature derived from the seismic relations?}

The relation between mode linewidths and the effective temperature is potentially able to provide us with a determination of effective temperature from the measurement of mode linewidths. This is very promising since it would provide a determination independent of stellar atmosphere models. 
Even if the prospect is attractive, a decisive preliminary step is to understand the underlying physics so as to have some control on the precision and accuracy of this relation. 

While a full computation of mode damping rates reproduces the observations, the dependence of $\eta$ on the effective temperature with such a large exponent is far from being obvious. Hence, to get a better insight into this relation, let us distinguish between the effect of the inertia and the work integral (see Eq.\ref{final}). 
For the latter, it is useful to consider Eq.~(\ref{damp_final}), which exhibits two terms related to the non-adiabatic part of the total pressure and the turbulent pressure. Since these correspond to a transfer of energy between the pulsation and convection, it can be assumed at first glance that the work integral varies dimensionally with the ratio $L/M$. Indeed, as verified by \cite{Belkacem2012}, one has 
\begin{equation}
\label{etaI}
\eta \, I \propto \left( \frac{L}{M} \right)^{2.7} \, .
\end{equation}
In contrast, mode inertia $I$ does not depend on the mode energy leakage but on the star's static structure, and more precisely on the properties of its uppermost layers. Hence, one can expect the mode inertia to scale with the surface gravity\footnote{Note that mode inertia also scales with the dynamical timescale $\sqrt(GM/R^3)$ with almost the same dispersion as for the surface gravity.}. More precisely, it has been shown in \cite{Belkacem2012} that 
\begin{equation}
  \label{Inertie}
  I \propto g^{-2.4} \, .
\end{equation}
Using \eq{etaI} and \eq{Inertie}, it finally turns out that 
\begin{equation}
  \eta \propto T_{\rm eff}^{10.8} \; g^{-0.3} \, .
\end{equation}
Such a crude analysis is unable to reproduce the precise shape of the mode line-width with effective temperature. However, it does allow us to explain qualitatively the strong dependence of mode damping rates on effective temperature. 

\subsection{The $\numax$--$\nuc$ relation}
\label{numax}

\begin{figure}[!]
\begin{center}
\includegraphics[height=11cm,width=9cm]{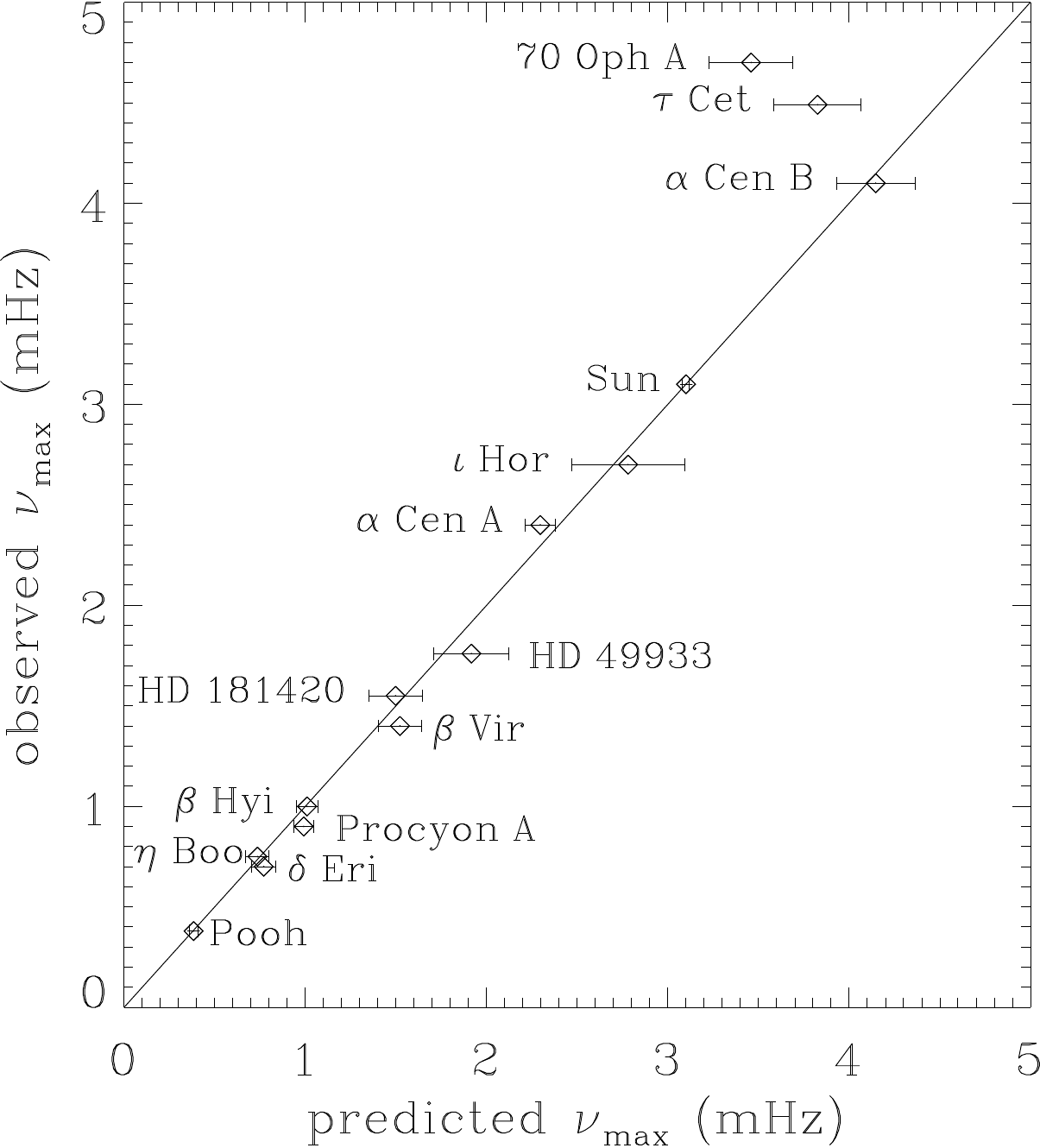}
\caption{Observed $\numax$ versus predicted $\numax$, computed using the ratio $g/\sqrt{T_{\rm eff}}$, for stars observed from the ground \citep[see][for details]{Bedding2011}. Figure from \cite{Bedding2011}.}
\label{fig_obs_bedding}
\end{center}
\end{figure}

The scaling relation that provides an estimate of the surface gravity derives from the proportionality between $\nu_{\rm max}$, which the the frequency at the maximum height in the oscillation power spectrum, and the cut-off frequency $\nu_{\rm c}$, which is the frequency beyond which there is no more reflection at the star surface (see Fig.~\ref{fig_obs_bedding}). In this section, we discuss the theoretical foundations of this scaling relation.  We follow the work of \cite{Belkacem2011}, who show that this relation can be explained by two intermediate relations, namely $\numax \propto \tau_{\rm th}^{-1}$ (where $ \tau_{\rm th}^{-1}$ is the thermal frequency, see Sect.~\ref{relation_nu_th} for a precise definition) and  $\tau_{\rm th}^{-1} \propto \nuc$, as displayed in Fig.~\ref{schema_numax}. 

\begin{figure}
\centerline{\includegraphics[height=6.cm,width=8cm]{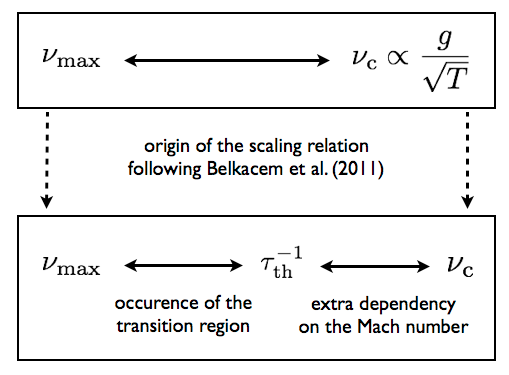}}
\caption{Sketch that illustrates the relation between $\numax$ and $\nuc$ (top panel). Following the work of \cite{Belkacem2011}, the bottom panel explicits the role of the thermal frequency. }
\label{schema_numax}
\end{figure}

\subsubsection{The transition region and the $\numax-\nu_{\rm th}$ relation}
\label{sect:transition}

The frequency $\numax$ is determined by the maximum height $H$ of the background-corrected power spectrum. For stochastically excited modes, the height of the mode profile in the power spectrum is given by \eq{H}. As shown for instance by \cite{Chaplin2008,Belkacem2011}, and confirmed with  observations of the solar-like stars by \emph{Kepler} \citep{Appourchaux12,Appourchaux14}, the maximum of $H$ is predominantly determined by the squared damping rates $(\Gamma^2)$ in Eq.~(\ref{H}). More precisely, $\numax$ arises from the depression (or plateau) of $\Gamma$ (or equivalently $\eta$). 

The depression of the damping rates occurs when the modal period nearly equals the thermal time-scale (or thermal adjustment time-scale) in the superadiabatic layers.  This was first mentioned by \cite{Balmforth92} (see his Sect. 7.2 and 7.3) and confirmed by \cite{Belkacem2011}, on the basis of two different non-adiabatic pulsation codes. In the context of classical pulsators, the layer in which this equality holds is referred to as the transition region and its occurrence in the ionization region is one of the necessary conditions for a mode to be excited by the $\kappa$-mechanism (see Sect.~\ref{self-excited} for details). In the context of solar-like pulsators, the situation is very similar, except that the destabilization by the perturbation of the opacity never dominates over damping terms \citep{Belkacem2012}. Moreover, the situation is complicated by the presence of convection which modifies the thermal time-scale \citep[see][for details]{Belkacem2011}. This is illustrated in Fig.~\ref{fig_transition}, which  displays the mode damping rates computed using the \cite{MAD05} formalism. It demonstrates that the perturbation of the opacity is the corner-stone of the relation between the modal period and the thermal time-scale. 

\begin{figure}
\centerline{\includegraphics[height=6cm,width=9cm]{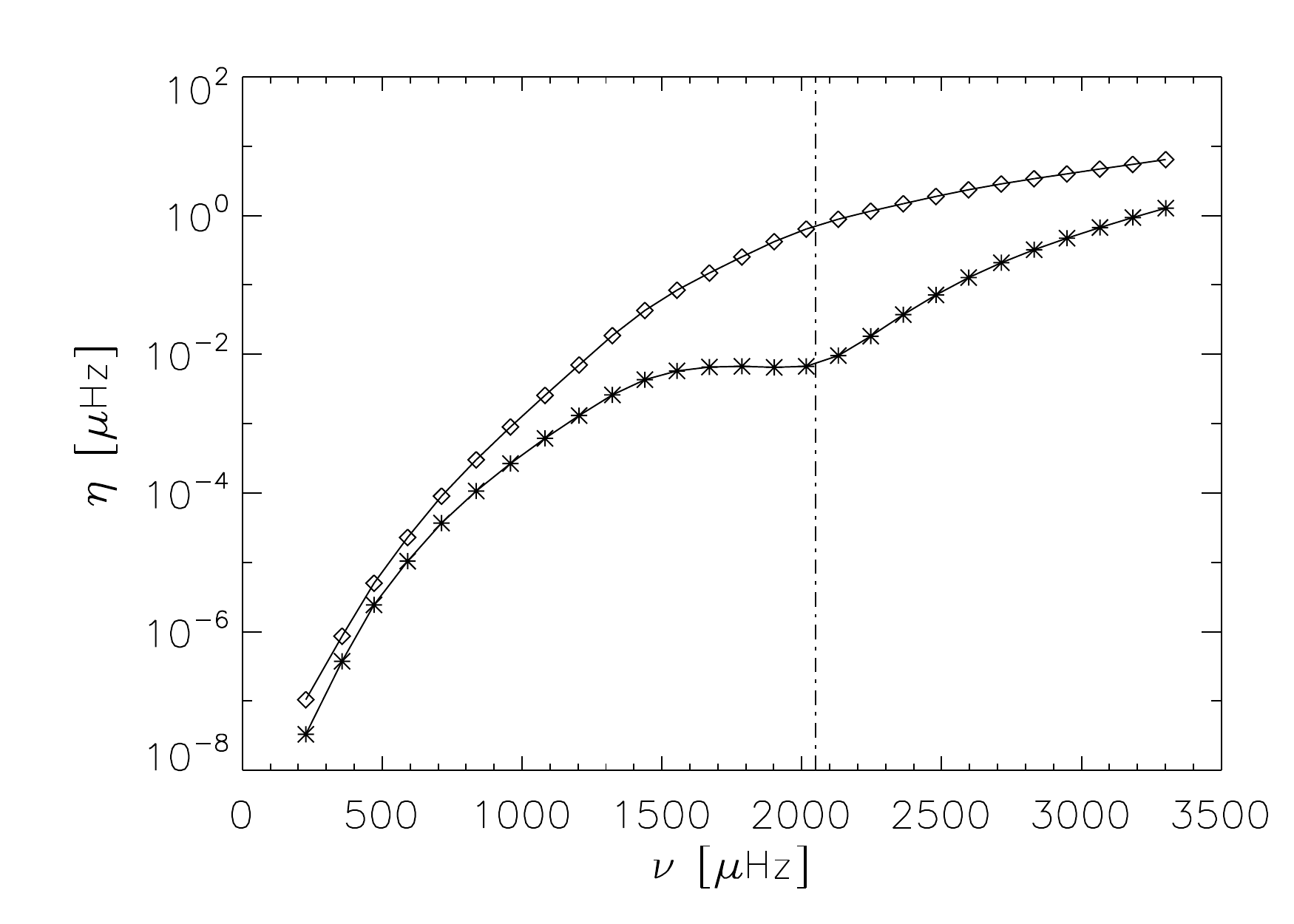}}
\caption{Mode damping rates versus mode frequency computed for a model of one solar mass on the main-sequence, using the \cite{MAD05} formalism as described in \cite{Belkacem2012}. The star symbols correspond to the full computation while the diamond symbols correspond to the computation for which we imposed $\delta \kappa / \kappa = 0$. 
The vertical dashed-dotted line corresponds to the frequency $\numax$ computed using the scaling relation.}
\label{fig_transition}
\end{figure}
 
\subsubsection{The $\nu_{\rm th}-\nuc$ relation}
\label{relation_nu_th}

As shown in the previous section, there is a linear relation between $\numax$ and the thermal frequency $\nu_{\rm th} \equiv \tau_{\rm th}^{-1}$. Let us now investigate the relation between $\nu_{\rm th}$ and $\nu_{\rm c}$. 

%The thermal adjustment time-scale has been extensively discussed by \cite{CoxGuili68,Cox80,Pesnell83}. It is defined as 
%\begin{eqnarray}
%\label{thermal_time}
%\tau_{\rm th} = \frac{1}{L} \int_{m_{\rm tr}}^{M} c_v T {\rm d}m
%\end{eqnarray}
%where $M$ is the total mass, $c_v$ is specific heat capacity at fixed volume, and $m_{\rm tr}$ is the mass at the transition region. 
%%Since our objective is to investigate the relation between $\numax$ and $\nuc$, it is not possible to use it to constrain $m_{\rm tr}$. Therefore, we adopt an alternative approach. 
%As already explained in Sect.~\ref{sect:transition}, the relation between $\numax$ and $\nu_{\rm th}$ holds due to the occurrence of the transition region in the ionization region. 

%\begin{figure}
%%\centerline{\includegraphics[height=7cm,width=9cm]{plot_Kepler_Corot}}
%\caption{$\numax$ as a function of the effective temperature ($T_{\rm eff}$). The filled red squares corresponds to the location of the 3D hydrodynamical models (see text for details), the filled blue circles to sub-giant and main-sequence targets observed by \emph{Kepler} \citep{Chaplin2011}, and the black ones to the red-giant stars observed by \emph{Kepler} \citep{Mathur2011}. Finally, the green square corresponds to the solar 3D model. }
%\label{fig_simu3D_HR}
%\end{figure}

The thermal adjustment time-scale has been extensively discussed in Sect.~\ref{timescales} and is given by \eq{tau_th4}. For the cut-off frequency, a general expression has been proposed by \cite{Balmforth90} 
\begin{equation}
\omega_c = 2\pi \nuc = \left( \frac{c_s}{2 H_\rho} \right) \sqrt{1-2 \deriv{H_\rho}{r}}
\end{equation}
with $c_s$ the sound speed, and $H_\rho = -({\rm d}\ln \rho / {\rm d}r)^{-1}$ the density scale height. For an isothermal atmosphere, $\omega_c$ reduces to 
\begin{equation}
\omega_c = \frac{c_s}{2 H_\rho} \, .
\end{equation}
To go further it is customary to use the pressure scale height $H_p$ as a proxy for the density scale height. This is based on the fact that both quantities are nearly equal at the photosphere (\emph{i.e.}, $H_p = H_\rho$). Finally, it can be shown that $H_p$ scales as the ratio between the surface gravity and the square root of the effective temperature. 
%At this step it is necessary to realize that the relation between $\numax$ and $\nuc$ is more artificial than physically grounded.  Indeed, \cite{Brown91} conjectured a relation between $\numax$ and $\nuc$ but from a physical point of view it would be more rigorous to mention the relation between $\numax$ and the ratio $g/\sqrt{T_{\rm eff}}$ or the thermal time-scale $\nu_{\rm th}$. 
Consequently, the cut-off frequency can be recast such as 
\begin{equation}
\omega_c = \frac{c_s}{2 H_p} \propto \frac{g}{\sqrt{T_{\rm eff}}} \, , 
\end{equation}
where we considered all the quantities at the photosphere, and we made use of the scaling relations $c_s^2 \propto T_{\rm eff}$, $H_p \propto T_{\rm eff} / g$. 

%\begin{figure}
%%\centerline{\includegraphics[height=7cm,width=9cm]{plot_nuth_nuc}}
%%\centerline{\includegraphics[height=7cm,width=9cm]{correction_mach}}
%\caption{\emph{Top panel:}  Thermal frequency $\tau_{\rm th}^{-1}$ as a function of the cut-off frequency $\nuc$. All the quantities are normalized by the values derived from the Solar 3D simulation. The filled squared correspond to the 3D models displayed in Fig.~\ref{fig_simu3D_HR}. The dashed-dotted line corresponds to the linear curve. \emph{Bottom panel:} As for the top panel, except that the thermal frequency is corrected by the term $\mathcal{M}_a^\alpha$ with $\alpha = 2.78$ (see \eq{numax_mach}).}
%\label{fig_simu3D_results}
%\end{figure}

To go further, one has to exhibit the relation between the thermal frequency and the cut-off frequency. To this end, using the mixing length formalism, it is possible to obtain \citep{Belkacem2011} 
%used the mixing-length formalism to derive a linear relation between $\nu_{\rm th}$ and $\nu_{\rm c}$. It gives
\begin{align}
\label{thermal_final}
\nu_{\rm th}  \propto \mathcal{M}^3 \, \nu_{\rm c} \, , 
\end{align}
where $\mathcal{M}$ is the Mach number. This result has been confirmed by \cite{Belkacem2013} using a set of 3D numerical simulation from the CIFIST grid \citep{Ludwig09b}. More precisely, it was found that 
\begin{equation}
\label{numax_mach}
\tau_{\rm th}^{-1} \propto \mathcal{M}^{2.78} \; \nu_{\rm c} \, . 
\end{equation}
To conclude, a crude and simple analytical approach based on the mixing length and more realistic 3D models give approximatively the same results.  The value of the exponent is quite a robust number since the dependence to the Mach number can be derived from simple energetical arguments that hardly depend on the assumptions related to the MLT.

\subsubsection{Effect of the Mach number on the scaling for stars on the red giant branch}

Figure~\ref{fig_nuth_nuc} illustrates the relation between $\nu_{\rm th}$ and $\nuc$ with a set of 1D stellar models. One can distinguish two regimes. First, for the main-sequence and subgiants there is  a relatively important dispersion related to the Mach number \citep{Belkacem2011,Belkacem2013}. Second, for the red-giant stars the dispersion is significantly reduced \citep[note that the same result is found from 3D models, see][]{Belkacem2013}. This dispersion is important because it translates into accuracy regarding the $\numax - \nuc$ relation. Indeed, from the effect of the Mach number on the scaling relation, one can deduce that the  $\numax - \nuc$ will be accurate for red-giants (as it will be shown below) but that important biases are to be expected for main-sequence stars. 

\begin{figure}[!]
\begin{center}
\includegraphics[width=10cm]{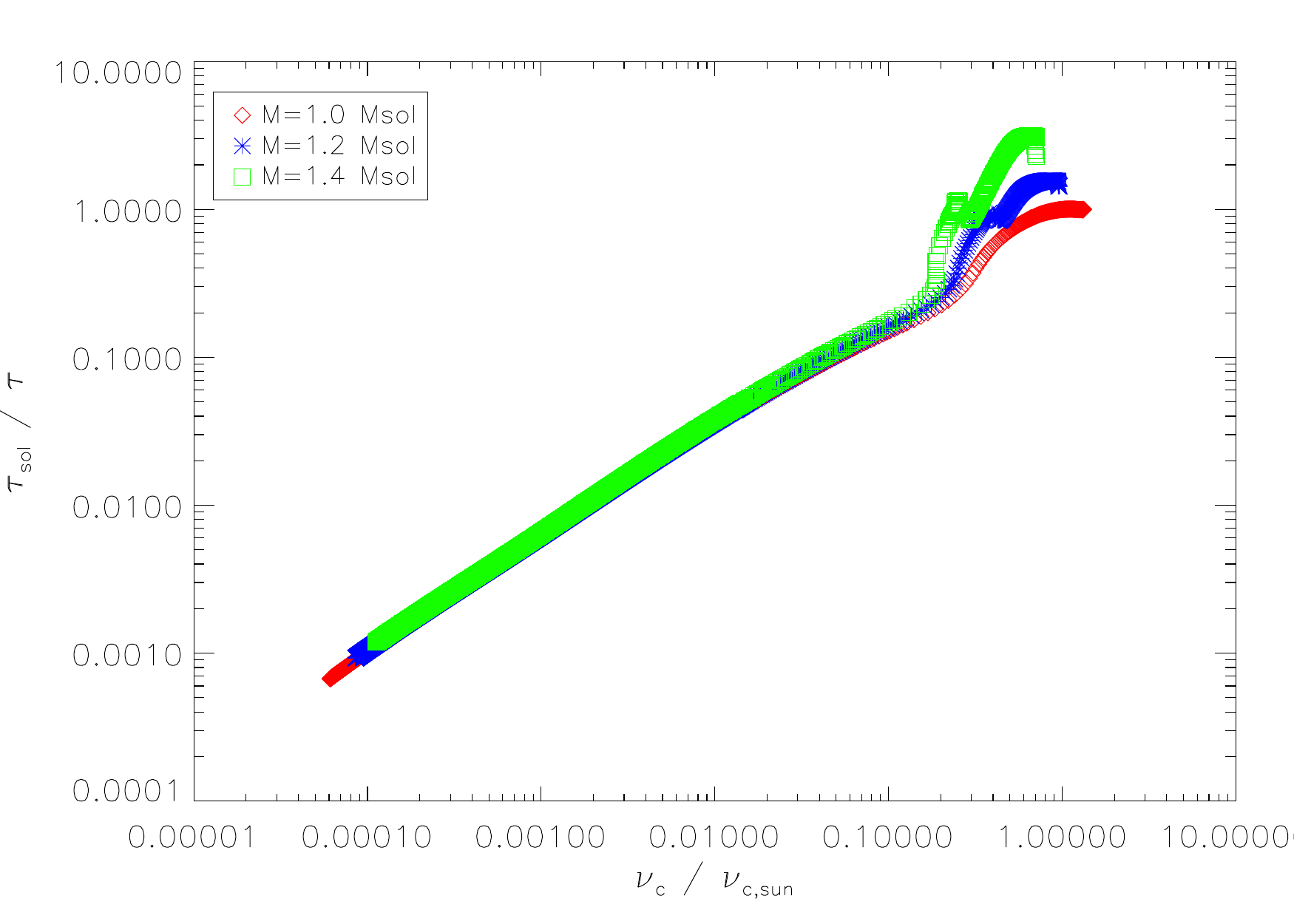}
\caption{Inverse of the thermal time-scale  ($1/\tau_{\rm th}$) versus the cut-off frequency (computed as the ratio $c_s/(2 Hp)$), normalized to the solar values, for models with masses from $1.0 \,M_\odot$ to $1.4 \,M_\odot$ from the ZAMS to the ascending vertical  branch. The inputs physics of the models can be found in \cite{Belkacem2011}. }
\label{fig_nuth_nuc}
\end{center}
\end{figure}

In the following, we explain qualitatively why the effect of the Mach number is negligible on the $\numax-\nuc$ relation for red-giant stars. To this end, we will first make several assumptions that will enable us to derive a scaling between the Mach number and stellar global parameters.  First we assume that the total flux is convective and proportional to the kinetic energy flux,  so that 
\begin{equation}
\label{scaling_mach_tmp}
\mathcal{M}_a \propto T_{\rm eff}^{5/6} \, \rho^{-1/3} \, , 
\end{equation}
where $\rho$ is the surface density. The latter is treated as in Sect.~\ref{amortissement_turbulent} so that \eq{scaling_mach_tmp} becomes
\begin{equation}
\label{scaling_mach}
\mathcal{M}_a \propto T_{\rm eff}^{3} \, g^{-2/9} \, , 
\end{equation}
where $g$ is the surface gravity. 

There is an additional relation between the surface gravity and the effective temperature of stars on the red giant branch. It reads
\begin{equation}
\label{Teff-g}
T_{\rm eff} \propto g^{0.07} \, .
\end{equation}
This is in very good agreement with the observations that give $T_{\rm eff} \propto \numax^{0.068}$ \citep{Mosser2013}. 
%Equation (\ref{Teff-g}) shows that the behavior of evolving red giants on the HR diagram are mainly dominated by the super-adiabatic convective layers near the surface. This is a classical result of stellar evolution that can be understood by considering the limiting case of fully convective stars \citep[{\it e.g.}][]{CoxGuili68} . 

It is then possible to understand why the $\numax-\nuc$ relation hardly depends on the Mach number on the red giant branch. Indeed, the Mach number becomes nearly  independent of both effective temperature and surface gravity. If we introduce \eq{scaling_mach} and \eq{Teff-g} into \eq{numax_mach}, we obtain
\begin{equation}
\nu_{\rm max} \propto \mathcal{M}_a^3 \, \nu_{\rm c} \propto  \frac{g^{0.988}}{\sqrt{T_{\rm eff}}} \approx {\rm cste} \;\times  \nu_{\rm c} \, .
\end{equation}
This result shows that for red-giant stars near the tip of the branch, the effect of the Mach number becomes negligible. In other words, one can conclude that the $\numax - \nuc$ relation is more accurate for red giants since the possible influence introduced by the Mach number becomes small. 

\subsection{Scaling relations for mode amplitudes}

In this section, we consider the relations between mode amplitudes, in terms of both velocity and intensity fluctuations, which, as we will highlight, provide information on turbulent convection, mode physics and stellar structure. 

\subsubsection{Theoretical scaling relation in terms of mode surface velocity}

On the basis of the theoretical calculations of \cite{JCD83b}, \cite{Kjeldsen95} have derived the first example of a scaling relation given in terms of the maximum of the mode surface velocity (hereafter $\vmax$). This scaling relation predicts that $\vmax$ varies as the ratio $(L/M)^{s}$ with a slope $s \simeq 1$. The theoretical calculations of \cite{JCD83b} were based on the assumption that there is an \emph{equipartition} between the energy carried by the most energetic eddies and the modes. As mentioned by \cite{Belkacem09} and \cite{Samadi11}, a necessary (but not sufficient) condition for having such equipartition is that turbulent viscosity is the dominant source of damping. However, there is currently no consensus as to the dominant physical processes contributing to the damping of $p$-modes and, furthermore, this assumption is not supported by observations (see Sect.~\ref{damping}). Following the scaling proposed by \cite{Kjeldsen95}, other theoretical scaling relations have been proposed and compared with ground-based Doppler measurements.  We present here the latest such proposals by \cite{Samadi07a} and \cite{Samadi12} and compare them with observational data.

We recall that the mean-squared surface velocity of a mode is given by \citep{Samadi11}
\begin{equation}
\label{v}
{\rm v}^2 (\nu_{\rm osc},r) = {\tau(\nu_{\rm osc}) \over 2} \, {{ {\cal P}(\nu_{\rm osc}) } \over { \mathcal{M}(\nu_{\rm osc,r }) }}
\end{equation}
where $\nu_{\rm osc} = \omega_{\rm osc}/2\pi$, $\tau$ is the mode life-time (which is equal to the inverse of the mode damping rate $\eta$),  $r$ is the radius in the atmosphere in which the mode velocity is measured, and $\mathcal{M}$ is the mode mass, which is defined for radial modes by the ratio  $ I / { \vert \xi_{\rm r} \vert^2 }$. Note that the mode mass must, in principle, be evaluated at the layer of the atmosphere where spectrographs dedicated to stellar seismology are the most sensitive. However, this layer is not well known \citep[see for a discussion][]{Samadi08}. Hence, for  sake of simplicity, $\cal M$ is evaluated at the photosphere (\ie, $T=\teff$).

For the Sun, the frequency $\numax$ at which $v$ reaches a maximum is shown to coincide with the frequency location of the plateau in the mode life-time  $\tau$ \citep[see][]{Belkacem2011}. It was also found by \cite{Samadi12} that $\cal P$, as well as the ratio $\cal P/ \cal M$, peaks at this frequency. As a consequence, the existence of a scaling relation for $\vmax$  simply relies on the existence of a scaling relation for $\tau_{\rm max}$, $\pmax$ and ${\cal M}_{\rm max}$ where $\tau_{\rm max}$, $\pmax$, and ${\cal M}_{\rm max}$ are respectively the values of $\tau$, $\cal P$, and $\cal M$ at $\nu = \numax$. Scaling relations for $\tau_{\rm max}$ (\ie, $1/\eta$) have been presented and discussed in Sect.~\ref{eta_scaling}. 

So, to derive a scaling relation for $v_{\rm max}$, the first step is to determine a relation between  $\pmax$ and ${\cal M}_{\rm max}$. \cite{Samadi07a} have established, on the basis of a small set of 3D models of the surface layers of main-sequence (MS) stars, that $\pmax$ scales as $\left (L /M \right )^{s}$, where the slope $s$ is found to depend significantly on the adopted prescription for the frequency factor $\chi_k(\omega)$ (see the definition given in Sect.~\ref{driving}). A Lorentzian $\chi_k(\omega)$ is the more realistic choice \citep{Samadi02II,Belkacem2010} and results in a slope $s=2.6$. More recently, this study was extended by \cite{Samadi12} to the case of sub- and red-giant stars. The authors found for $\pmax$ the same scaling relation as the one found for MS. 

The dependence of $\pmax$ on $L$ and $M$ can be  explained on the basis of simple theoretical considerations \citep{Samadi11}. We first point out that the ratio $L/M$ is equivalent to the ratio $\teff^4/g$.  In turn, the dependence on $\teff$ and $g$ can be roughly explained as follows. Starting from Eq.~(\ref{P_5}), also assuming a propagating wave, it is shown that $ \left ({\rm d} \xi_{\rm r} / {\rm d}r \right ) ^2 = \omega_{\rm osc}^2 / c_s^2 \, \xi_{\rm r}^2$ where $c_s$ is the sound speed. Accordingly, Eq.~(\ref{P_5}) simplifies to
\eqna{
 {\cal P} & \propto & { \nu_{\rm osc}^2 \over {I} } \, \int \, \left ( { \xi_{\rm r}
   \over c_s } \right )^2  \, F_{\rm kin} \, \Lambda^4  \, {\rm d} m \; . 
\label{P_6}
}
The integrand of Eq.~(\ref{P_6})  is  evaluated --- for  the sake of simplicity ---  at a single layer of the surface where mode driving is predominant. This layer being close to the photosphere, the term $\Lambda^4 \,  F_{\rm kin}$ that appears in the integrand of Eq.~(\ref{P_6}) is evaluated at the photosphere (\ie, at $T=\teff$). This yields
\eqna{
 {\cal P} & \propto & F_{\rm kin} \, \Lambda^4 \, { \left ( \nu_{\rm osc} \over {c_s} \right )^2 } \, {1 \over I} \,  \int {\rm d} m \; \xi_{\rm r} ^2 \; .
\label{P_7}
}
Finally, with the help of the definition of the inertia (Eq.~\ref{inertia}), the above relation reduces to 
\eqna{
{\cal P} & \propto & F_{\rm kin} \, \Lambda^4 \, { \left ( \nu_{\rm osc} \over {c_s} \right )^2 } \; .
\label{P_8}
}
As already noticed in Sect.~\ref{damping}, $F_{\rm kin}$ scales approximately with the convective flux $F_c$, which is proportional to $\teff^4$ (in the upper part of the convective zone). We recall that $\numax$ is shown to scale as $g / \teff^{1/2}$ and $c_s$ as $\teff^{1/2}$ (see Sect.~\ref{numax}).  Finally, the characteristic size $\Lambda$ scales in turn as $\teff/g$. When we combine all these scaling relations in Eq.~(\ref{P_8}), we establish  that $\pmax$ scales approximately as $\teff ^6 \, g^{-2}$. This crude result then qualitatively explains the theoretical scaling laws found for $\pmax$ by \cite{Samadi07a} and \cite{Samadi12}. 

We now turn to ${\cal M}_{\rm max}$.  In \cite{Samadi12} it is found that for sub- and red-giant stars ${\cal M}_{\rm max}$ scales as $ (M/R^3)^{-p/2.}$ with $p = 2.0$.  Accordingly, ${\cal M}_{\rm max}$ scales  as the inverse of the star mean density, \ie, $\langle \rho \rangle \propto (M/R^3)$.  For MS stars, the calculations performed by \cite{Samadi07a} yield a different slope, $p = 1.3$. These two different relations between ${\cal M}_{\rm max}$  and $\langle \rho \rangle$ are not yet understood and call for some theoretical support. When we combine the scaling relations for $\tau_{\rm max}$ (see Sect.~\ref{damping}), $\pmax$ and ${\cal M}_{\rm max}$ into Eq.~(\ref{v}), we obtain for $\vmax$ the following relation:
\eqn{
{\rm v}_{\rm max} \propto   \teff  ^{-5.4} \, g^{0.15}   \,  \left ( {L} \over {M} \right )^{1.3} \, \left ( M \over R^3 \right )^{p/4} \; .
\label{vmax}
}
Given the fact that the large separation $\Delta \nu$ typically scales as $ \left ( M / R^3 \right )^{1/2}$ \citep[{e.g.},][]{White2011}, $\numax$ scales as $g/\teff^{1/2}$, and since $\left ( {L} / {M} \right )$ is proportional to  $ { {\teff^{7/2}} / {\nu_{\rm max}} }$ \citep{Baudin11,Baudin11b}, Eq.~(\ref{vmax}) can be reformulated to include only the seismic indices $\numax$ and $\Delta \nu$ and $\teff$ yielding to
\eqn{
{\rm v}_{\rm max} \propto   \teff  ^{-0.77}    \,  \numax^{-1.15} \,  \Delta \nu^{p/2}\; .
\label{vmax_2}
}
The scaling relation given by Eq.~(\ref{vmax_2}) is compared in Fig.~\ref{fig:vmax} with the ground-based Doppler velocity measurements obtained to the present. The amplitudes of the solar-like oscillations measured in MS are rather well reproduced by the theoretical scaling relation. This is not the case for the sub- and red-giant stars for which the predictions are found to be systematically below the observations. Such a discrepancy is mainly attributed to non-adiabatic effects \cite[see][ for a detailed discussion]{Samadi12,Samadi13c}.

\begin{figure}
\begin{center}
\includegraphics[width=10cm,angle=0]{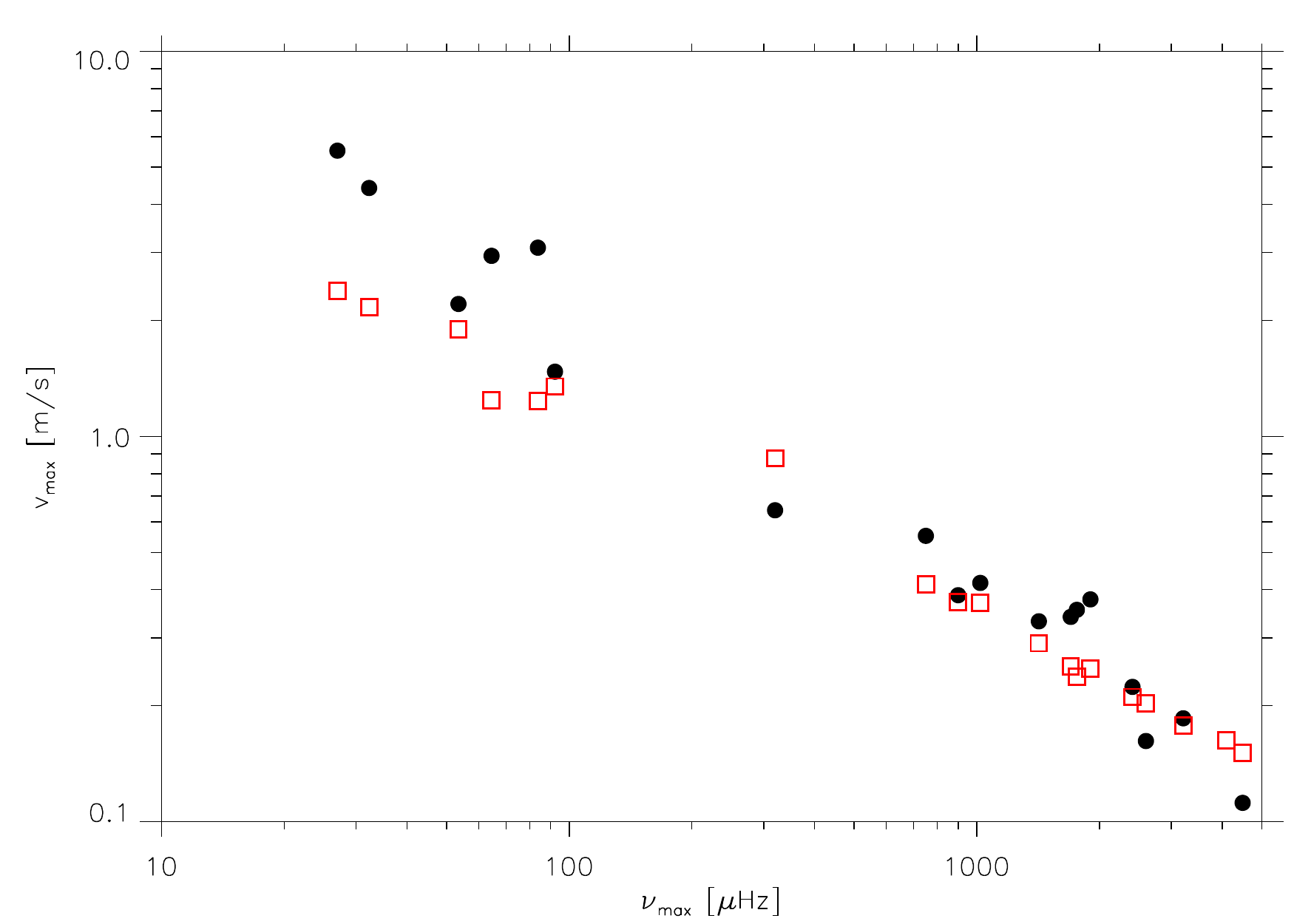}
\caption{Maximum of the mode velocity  $v_{\rm max}$ as a function of $\nu_{\rm max}$. The filled circles correspond to observations, while the open red squares correspond to the scaling relation given by Eq.~(\ref{vmax_2}). For sub- and red-giants ($\numax \lesssim 200\,\mu$Hz), we have taken $p=2.0$ and for MS stars ($\numax \gtrsim 200\,\mu$Hz) $p=1.3$.}
\label{fig:vmax}
\end{center}
\end{figure}

\subsubsection{From velocity to bolometric amplitudes} 

Current space-based missions detect and measure solar-like oscillations in numerous stars, using methods of high-precision photometry. Therefore, in order to compare predicted with measured mode amplitudes, it is necessary to convert mode-velocity amplitudes to intensity amplitudes.  

The instantaneous bolometric mode amplitude is deduced at the photosphere according to %  \cite{Dziembowski77a,Pesnell90}   
\begin{equation}
  \frac{\delta L (t)}{L}  = 4\, \frac{\delta \teff(t)}{\teff} 
  + 2\, \frac{\delta R_* (t)}{R_*} \, , 
  \label{dL}
\end{equation}
where $\delta L(t)$ is the mode Lagrangian (bolometric) luminosity perturbation, $\delta \teff(t)$ the  effective temperature fluctuation, and $\delta R_*(t)$ the variation of the stellar radius. For solar-like oscillations, the second term of \eq{dL} is negligible compared to $\delta \teff(t)$, so the rms bolometric  amplitudes are given by 
\begin{equation} 
\left(  {{\delta L} \over {L }} \right )_{\rm rms}  =   4  \left (\displaystyle \frac{\delta \teff}{\teff} \right )_{\rm  rms} 
\label{dLrms0}
\end{equation}
where the subscript rms denotes the root mean-square. 

We need a relationship between  $\left(  {{\delta \teff }  / \teff } \right )_{\rm rms}$ (or equivalently $\dLrms$) and the rms mode velocity ${\rm v}_{\rm rms}$. To this end, we introduce the dimensionless coefficient $\zeta$ defined by 
\eqn{
\dLrms =  4  \left (\displaystyle \frac{\delta \teff}{\teff} \right )_{\rm
  rms} =  \zeta \,  \dLrms^\odot \, \left ( { {\rm v}_{\rm rms} \over  {\rm v}_\odot }   \right )\, , 
\label{dLrms}
}
where  $\dLrms^\odot = 2.53$~$\pm$0.11~ppm is the maximum of the solar bolometric mode amplitude  \citep{Michel09}, $\teff^\odot=5777$~K the effective temperature of the Sun, and  $v^\odot_{\rm rms}=18.5~\pm~1.5$~cm/s is the maximum of the solar mode (intrinsic) surface velocity evaluated at the photosphere as explained in \cite{Samadi09b}.

Let us now define $\dLmax$ to be the maximum of $(\delta L / L)_{\rm rms}$. We want to establish a scaling for $\dLmax$. As seen in Eq.~(\ref{dLrms}), this requires a scaling relation for $\zeta$ since that for ${\rm v}_{\rm rms}$ is given by Eq.~(\ref{vmax}) (or equivalently by \eq{vmax_2}). Consistent calculation of $\zeta$ requires us to take into account the energy lost by the pulsation. This can be estimated using a non-adiabatic pulsation code that takes  into account coupling between oscillation, radiation and turbulent convection (as described in Sect.~\ref{damping}). Due to the difficulties of consistently treating the underlying mechanisms, the use of the quasi-adiabatic relation has been proposed by \cite{Kjeldsen95} and is adopted for converting mode surface velocity into intensity amplitude. 
Indeed, adopting quasi-adiabatic pulsation and assuming an isothermal atmosphere\footnote{A more sophisticated quasi-adiabatic approach has been proposed by \cite{Severino08}. These authors go beyond the approximation of isothermal atmosphere by taking into account the temperature gradient as well as the fact that the intensity is measured at constant instantaneous optical depth. Both effects are  taken into account by the non-adiabatic pulsation code MAD.}, one can easily relate mode surface velocity to intensity perturbations \citep[{e.g.},][]{Kjeldsen95}. These approximations yield the following simple expression for $\zeta$ \citep{Kjeldsen95}
\eqn{
\zeta_{\rm K95} = \sqrt{ \teff^\odot  \over   \teff }  
\label{zeta_K95}
}
Derivation of Eq.~(\ref{zeta_K95}) supposes that the modes propagate at the surface where they are measured. However, the acoustic modes are evanescent at the surface. Combining Eq.~(\ref{zeta_K95}) and \eq{dLrms} gives for $\dLmax$  the (quasi)-adiabatic scaling relation 
\eqn{
\dLmax \propto  \teff  ^{-0.5}  \, \vmax  \, , 
\label{dLmax_ad}
}
where the scaling relation of $\vmax$ is given by Eq.~(\ref{vmax_2}).

Bolometric mode amplitudes computed on the basis of Eq.~(\ref{dLmax_ad}) are compared with the bolometric mode amplitudes measured by \cite{Baudin11,Baudin11b} on a set of CoRoT red-giant stars in Fig.~\ref{fig:dLmax}. The adiabatic relation of Eq.~(\ref{dLmax_ad}) results, for red-giant stars, in a significant under-estimation compared to the CoRoT seismic data. 

\begin{figure}
\begin{center}
\includegraphics[width=10cm,angle=0]{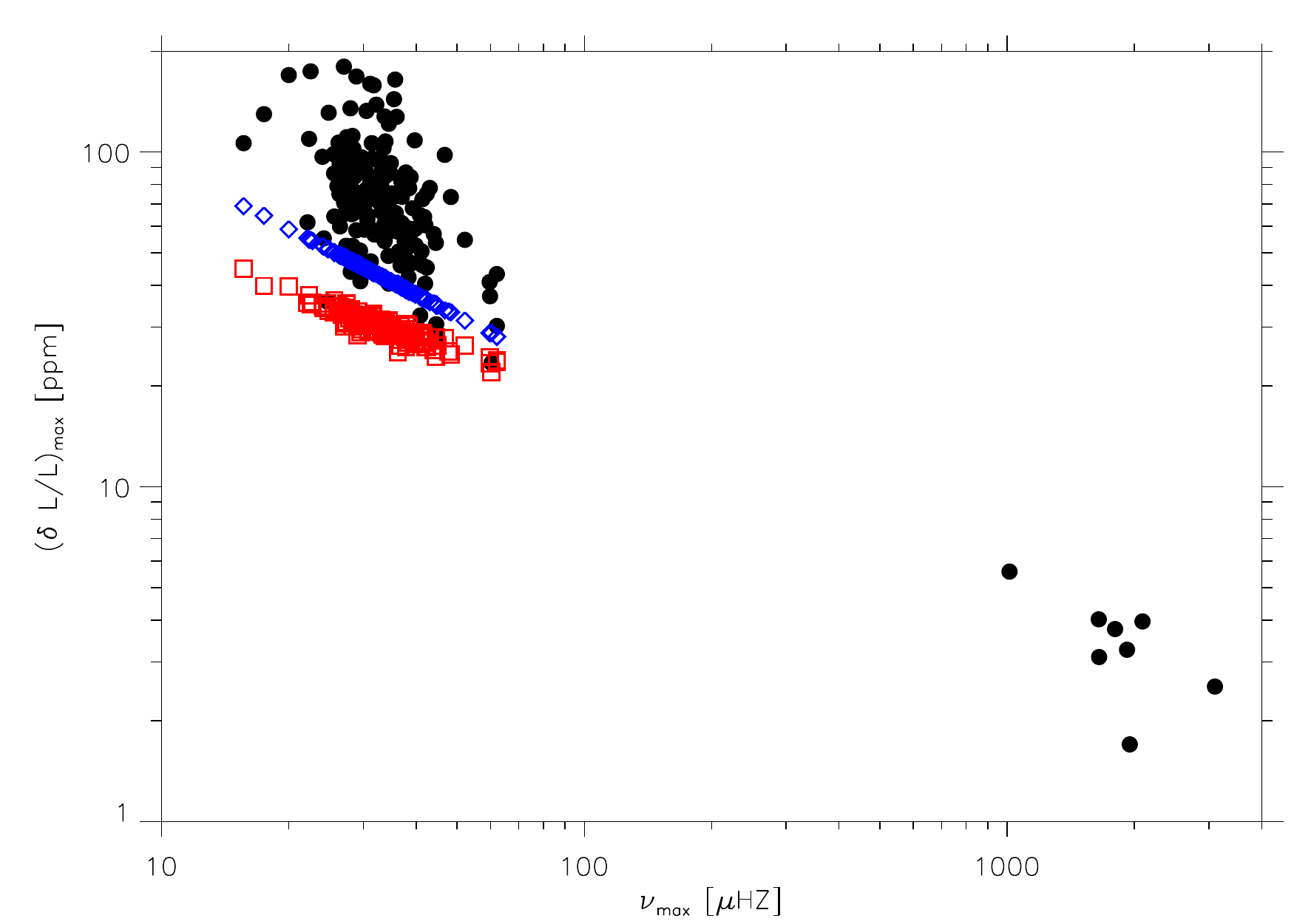}
\caption{Maximum of the mode intensity fluctuation $(\delta L/L)_{\rm max}$ as a function of $\numax$. The filled circles located below $\numax = 200~\mu$Hz correspond to the seismic measures on a large number of CoRoT red-giant stars while those located above  $\numax = 200~\mu$Hz  correspond to the MS stars observed so far by CoRoT \citep{Baudin11,Baudin11b}. The red squares are the theoretical amplitudes obtained with the adiabatic scaling law of Eq.~(\ref{dLmax_ad}) while the blue diamonds to those computed on the basis of the non-adiabatic  scaling law given by Eq.~(\ref{dLmax_nad}). }
\label{fig:dLmax}
\end{center}
\end{figure}

\cite{Samadi12} have computed the coefficient $\zeta$ (see Eq.~(\ref{dLrms})) using the  MAD non-adiabatic pulsation code \citep{Grigahcene05} and established that, for sub- and red-giants, $\zeta$ scales as 
\eqn{
\zeta_{\rm nad} = \zeta_0 \, \left ( {L \over L_\odot} \, { M_\odot \over M} \right )^{0.25} \; ,
\label{zeta_nad}
}
where $\zeta_0= 0.59$. The increase of $\zeta$ with the ratio $L/M$ is not surprising. Indeed, energy losses scale dimensionally as $L/M$. Red-giants stars are characterised by high luminosities. As a consequence we expect, for red-giants, large differences between $\zeta_{\rm nad}$ and $\zeta_{K95}$. 
We recall that $L/M$ is proportional to $\teff^{7/2}/\numax$. Accordingly, substituting Eq.~(\ref{zeta_nad}) into  Eq.~(\ref{dLrms}) yields,  for $\dLmax$, the non-adiabatic scaling relation\footnote{This scaling is only valid for sub- and red-giant stars.}
\eqn{
\dLmax \propto  \teff  ^{1.75}  \, \numax^{-0.25} \,  \vmax
\label{dLmax_nad}
}
where  $\vmax$ is given by Eq.~(\ref{vmax_2}) with $p=2$. This scaling relation is compared in Fig.~\ref{fig:dLmax} with CoRoT observations. As seen in the figure, the differences that were found between the adiabatic scaling relation and the CoRoT observations are reduced using the non-adiabatic scaling relation. However, the remaining differences are still important. Their possible origins are discussed in \cite{Samadi12} and \citet{Samadi13c}. 

%%-----------------------------
%%      your bibliography
%%-----------------------------
%\begin{thebibliography}{99}
%\end{thebibliography}

\bibliographystyle{astron}
%\bibliography{biblio.bib}

\end{document}